\pdfoutput=1

\documentclass[11pt,twoside,a4paper,cmspaper,final,collab]{cms-tdr}
\def\svnVersion{9df661a-D}\def\svnDate{2020/09/17}\def\cmsCernNoTag{CERN-EP-2020-017}\def\cmsCernDate{\today}\def\cmsMessage{"Published in the Journal of Instrumentation as \href{http://dx.doi.org/10.1088/1748-0221/15/09/P09018}{\doi{10.1088/1748-0221/15/09/P09018}.}"}

\begin{document}\cmsNoteHeader{JME-18-001}

\hyphenation{had-ron-i-za-tion}
\hyphenation{cal-or-i-me-ter}
\hyphenation{de-vices}
\newcommand{\upar}{\ensuremath{u_{||}\xspace}}
\newcommand{\uper}{\ensuremath{u_{\perp}\xspace}}
\newcommand{\Zmumu}{\ensuremath{\PZ \to \PGm \PGm}\xspace}
\newcommand{\Zee}{\ensuremath{\PZ \to \Pe\Pe}\xspace}
\newcommand{\LV}{\ensuremath{\text{LV}\xspace}}
\newcommand{\PU}{\ensuremath{\text{PU}\xspace}}
\newcommand{\RMS}{\ensuremath{\text{RMS}\xspace}}
\newcommand{\deltabeta}{\ensuremath{\delta\beta}\xspace}
\newcommand{\deltabetacorrected}{\ensuremath{\delta\beta\text{-corrected}}\xspace}
\newcommand{\mb}{\unit{mb}}
\newcommand{\vptmiss}{\ensuremath{\vec{p}_{T}^{\text{miss}}}\xspace}

\newcommand{\secref}[1]{Section~\ref{#1}}
\newcommand{\figref}[1]{Fig.~\ref{#1}}
\newcommand{\figrefb}[1]{Figure~\ref{#1}}

\cmsNoteHeader{JME-18-001} 
\title{Pileup mitigation at CMS in 13\TeV data}

\date{\today}

\abstract{
With increasing instantaneous luminosity at the LHC come additional reconstruction challenges. At high luminosity, many collisions occur simultaneously within one proton-proton bunch crossing. The isolation of an interesting collision from the additional ``pileup'' collisions is needed for effective physics performance. In the CMS Collaboration, several techniques capable of mitigating the impact of these pileup collisions have been developed. Such methods include charged-hadron subtraction, pileup jet identification, isospin-based neutral particle ``$\delta\beta$'' correction, and, most recently, pileup per particle identification. This paper surveys the performance of these techniques for jet and missing transverse momentum reconstruction, as well as muon isolation. 
The analysis makes use of data corresponding to 35.9\fbinv collected with the CMS experiment in 2016 at a center-of-mass energy of 13\TeV. The performance of each algorithm is discussed for up to 70 simultaneous collisions per bunch crossing. Significant improvements are found in the identification of pileup jets, the jet energy, mass, and angular resolution, missing transverse momentum resolution, and muon isolation when using pileup per particle identification. 
}

\hypersetup{
pdfauthor={CMS Collaboration},
pdftitle={Pileup mitigation at CMS in 13 TeV data},
pdfsubject={CMS},
pdfkeywords={CMS, physics, pileup, jets, missing transverse momentum}}

\maketitle

\section{Introduction}
At the CERN LHC, instantaneous luminosities of up to $1.5 \times 10^{34}\cm^{-2}\unit{s}^{-1}$~\cite{CMS-PAS-LUM-17-001} are sufficiently large for multiple proton-proton ($\Pp\Pp${}) collisions to occur in the same time window in which proton bunches collide. This leads to overlapping of
particle interactions in the detector. To study a specific $\Pp\Pp$ interaction, it is necessary to separate this single interaction from the overlapping ones. The additional collisions, known as pileup (PU), will result in additional particles throughout the detector that confuse the desired measurements. With PU mitigation techniques, we can minimize the impact of PU and better isolate the single collision of interest.
With increasing beam intensity over the past several years,  identification of interesting $\Pp\Pp$ collisions has become an ever-growing challenge at the LHC. The number of additional collisions that occur when two proton bunches collide was, on average, 23 in 2016 and subsequently increased to 32 in 2017 and 2018. At this level of collision density, the mitigation of the PU effects is necessary to enable physics analyses at the LHC.

The CMS Collaboration has developed various widely used techniques for PU mitigation.
One technique, charged-hadron subtraction (CHS)~\cite{Sirunyan:2017ulk}, has been the standard method to mitigate the impact of PU on the jet reconstruction for the last few years.
It works by excluding charged particles associated with reconstructed vertices from PU collisions from the jet clustering procedure.
In this technique, to mitigate the impact of neutral PU particles in jets, an event-by-event jet-area-based correction~\cite{jetarea_fastjet,jetarea_fastjet_pu,Khachatryan:2016kdb} is applied to the jet four-momenta.
Further, a PU jet identification (PU jet ID) technique~\cite{CMS-PAS-JME-16-003} is used to reject jets largely composed of particles from PU interactions.

These techniques have limitations when attempting to remove PU contributions due to neutral particles.
For the jet-area-based correction, the jet four-momentum correction acts on a whole jet and is therefore not capable of removing PU contributions from jet shape or jet substructure observables. 
To overcome this limitation, a new technique for PU mitigation, pileup per particle identification (PUPPI)~\cite{Bertolini:2014bba}, is introduced that operates at the particle level.
The PUPPI algorithm builds on the existing CHS algorithm. In addition, it calculates a probability that each neutral particle originates from PU and scales the energy of these particles based on their probability.
As a consequence, objects clustered from hadrons, such as jets, missing transverse momentum (\ptmiss), and lepton isolation are expected to be less susceptible to PU when PUPPI is utilized.

In this paper, the performance of PU mitigation techniques, including the commissioning of PUPPI in $\Pp\Pp$ collision data, is summarized.
After a short description of the CMS detector in \secref{sec_CMSgeneraldescription} and 
definitions of the data set and Monte Carlo (MC) simulations used in these studies in \secref{sec_environment}, 
the CHS and PUPPI algorithms are described in \secref{sec_puppibasic}.
In \secref{sec_jet} performance in terms of
 jet resolution at a high number of interactions is presented.
\secref{sec_noiseid} summarizes the impact on noise rejection of PU mitigation techniques.
\secref{sec_puid} presents the rejection of jets originating from PU with PU jet ID and PUPPI.
Jets reconstructed with a larger cone size are often used to identify the decay of Lorentz-boosted heavy particles such as \PW{}, \PZ{}, and Higgs bosons, and top quarks.
Pileup significantly degrades the reconstruction performance, and the gain from PU mitigation techniques for such large-size jets is discussed in \secref{sec_fatjet}.
The measurement of \ptmiss  also benefits from PU mitigation techniques, which is discussed in \secref{sec_met}.
Mitigation of PU for muon isolation variables is presented in \secref{muoniso}. 

\section{The CMS detector}
\label{sec_CMSgeneraldescription}

The central feature of the CMS apparatus is a superconducting solenoid of 6\unit{m} internal diameter, providing a magnetic field of 3.8\unit{T}. Within the solenoid volume are a silicon pixel and strip tracker, a lead tungstate crystal electromagnetic calorimeter (ECAL), and a brass and scintillator hadron calorimeter (HCAL), each composed of a barrel and two endcap sections. The ECAL covers the pseudorapidity range $\abs{\eta}<3$, while the HCAL is extended with forward calorimeters up to $\abs{\eta}<5$.
 Muons are detected in gas-ionization chambers embedded in the steel flux-return yoke outside the solenoid.
The silicon tracker measures charged particles within $\abs{\eta} < 2.5$. It consists of 1440 silicon pixel and 15\,148 silicon strip detector modules. For nonisolated particles with transverse momentum of $1 < \pt < 10\GeV$ and $\abs{\eta} < 1.4$, the track resolutions are typically 1.5\% in \pt and 25--90 (45--150)\mum in the transverse (longitudinal) impact parameter~\cite{Chatrchyan:2014fea}.
A more detailed description of the CMS detector, together with a definition of the coordinate system used and the relevant kinematic variables, can be found in Ref.~\cite{Chatrchyan:2008zzk}.

The particle-flow (PF) event reconstruction~\cite{Sirunyan:2017ulk} reconstructs and identifies each individual particle in an event, with an optimized combination of all subdetector information. In this process, the identification of the particle type (photon, electron, muon, charged or neutral hadron) plays an important role in the determination of the particle direction and energy. Photons (\eg, coming from \Pgpz\ decays or from electron bremsstrahlung) are identified as ECAL energy clusters not linked to the extrapolation of any charged particle trajectory to the ECAL. Electrons (\eg, coming from photon conversions in the tracker material or from \PB hadron semileptonic decays) are identified as a primary charged-particle track and potentially many ECAL energy clusters corresponding to this track extrapolation to the ECAL and to possible bremsstrahlung photons emitted along the way through the tracker material. Muons are identified as tracks in the central tracker consistent with either tracks or several hits in the muon system, and associated with calorimeter deposits compatible with the muon hypothesis. Charged hadrons are identified as charged particle tracks neither identified as electrons, nor as muons. Finally, neutral hadrons are identified as HCAL energy clusters not linked to any charged-hadron trajectory, or as a combined ECAL and HCAL energy excess with respect to the expected charged-hadron energy deposit.

The energy of photons is obtained from the ECAL measurement, corrected for zero-suppression effects. The energy of electrons is determined from a combination of the track momentum at the main interaction vertex, the corresponding ECAL cluster energy, and the energy sum of all bremsstrahlung photons attached to the track. The energy of muons is obtained from the corresponding track momentum. The energy of charged hadrons is determined from a combination of the track momentum and the corresponding ECAL and HCAL energy, corrected for zero-suppression effects and for the response function of the calorimeters to hadronic showers. Finally, the energy of neutral hadrons is obtained from the corresponding corrected ECAL and HCAL energy.

The collision rate is 40\unit{MHz}, and the events of interest  are selected using a two-tiered trigger system~\cite{Khachatryan:2016bia}. The first level (L1), composed of custom hardware processors, uses information from the calorimeters and muon detectors to select events at a rate of around 100\unit{kHz} within a fixed time interval of less than 4\unit{$\mu$s}. The second level, known as the high-level trigger (HLT), consists of a farm of processors running a version of the full event reconstruction software optimized for fast processing, and reduces the event rate to around 1\unit{kHz} before data storage.

All detector subsystems have dedicated techniques to reject signals from electronic noise or from particles that do not originate from the $\Pp\Pp$ collisions in the bunch crossing of interest, such as particles arriving from $\Pp\Pp$ collisions that occur in adjacent bunch crossings before or after the bunch crossing of interest (so called out-of-time PU).
While these rejection techniques are not the focus of this paper, some false signals can pass these filters and affect the PF reconstruction.
Particularly relevant is residual noise from ECAL and HCAL electronics that may add to the energy of reconstructed photons, electrons, and hadrons.
Algorithms for the rejection of this noise are further discussed in \secref{sec_noiseid}.

\section{Data and simulated samples}
\label{sec_environment}

\begin{figure}[hbtp]
  \centering
    \includegraphics[width=0.45\textwidth]{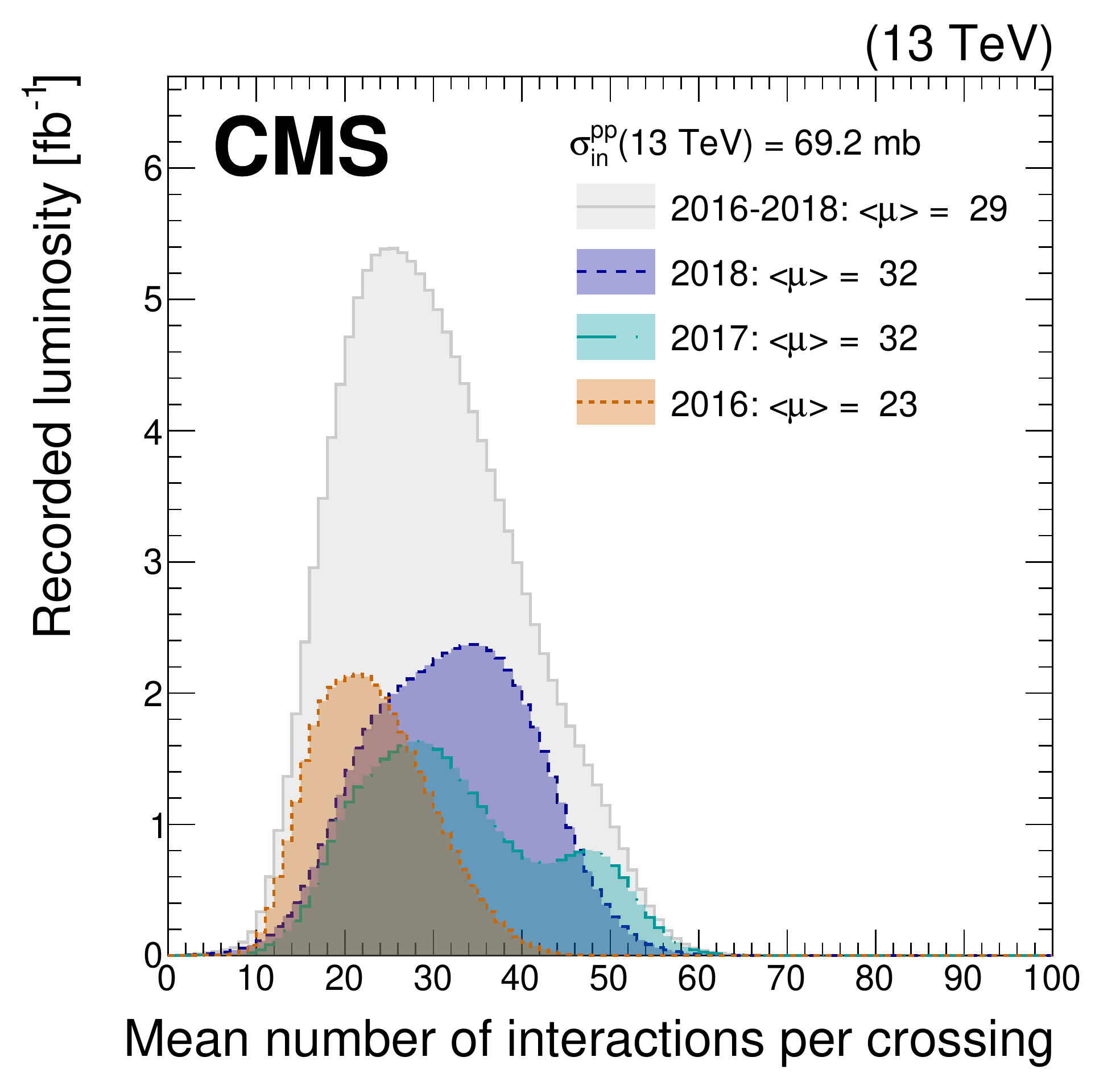}
    \caption{Distribution of the mean number of inelastic interactions per crossing (pileup) in data for $\Pp\Pp$ collisions in 2016 (dotted orange line), 2017 (dotted dashed light blue line), 2018 (dashed navy blue line), and integrated over 2016--2018 (solid grey line). 
    A total inelastic $\Pp\Pp$ collision cross section of 69.2\mb is chosen. The mean number of inelastic interactions per bunch crossing is provided  in the legend for each year.}
    \label{fig:pu_conditions}
\end{figure}

In this paper, data corresponding to an integrated luminosity of 35.9\fbinv~\cite{CMS-PAS-LUM-17-001} taken in 2016 are used.
\figrefb{fig:pu_conditions} shows the PU conditions in the years 2016--2018.
The number of $\Pp\Pp$ interactions is calculated from the instantaneous luminosity based on an estimated inelastic $\Pp\Pp$ collision cross section of 69.2\mb.
This number is obtained using the PU counting method described in the inelastic cross section measurements~\cite{Chatrchyan:2012gwa, CMS-PAS-QCD-11-002}.
In the following sections of this paper, we distinguish between two definitions: ``mean number of interactions per crossing" (abbreviated ``number of interactions" and denoted $\mu$) and ``number of vertices" (denoted $N_{\text{vertices}}$). 
Vertices are reconstructed through track clustering using a deterministic annealing algorithm~\cite{Chatrchyan:2014fea}.
The number of interactions is used to estimate the amount of PU in simulation. 
The number of vertices can be determined in both data and simulation. Further details on the relationship between $\mu$ and $N_{\text{vertices}}$ are provided in Section~\ref{sec_puid}.
The studies presented in this paper focus on the PU conditions in 2016, though the trends towards higher PU scenarios with up to 70 simultaneous interactions are explored as well.
The trigger paths used for the data taking are mentioned in each section.

Samples of simulated events are used to evaluate the
performance of the PU mitigation techniques discussed in this paper.
The simulation of standard model events composed uniquely of jets produced through the strong interaction, referred to as quantum chromodynamics (QCD) multijet events, is performed with \PYTHIA v8.212~\cite{Sjostrand:2014zea} in standalone mode using the Lund string fragmentation model~\cite{Andersson:1983ia,Sjostrand:1984iu} for jets.
For studies of lepton isolation, dedicated QCD multijet samples that are enriched in events containing electrons or muons (\eg, from heavy-flavor meson decays) are used.
The \PW and \PZ boson production in association with jets is simulated at leading-order (LO) with the \MGvATNLO v2.2.2~\cite{Alwall:2014hca} generator.
Production of top quark-antiquark pair (\ttbar{}) events is simulated with \POWHEG (v2)~\cite{powheg1, powheg2, powheg3}.
Single top quark production via the $s$- and $t$-channels, and \PQt{}\PW processes are simulated at next-to-leading-order (NLO) with \MGvATNLO that is interfaced with \PYTHIA{}.
For Lorentz-boosted $\PW$ boson studies~\cite{Khachatryan:2014vla}, MC simulation of high mass bulk graviton resonance~\cite{Agashe:2007zd,Fitzpatrick:2007qr,Antipin:2007pi} decaying to $\PW\PW$ boson pairs 
are generated at LO with \MGvATNLO.
All parton shower simulations are performed using \PYTHIA. For \PZ{}+jets production, an additional sample is generated using \MGvATNLO interfaced
with \HERWIGpp v2.7.1 \cite{Bahr:2008pv,Bellm:2013hwb} with the UE-EE-5C underlying event tune~\cite{Seymour:2013qka}
to assess systematic uncertainties related to the modeling of the parton showering and hadronization.

The LO and NLO NNPDF 3.0~\cite{nnpdf} parton distribution functions (PDF) are used in all generated samples matching the QCD order of the respective process.
The \PYTHIA parameters for the underlying event are set according to the CUETP8M1 tune~\cite{Skands:2014pea,Khachatryan:2015pea}, except for the $\ttbar$ sample, which uses CUETP8M2~\cite{CMS-PAS-TOP-16-021}.
All generated samples are passed through a detailed simulation of the CMS detector using \GEANTfour~\cite{Agostinelli:2002hh}.
To simulate the effect of additional $\Pp\Pp$ collisions within the same or adjacent bunch crossings, additional inelastic events are generated using \PYTHIA with the same underlying event tune as the main interaction and superimposed on the hard-scattering events.
The MC simulated events are weighted to reproduce the distribution of the number of interactions observed in data.

\section{The CHS and PUPPI algorithms}
\label{sec_puppibasic}

A detailed description of the CHS algorithm and its performance is found in Ref.~\cite{Sirunyan:2017ulk}.
In the following, we summarize the salient features and differences with respect to the PUPPI algorithm.
Both algorithms use the information of vertices reconstructed from charged-particle tracks.
The physics objects considered for selecting the primary $\Pp\Pp$ interaction vertex are track jets, clustered using the anti-\kt algorithm~\cite{Cacciari:2008gp,Cacciari:2011ma} with the tracks assigned to the vertex as inputs, and the associated $\vec{p}_{\mathrm{T}, \text{tracks}}^{\text{miss}}$, which is the negative vector \pt sum of those jets.  The reconstructed vertex with the largest value of summed physics-object $\pt^2$ is selected as the primary $\Pp\Pp$ interaction vertex or ``leading vertex" (LV).
Other reconstructed collision vertices are referred to as PU vertices.

The CHS algorithm makes use of tracking information to identify particles originating from PU after PF candidates have been reconstructed and before any jet clustering.
The procedure removes charged-particle candidates that are associated with a reconstructed PU vertex. A charged particle is associated with a PU vertex if it has been used in the fit to that PU vertex~\cite{Chatrchyan:2014fea}.
Charged particles not associated with any PU vertex and all neutral particles are kept.

The PUPPI~\cite{Bertolini:2014bba} algorithm aims to use information related to local particle distribution, event PU properties, and tracking information to mitigate the effect of PU on observables of clustered hadrons, such as jets, \ptmiss, and lepton isolation.  
The PUPPI algorithm operates at the particle candidate level, before any clustering is performed.
It calculates a weight in a range from 0 to 1 
for each particle, exploiting information about the surrounding particles,
 where a value of 1 is assigned to particles considered to originate from the LV.
These per-particle weights are used to rescale the particle four-momenta to correct for PU at particle-level, and thus reduces the contribution of PU to the observables of interest.

For charged particles, the PUPPI weight is assigned based on tracking information.
Charged particles used in the fit of the LV are assigned a weight of 1,
 while those associated with a PU vertex are assigned a weight of 0.
A weight of 1 is assigned to charged particles not associated with any vertex provided the distance of closest approach to the LV along the $z$ axis ($d_z$) is smaller than 0.3\cm; a weight of 0 is applied in all other scenarios.
The threshold of 0.3\cm corresponds to about 15 standard deviations of the vertex reconstruction resolution in the $z$ direction at an average PU of 10~\cite{Chatrchyan:2014fea}, 
 and it works as an additional filter against undesirable
 objects, such as accidentally reconstructed particles from detector noise.

Neutral particles are assigned a weight based on a discriminating variable $\alpha$.
In general, the $\alpha$ variable is used to calculate a weight, which encodes the probability that an individual particle originates from a PU collision. 
As discussed in Ref.~\cite{Bertolini:2014bba}, various definitions of $\alpha$ are possible. 
Within CMS, the $\alpha$ variable for a given particle $i$ is defined as
\begin{linenomath}
\begin{equation}
 \label{eq:metricalpha}
  \alpha_i =    \log \sum_{j \neq i,\, \Delta R_{ij}<R_0 } \left(\frac{p_{\mathrm{T},\, j}}{\Delta R_{ij}}\right)^{2} 
  \begin{cases}
    \text{for } \abs{\eta _i } < 2.5, & j \text{ are charged particles from LV,} \\
    \text{for } \abs{\eta _i } > 2.5, & j \text{ are all kinds of reconstructed particles,} \\
  \end{cases}
\end{equation}
\end{linenomath}
where 
$i$ refers to the particle in question,
$j$ are other particles, 
$p_{\mathrm{T},\, j} $ is the transverse momentum of particle $j$ in \GeV, 
and $\Delta R_{ij} = \sqrt{\smash[b]{(\Delta \eta_{ij}) ^2 + (\Delta \phi_{ij}) ^ 2}}$ (where $\phi$ is the azimuthal angle in radians)
is the distance between the particles $i$ and $j$ in the $\eta$-$\phi$ plane.
The summation runs over the particles $j$ in the cone of particle $i$ with a radius of $R_0=0.4$.
A value of $\alpha_i=0$ is assigned when there are no particles in the cone.
The choice of the cone radius $R_0$ in the range of 0.2--0.6 has a weak impact on the performance.
The value of 0.4 was chosen as a compromise between the performance when used in the definition of the isolation variable (preferring larger cones) and jet performance (preferring smaller cones).
In $\abs{\eta}<2.5$, where tracking information is available, 
only charged particles associated with the LV are included as particle $j$,
whereas all particles with $\abs{\eta}>2.5$  are included.
The variable $\alpha$ contrasts the collinear structure of QCD in parton showers with the soft diffuse radiation coming from PU interactions.
A particle from a shower is expected to be close to other particles from the same shower, whereas PU particles can be distributed more homogeneously.
The $\alpha$ variable is designed such that a particle 
 gets a large value of $\alpha$ if 
 it is close to either particles from the LV or, in $\abs{\eta}>2.5$, close to highly energetic particles. 

To translate $\alpha_i$ of each particle into a probability, charged particles assigned to PU vertices are used to generate the expected PU distribution in an event. From this expected distribution a median and root-mean-square (RMS) of the $\alpha$ values are computed.
The $\alpha_i$ of each neutral particle is compared with the computed median and RMS of the $\alpha$ distribution of the charged \PU\ particles using a signed $\chi^{2}$ approximation:
\begin{linenomath}
\begin{equation}
\text{signed } \chi^{2}_{i} = \frac{(\alpha_i - \overline{\alpha}_{\PU})\abs{\alpha_i - \overline{\alpha}_{\PU}}}{(\alpha_{\PU}^{\RMS})^{ 2}},
\end{equation}
\end{linenomath}
where $\overline{\alpha}_{\PU}$ is the median value of the $\alpha_i$ distribution for charged PU particles in the event and $\RMS_{\PU}$ is the corresponding RMS.
If signed $\chi^{2}_{i}$ is large, the particle most likely originates from the LV. The sign of the numerator is sensitive to the direction of the deviation of $\alpha _i$ from $ \overline{\alpha}_{\PU} $.
For the detector region where $\abs{\eta}>2.5$ and tracking is not available,
 the values $\overline{\alpha}_{\PU}$ and $\RMS_{\PU}$ can not be calculated directly. Therefore, $\overline{\alpha}_{\PU}$ and $\RMS_{\PU}$ are taken from the detector region where $\abs{\eta}<2.5$ and extrapolated to the  region where $\abs{\eta}>2.5$ by multiplying with transfer factors (see Tab.~\ref{table_tunableparameter}) derived from MC simulation.
The transfer factors are necessary, since the granularity of the detector varies with $\eta$ and leads to a variation of $\alpha$ with $\eta$, particularly outside of the tracker coverage ($\abs{\eta}=2.5$) and ECAL coverage ($\abs{\eta}=3.0$). 
Lastly, to compute the \pt weight of the particles, the signed $\chi^{2}_{i}$ for PU particles is assumed to be approximately distributed according to a $\chi^2$ distribution for $\chi^{2}_{i}>0$.
The \pt weight is given by $w_i = F_{\chi^2,\,\text{NDF}=1}(\text{signed }\chi^2_i)$ where $F_{\chi^2,\,\text{NDF}=1}$ is the cumulative distribution function of the $\chi^2$ distribution with one degree of freedom.
Particles with weights $w_i$ smaller than 0.01, \ie, those with a probability greater than 99\% to originate from PU are rejected; this last rejection removes remaining high-energy noise deposits. 
In addition, neutral particles that fulfill the following condition:
$w_i \,  p_{\mathrm{T},\, i} < (A + B \,  N_{\text{vertices}}) \GeV$, 
where $N_{\text{vertices}}$ is the number of vertices in the event, get a weight of 0.
This selection reduces the residual dependence of jet energies on the number of interactions.
The parameters $A$ and $B$ are tunable parameters. 
To perform the tuning of these parameters, jets clustered from PUPPI-weighted particles in the regions $\abs{\eta}<2.5$ and $2.5<\abs{\eta}<3.0$ are adjusted to have near-unity jet response, as a function of the number of interactions, i.e., the reconstructed jet energy matches the true jet energy regardless of the amount of PU.
In the region $\abs{\eta}>3$, the parameters are chosen such that \ptmiss resolution is optimized. 
Table~\ref{table_tunableparameter} summarizes the resulting
parameters that have been obtained using QCD multijet simulation 
with an average number of interactions of 23 and a significant amount of events beyond 30 interactions reflecting the 2016 data (orange curve in \figref{fig:pu_conditions}). 
The parameters $A$ and $B$ are smaller in $\abs{\eta} < 2.5$ (where the majority of particles are reconstructed with the tracker) than in $\abs{\eta}>2.5$ (where the measurement comes solely from the calorimeters that have a coarser granularity and thus collect more PU energy per cell).

\begin{table}[hbtp]
  \topcaption{ The tunable parameters of PUPPI optimized for application in 2016 data analysis. The transfer factors used to extrapolate the $\overline{\alpha}_{\PU}$ and $\alpha_{\PU}^{\text{RMS}}$ to $\abs{\eta} > 2.5$ are denoted TF.}
  \label{table_tunableparameter}
  \centering
    \begin{tabular}{ccccc}
      $\abs{\eta}$ of particle & $A$ [\GeV] & $B$ [\GeV] & TF $\overline{\alpha}_{\PU}$ & TF $\alpha_{\PU}^{\text{RMS}}$\\ \hline
      $[0, 2.5]$  & 0.2 & 0.015 & 1 & 1\\
      $[2.5, 3]$  & 2.0 & 0.13 & 0.9& 1.2\\
     $[3, 5]$  & 2.0 & 0.13 & 0.75&0.95\\
  \end{tabular}
\end{table}

\subsection{Data-to-simulation comparison for variables used within PUPPI}

\begin{figure}[hbtp]
  \centering
    \includegraphics[width=0.45\textwidth]{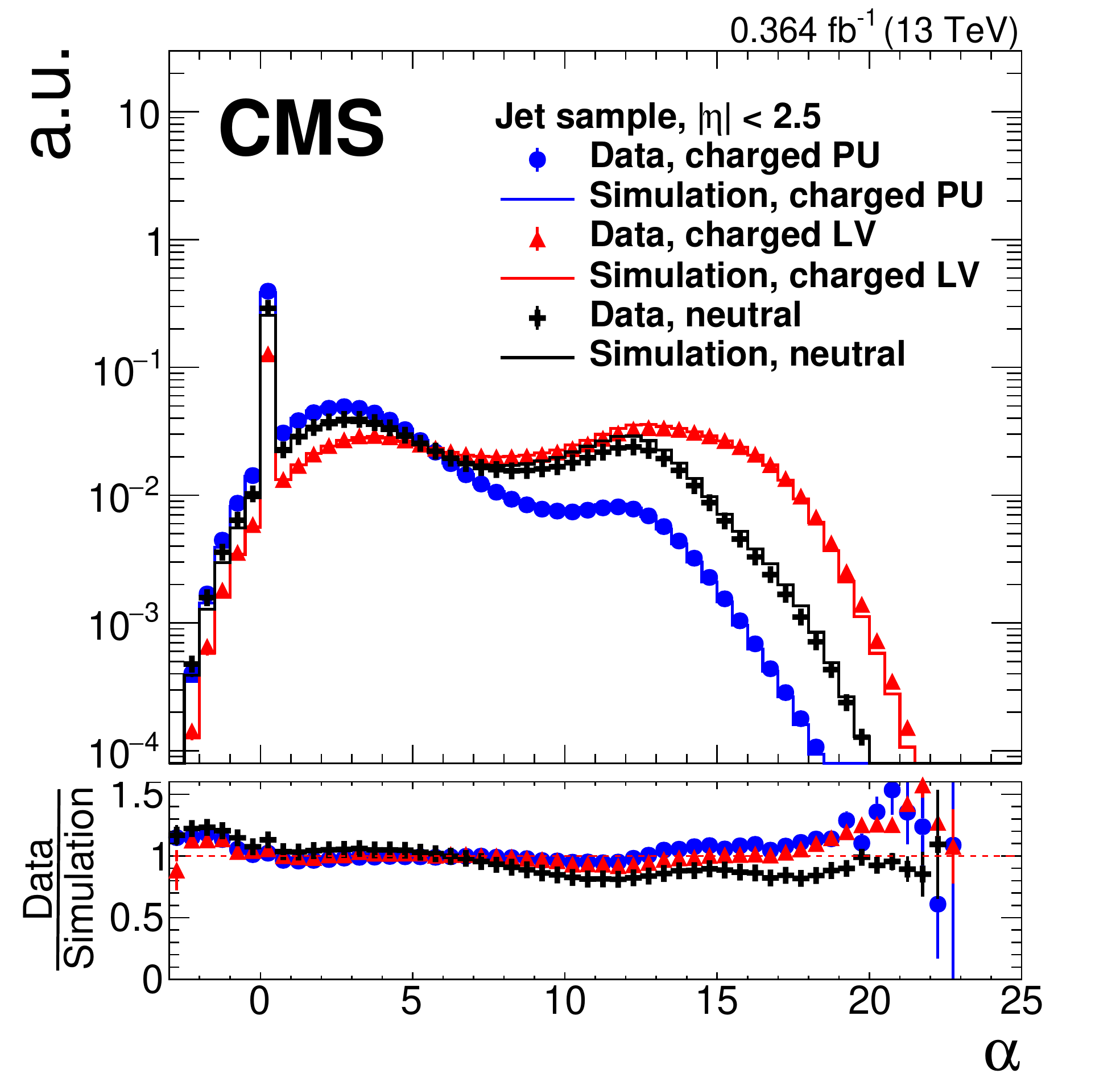} 
    \includegraphics[width=0.45\textwidth]{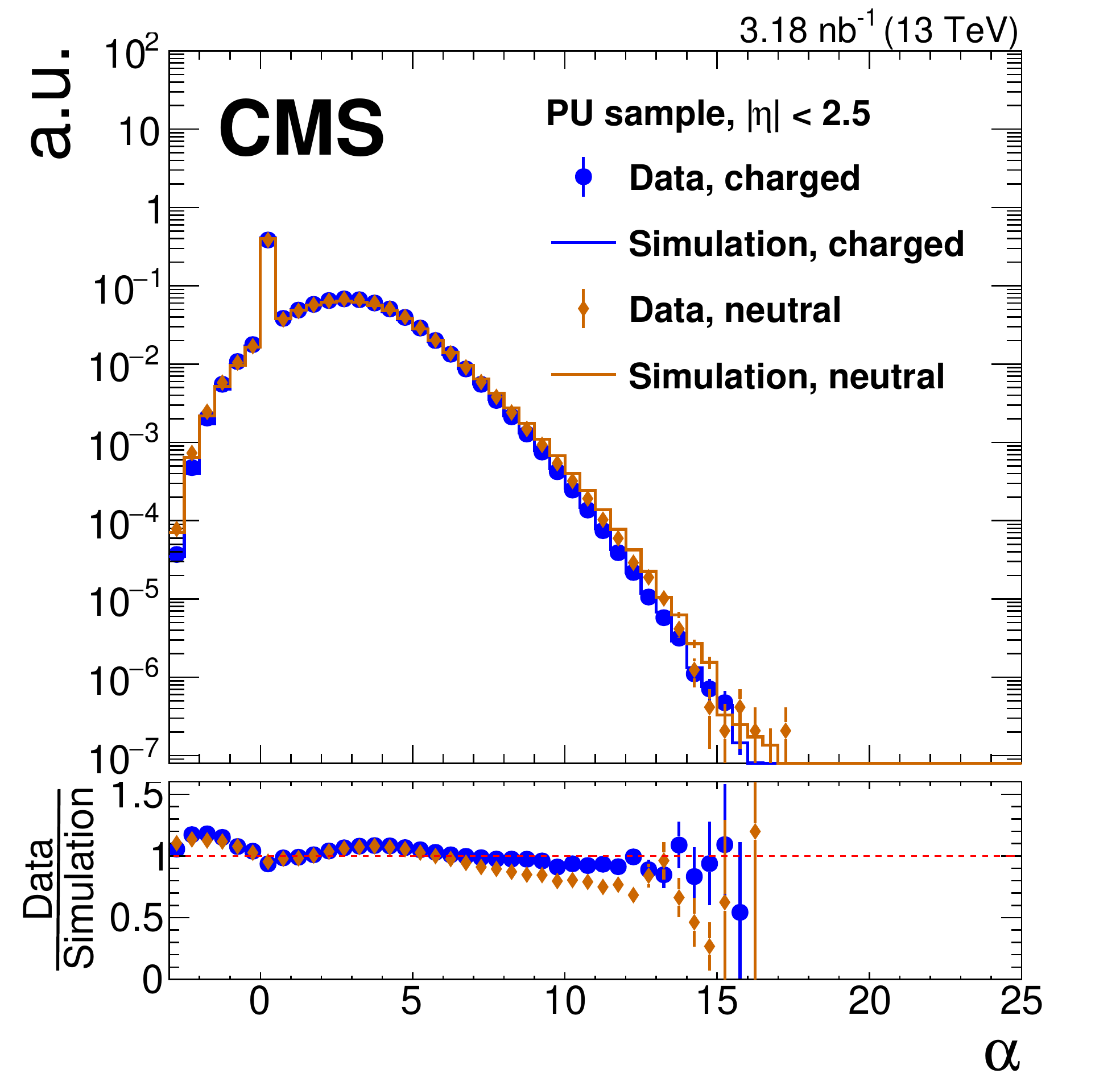}\\
    \includegraphics[width=0.45\textwidth]{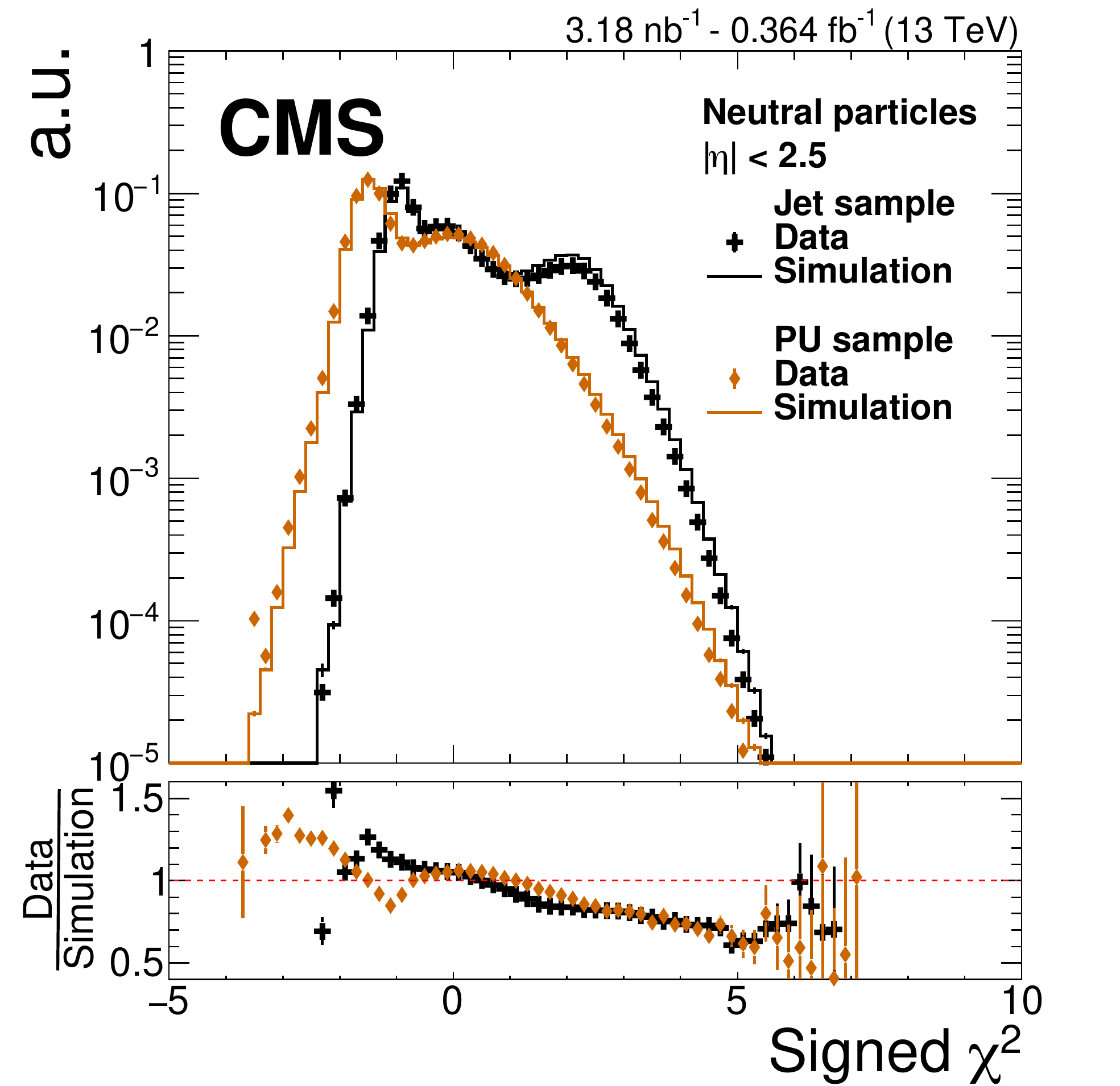}
    \includegraphics[width=0.45\textwidth]{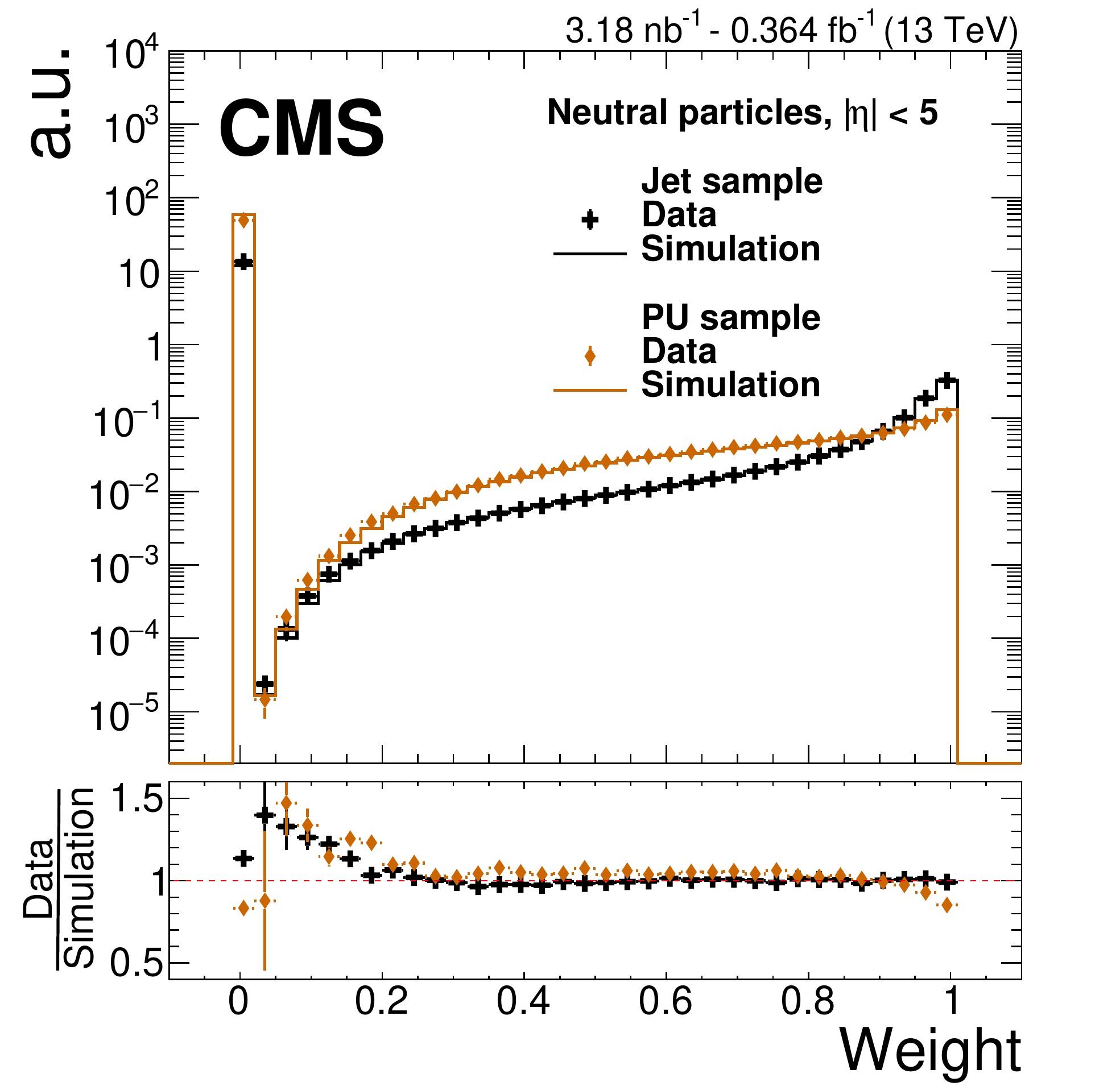}
    \caption{Data-to-simulation comparison for three different variables of the PUPPI algorithm. The markers show a subset of the data taken in 2016 of the jet sample and the PU sample, while the solid lines are QCD multijet simulations or PU-only simulation. The lower panel of each plot shows the ratio of data to simulation. Only statistical uncertainties are displayed. 
    The upper left plot shows the $\alpha$ distribution in the jet sample for charged particles associated with the LV (red triangles), charged particles associated with PU vertices (blue circles), and neutral particles (black crosses) for $\abs{\eta}<2.5$.
    The upper right plot shows the $\alpha$ distribution in the PU sample for charged (blue circles) and neutral (orange diamond) particles.
The lower left plot shows the
signed $\chi^{2}=(\alpha -  \overline{\alpha}_{\PU})\abs{\alpha -  \overline{\alpha}_{\PU}}/(\alpha_{\PU}^{\RMS})^{ 2}$ for neutral particles with $\abs{\eta}<2.5$ in the jet sample (black crosses) and in the PU sample (orange diamonds).
The lower right plot shows the
PUPPI weight distribution for neutral particles in the jet sample (black crosses) and the PU sample (orange diamonds). 
The error bars correspond to the statistical uncertainty.}
    \label{fig:puppi_metric}
\end{figure}

The behavior of the variables used in PUPPI has been studied in two complementary data samples. A subset of the data taken in 2016, corresponding to an integrated luminosity of 0.36\fbinv and selected using trigger paths based on the scalar sum (\HT) of the \pt of jets with $\pt>30\GeV$ and $\abs{\eta} < 3$, requiring an offline selection of $\HT > 1500\GeV$, is referred to as the jet sample.
The details of jet reconstruction and performance are discussed in \secref{sec:Jetreco}.
Here, we present comparisons of data and QCD multijet simulation based on all PF candidates in the event, rather than clustered jets.
As a reference, a data sample enriched in events containing mainly particles from PU collisions is compared with PU-only simulation and is referred to as the PU sample. The PU data sample is recorded with a zero-bias trigger that randomly selects a fraction of the collision events, corresponding to an integrated luminosity of 3.18\nbinv. The distribution of the number of PU interactions in both subsets of data is comparable to the one in the whole data sample collected in 2016.

\figrefb{fig:puppi_metric} shows the distribution of the three main variables used in PUPPI for data and simulation. The upper left plot presents the distribution of $\alpha$ for charged particles from the LV and the PU vertices and for neutral particles with $\abs{\eta}<2.5$ in the jet sample.
The separation power of the variable $\alpha$ between particles from the LV and PU vertices for charged particles can be deduced from this figure.
The majority of the charged particles from PU vertices have an $\alpha$ value below 8, whereas only a small fraction of particles have higher values. Charged particles from the LV exhibit a double-peak structure. The first peak at large $\alpha$ is characteristic of particles within jets originating from the LV. The second peak at lower $\alpha$ consists of charged particles that are isolated from other particles originating from the LV. With the exception of particles from lepton decays, which are directly addressed later, isolated particles have limited physics impact and consequently a low $\alpha$ value has a negligible impact on the algorithm performance on physics objects. 

The $\alpha$ distribution of neutral PU particles can be compared to charged PU particles in the PU sample shown in \figrefb{fig:puppi_metric} (upper right).
It becomes clear that the median and RMS of the $\alpha$ distribution are similar for charged and neutral particles originating from PU.
 This similarity confirms one of the primary assumptions of PUPPI, namely that $\overline{\alpha}_\mathrm{PU}$ and $\RMS_\mathrm{PU}$, which are computed for charged particles, can be used to compute weights for neutral particles with a discrimination power between PU and LV particles.
Although the qualitative features of the $\alpha$ distribution in data are reproduced by the simulation, a disagreement between data and simulation is observed, which is most pronounced for neutral particles from PU with large values of $\alpha$.

The $\chi^{2}$ distribution shown in \figref{fig:puppi_metric} (lower left) shows two peaks for both the jet sample and the PU sample. The first peak results from particles without any neighbor and an $\alpha$ value of zero. The second peak at zero represents all PU particles. The jet sample (black curve) shows a third peak for all LV particles.
Additionally, the shape of the resulting PUPPI weight distribution, shown in \figref{fig:puppi_metric} (lower right) is well modeled by simulation for particles with high weights (\ie, those likely originating from the \LV).
A considerable mismodeling is observed at low values of PUPPI weight, where low-\pt particles from PU interactions dominate.
This mismodeling does not propagate to further observables, because these particles receive small weights, and as a consequence have a negligible contribution. 
Although both samples have a similar distribution of number of interactions, the weight distribution of the jet sample has more events at higher values of the weight compared to the PU sample because of the selection of a high \pt{} jet.

\section{Jet reconstruction}
\label{sec:Jetreco}
Jets are clustered from PF candidates using the anti-\kt algorithm~\cite{Cacciari:2008gp} with the \textsc{FastJet} software package~\cite{Cacciari:2011ma}.
Distance parameters of 0.4 and 0.8 are used for the clustering.
While jets with $R = 0.4$ (AK4 jets) are mainly used in CMS for reconstruction of showers from light-flavor quarks and gluons, jets with $R = 0.8$ (AK8 jets) are mainly used for reconstruction of Lorentz-boosted \PW{}, \PZ{}, and Higgs bosons, and for top quark identification, as discussed in detail in \secref{sec_fatjet}. Before jet clustering, CHS- or PUPPI-based PU mitigation is applied to the PF candidates.
Reconstructed jets with the respective PU mitigation technique applied are referred to as CHS and PUPPI jets, respectively.
 
Jet momentum is determined as the vectorial sum of all particle momenta in the jet, and from simulation is, on average, within 5 to 20\% of the true momentum over the whole \pt spectrum and detector acceptance.
For CHS jets, an event-by-event jet-area-based correction~\cite{jetarea_fastjet,jetarea_fastjet_pu,Khachatryan:2016kdb} is applied to the jet four-momenta to remove the remaining energy due to neutral and charged particles originating from PU vertices, while no such correction is necessary for PUPPI jets.
Although CHS removes charged particles associated with a PU vertex, charged particles not associated with any vertex are kept and can add charged PU energy to the jet.
The remaining energy from PU particles subtracted from the jet energy is assumed proportional to the jet area and parametrized as a function of the median energy density in the event, the jet area, $\eta$, and \pt.
In addition, jet energy corrections are derived from simulation for CHS and PUPPI to bring the measured response of jets to that of generated particle-level jets on average.
In situ measurements of the momentum balance in dijet, photon+jets, \PZ{}+jets, and multijet events are used to correct any residual differences in jet energy scale between data and simulation~\cite{Khachatryan:2016kdb}.

In the following, only jets with $\pt > 15\GeV$ are used, which is the lowest jet \pt used in physics analysis in CMS.
The presentation of jet performance focuses on $\abs{\eta}<2.5$, covered by the tracking detector, ECAL, and HCAL, and the forward region, $\abs{\eta}>3$, where only the hadron forward calorimeter is present.
The intermediate region, $2.5<\abs{\eta}<3.0$, which is covered by ECAL and HCAL resembles the forward region in sensitivity to PU and is not discussed in this paper. For Sec.~\ref{sec_jet} the focus is set on $\abs{\eta}<0.5$, as the region $0.5<\abs{\eta}<2.5$ provides no further information and shows a similar performance.

\subsection{Jet energy and angular resolutions}
\label{sec_jet}

\begin{figure}[hbtp]
  \centering
    \includegraphics[width=0.45\textwidth]{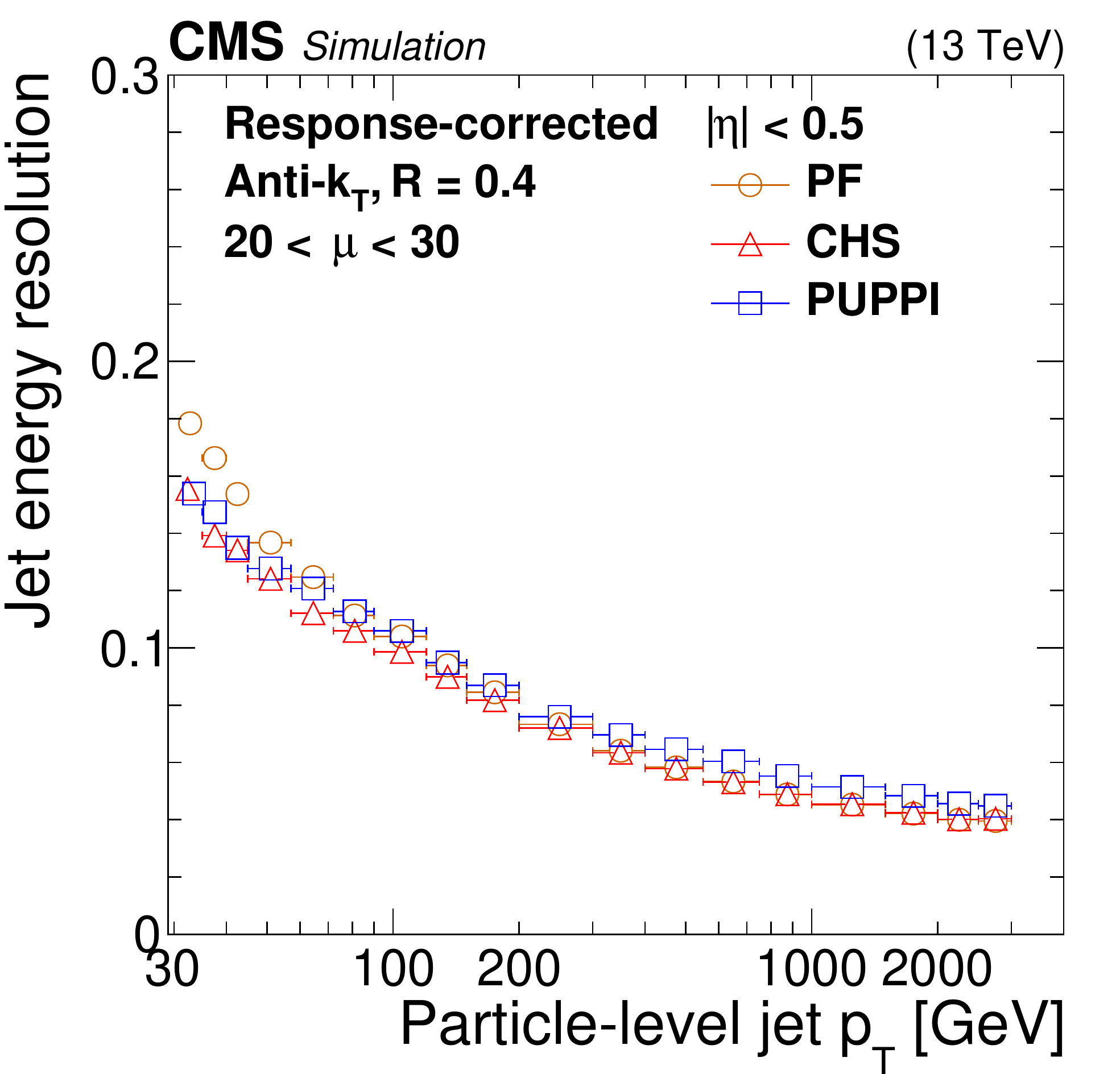}
        \includegraphics[width=0.45\textwidth]{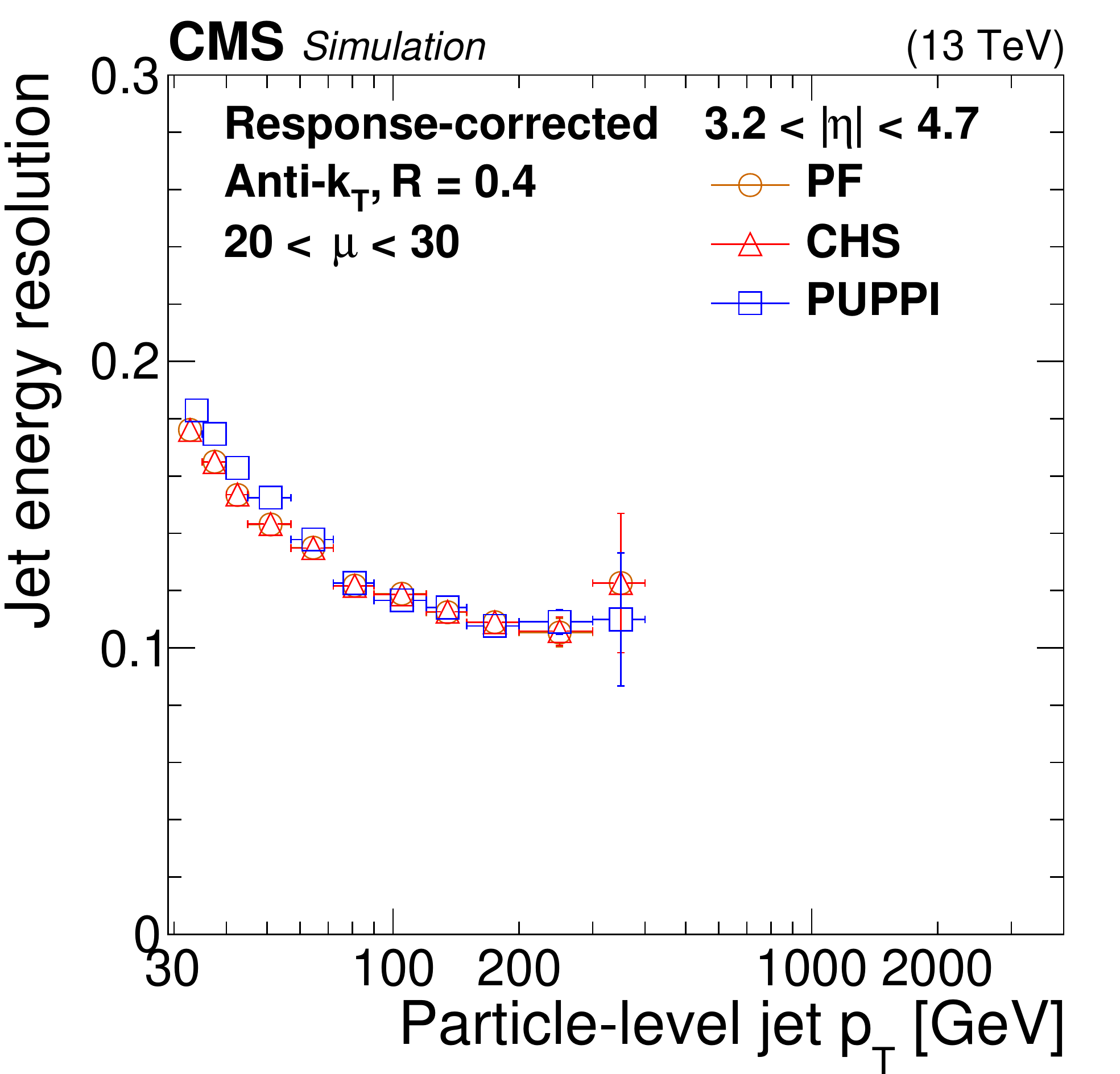}
    \includegraphics[width=0.45\textwidth]{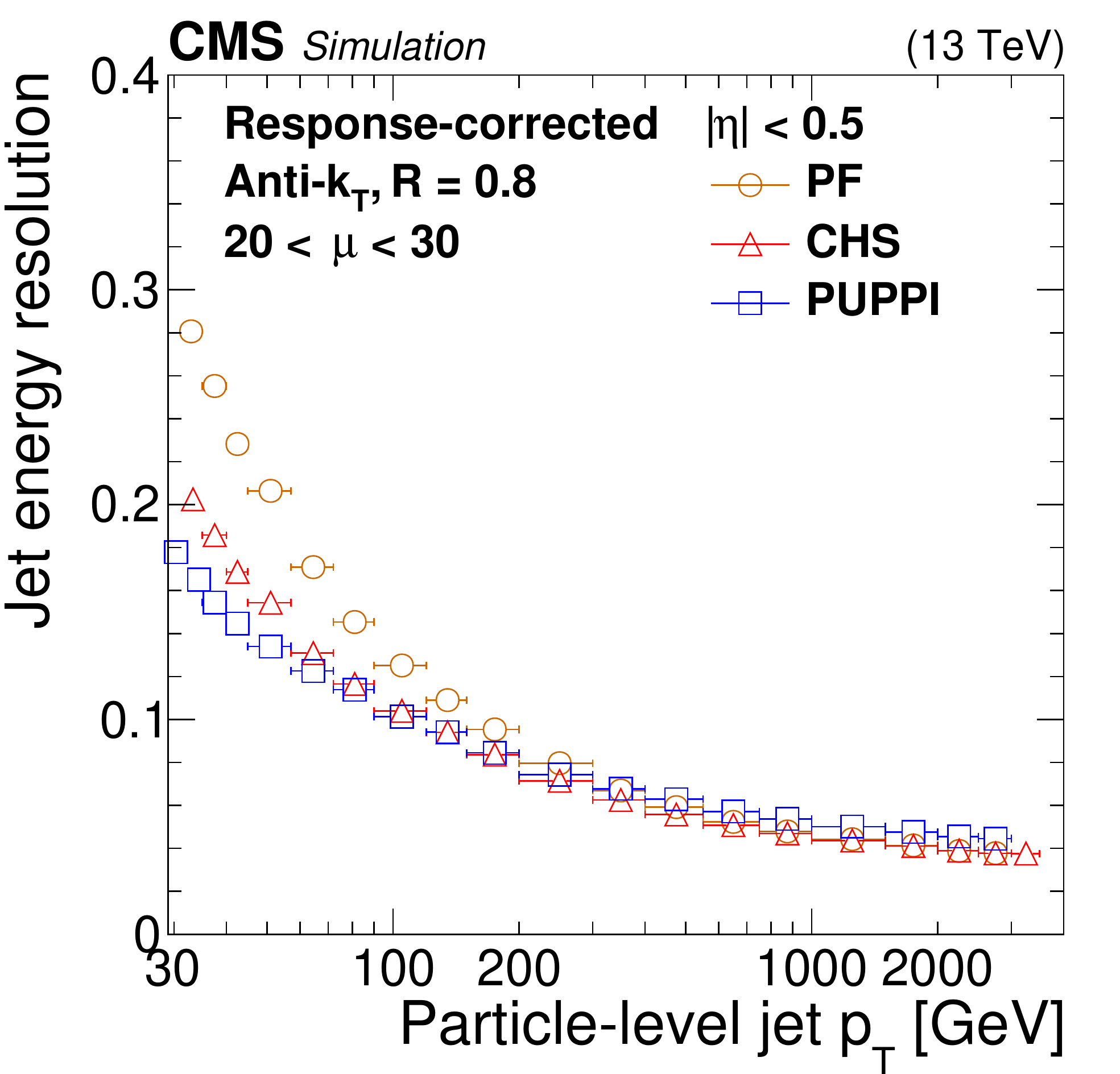}

    \caption{Jet energy resolution as a function of the particle-level jet \pt for PF jets (orange circles), PF jets with CHS applied (red triangles), and PF jets with PUPPI applied (blue squares) in QCD multijet simulation. The number of interactions is required to be between 20 and 30. The resolution is shown for AK4 jets with $\abs{\eta}<0.5$ (upper left) and $3.2<\abs{\eta}<4.7$ (upper right), as well as for AK8 jets with $\abs{\eta}<0.5$ (lower). The error bars correspond to the statistical uncertainty in the simulation.}
    \label{fig:jet_resolution1}
\end{figure}

The performance of the jet four-momentum reconstruction is evaluated in QCD multijet simulation
by comparing the kinematics of jets clustered from reconstructed PF candidates (recon\-struc\-tion-level jets) to jets clustered from
stable (lifetime $c \tau > 1\cm$) particles excluding neutrinos before any detector simulation (particle-level jets).
Particle-level jets are clustered without simulation of PU collisions whereas 
the reconstruction-level jets include simulation of PU collisions.
Jet energy corrections are applied to the reconstruction-level jets such that the ratio of reconstruction and particle-level jet \pt (the response) is on average 1.
The jet energy resolution (JER) is defined as the spread of the response distribution, which is Gaussian to a good approximation.
The resolution is defined as the $\sigma$ of a Gaussian fit to the distribution in the range $[m - 2\sigma,m + 2\sigma]$, where $m$ and $\sigma$ are the mean and width of the Gaussian fit, determined with an iterative procedure.
The cutoff at $\pm 2 \sigma$ is set so that the evaluation is not affected by outliers in the tails of the distribution.
\figrefb{fig:jet_resolution1} shows the JER
 as a function of jet \pt for jets reconstructed from all of the PF candidates (PF jets), CHS jets, and PUPPI jets, simulated with on average 20--30 PU interactions.
For AK4 jets, the performance of the CHS and PUPPI algorithms is similar. 
Jet resolution for PUPPI is slightly degraded below 30 PU, since PUPPI has been optimized for overall performance, including \ptmiss resolution and stability, beyond 30 PU interactions. This behavior at low PU can in principle be overcome through a special treatment in the limit of small amount of PU, where the number of particles to compute $\overline{\alpha}_{\text{PU}}$ and $\text{RMS}_{\text{PU}}$ is limited.
The PF jets in the detector region of $\abs{\eta}<0.5$ exhibit a worse performance, particularly at low \pt, since these jets are more affected by PU. In the region of $3.2<\abs{\eta}<4.7$, PF jets show the same performance as CHS jets, because no tracking is available. 
For AK8 jets, PUPPI provides better performance than the CHS and PF algorithms, since neutral particles from PU interactions contribute significantly to such jets.

\begin{figure}[hbtp]
  \centering
    \includegraphics[width=0.45\textwidth]{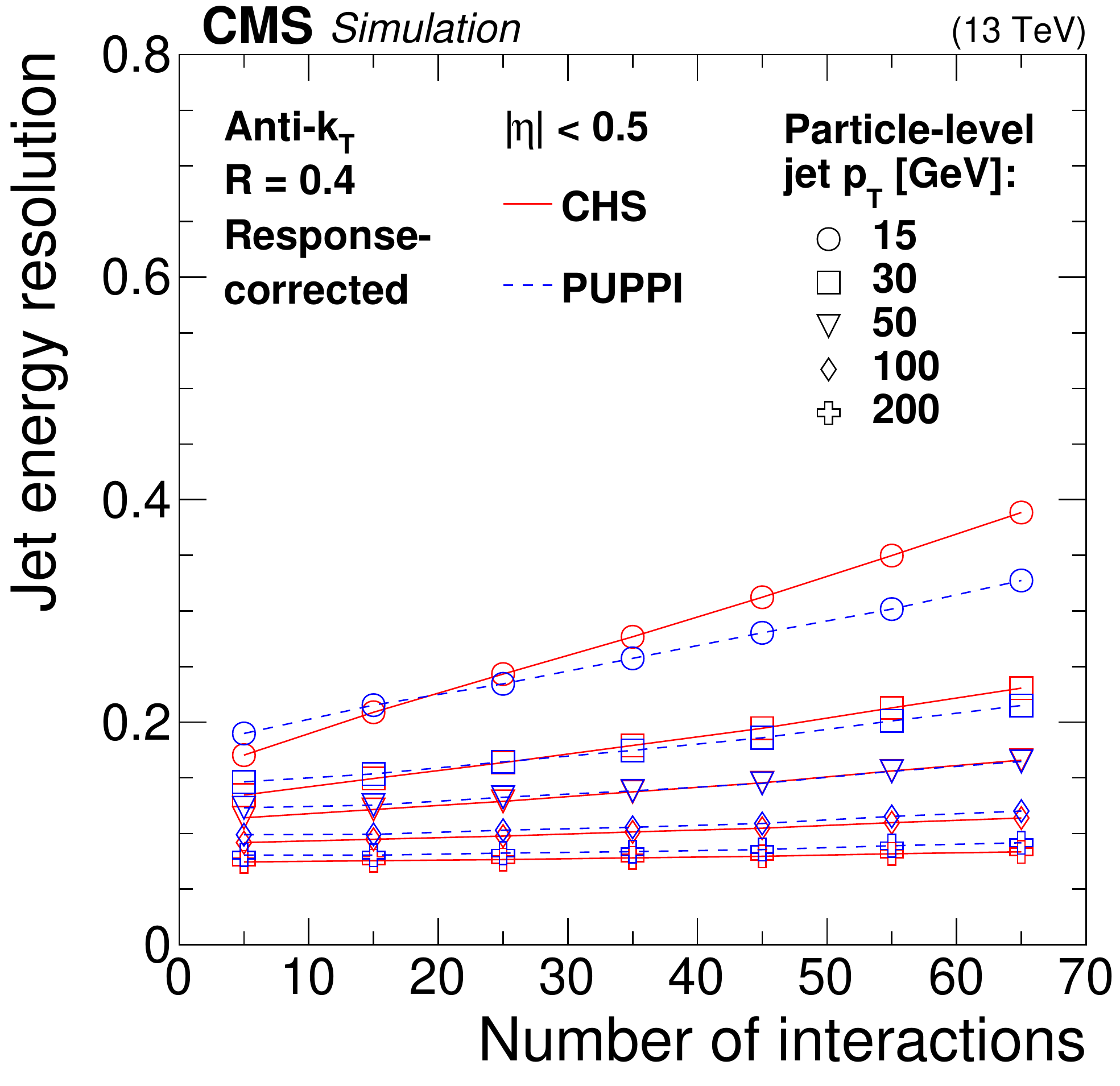}
    \includegraphics[width=0.45\textwidth]{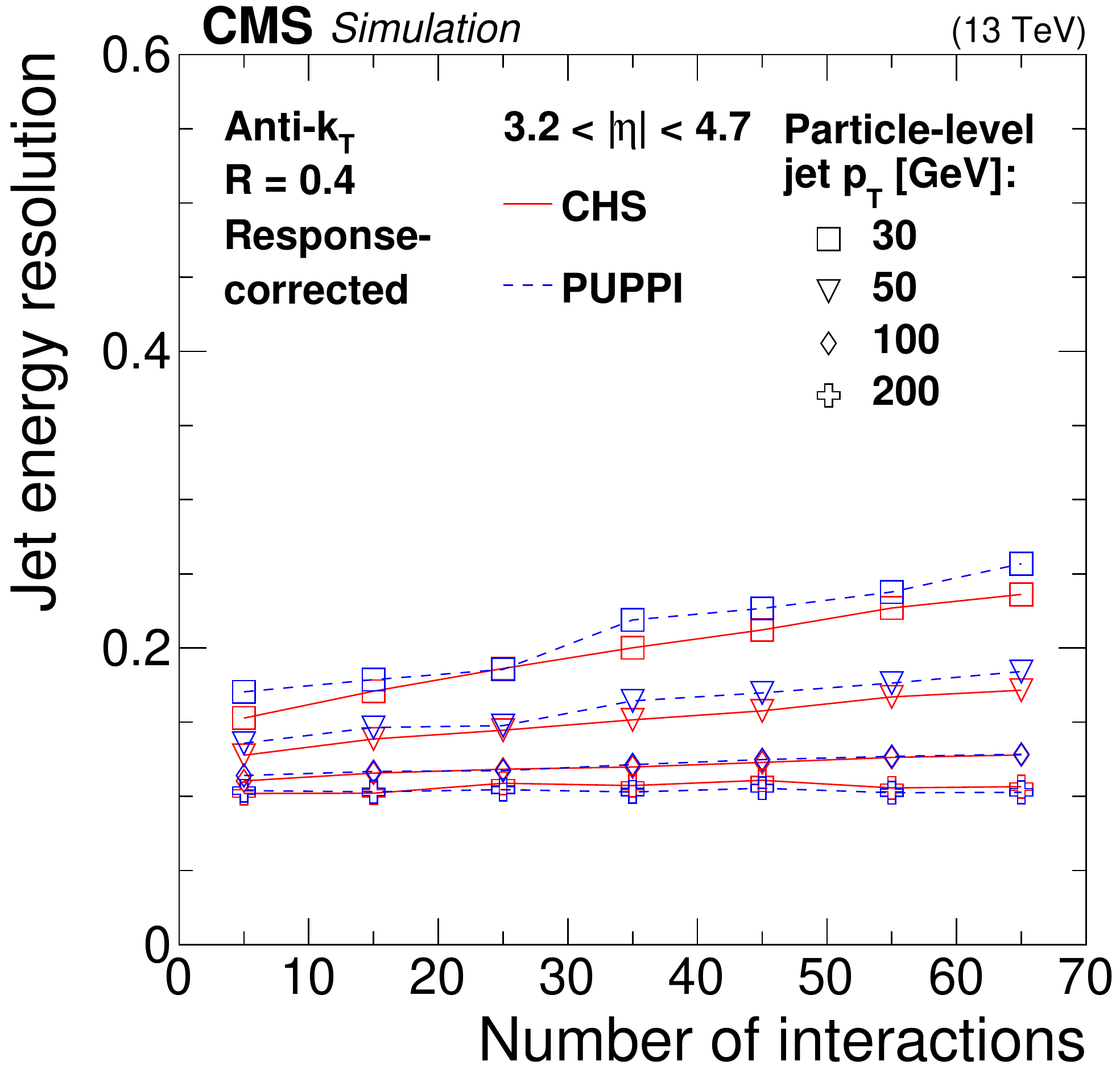}
    \includegraphics[width=0.45\textwidth]{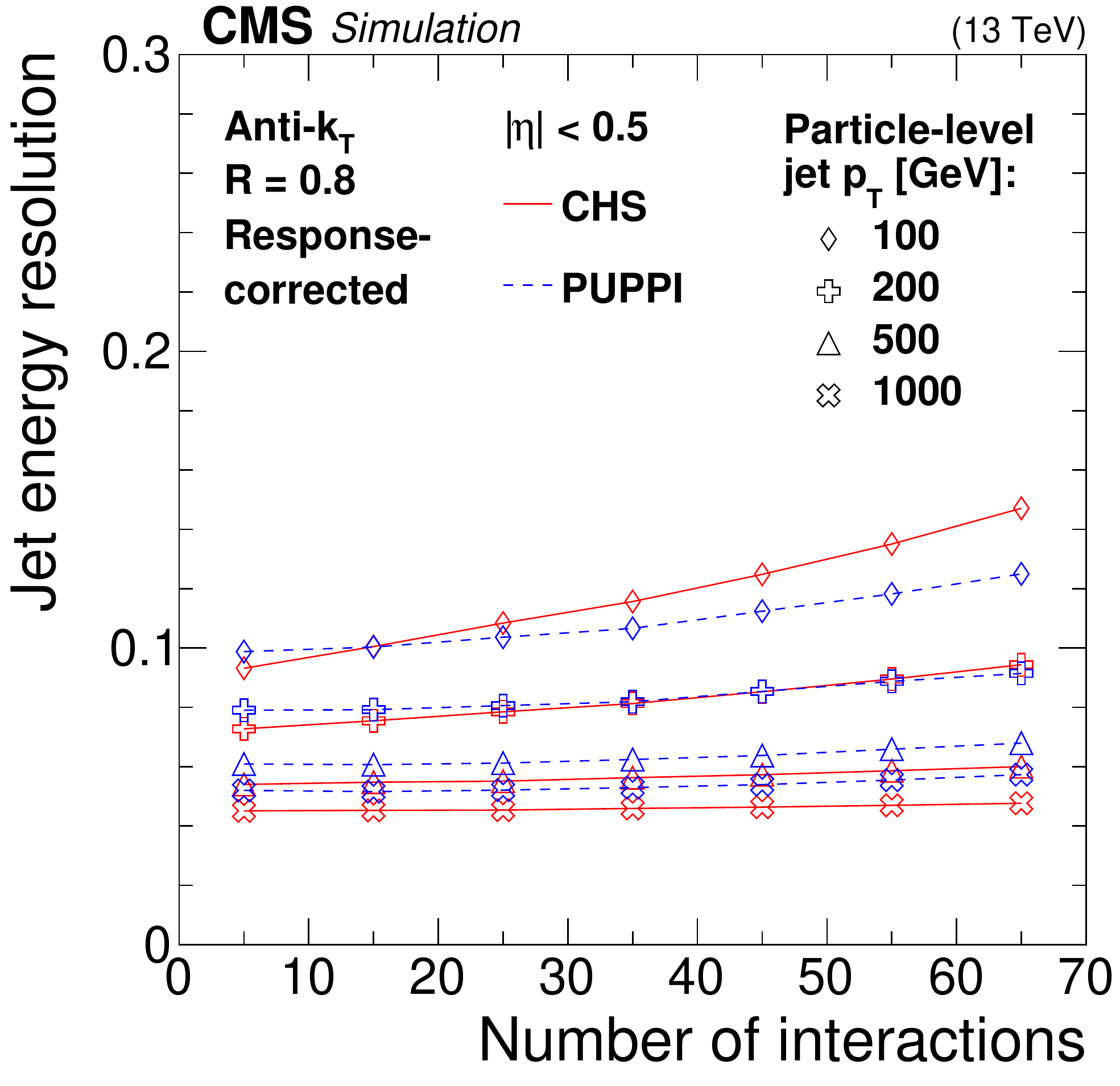}
    \caption{Jet energy resolution as a function of the number of interactions for jets with CHS (solid red line) and with PUPPI (dashed blue line) algorithms applied in QCD multijet simulation for different jet \pt values (different markers).
    The resolution is shown for AK4 jets with $\abs{\eta}<0.5$ (upper left) and $3.2<\abs{\eta}<4.7$ (upper right), as well as for AK8 jets with $\abs{\eta}<0.5$ (lower). The error bars correspond to the statistical uncertainty in the simulation.}
    \label{fig:jet_resolution2}
\end{figure}

\figrefb{fig:jet_resolution2} demonstrates how the JER scales with the number of interactions.
At more than 30 interactions, JER for AK4 jets with $\abs{\eta}<0.5$ and $\pt=30\GeV$ is better with the PUPPI than with the CHS PU mitigation.
However, JER for AK4 jets with $3.2<\abs{\eta}<4.7$ and $\pt=30\GeV$ is better with the CHS than with the PUPPI PU mitigation, which is a result of the PUPPI algorithm  being tuned to yield the best \ptmiss resolution rather than the best jet  resolution in the  $\abs{\eta} > 3$ region. This is achieved with a low PU particle rate, rather than the best jet resolution, achieved by high LV particle efficiency. 
 At $\pt>100 \GeV$, PUPPI jets have a resolution that is slightly worse than that of CHS jets with $\abs{\eta}<0.5$, while in $3.2<\abs{\eta}<4.7$ PUPPI and CHS performances are comparable.
For AK8 jets at low \pt, PUPPI yields a better JER than CHS; this improvement is present through the high-PU scenarios, \eg, at 50 or 60 interactions. The jet energy resolution becomes worse with PUPPI than with CHS for jets with $\pt > 200 \GeV$.
The behavior of PUPPI at high \pt is to a large extent limited by the quality of track-vertex association using $d_z$ for high-\pt charged hadrons. The effect is not visible in CHS because the $d_z$ requirement for charged particles that are not associated to any vertex is not used, but instead CHS keeps all charged particles not associated with any vertex.

\begin{figure}[hbtp]
  \centering
    \includegraphics[width=0.45\textwidth]{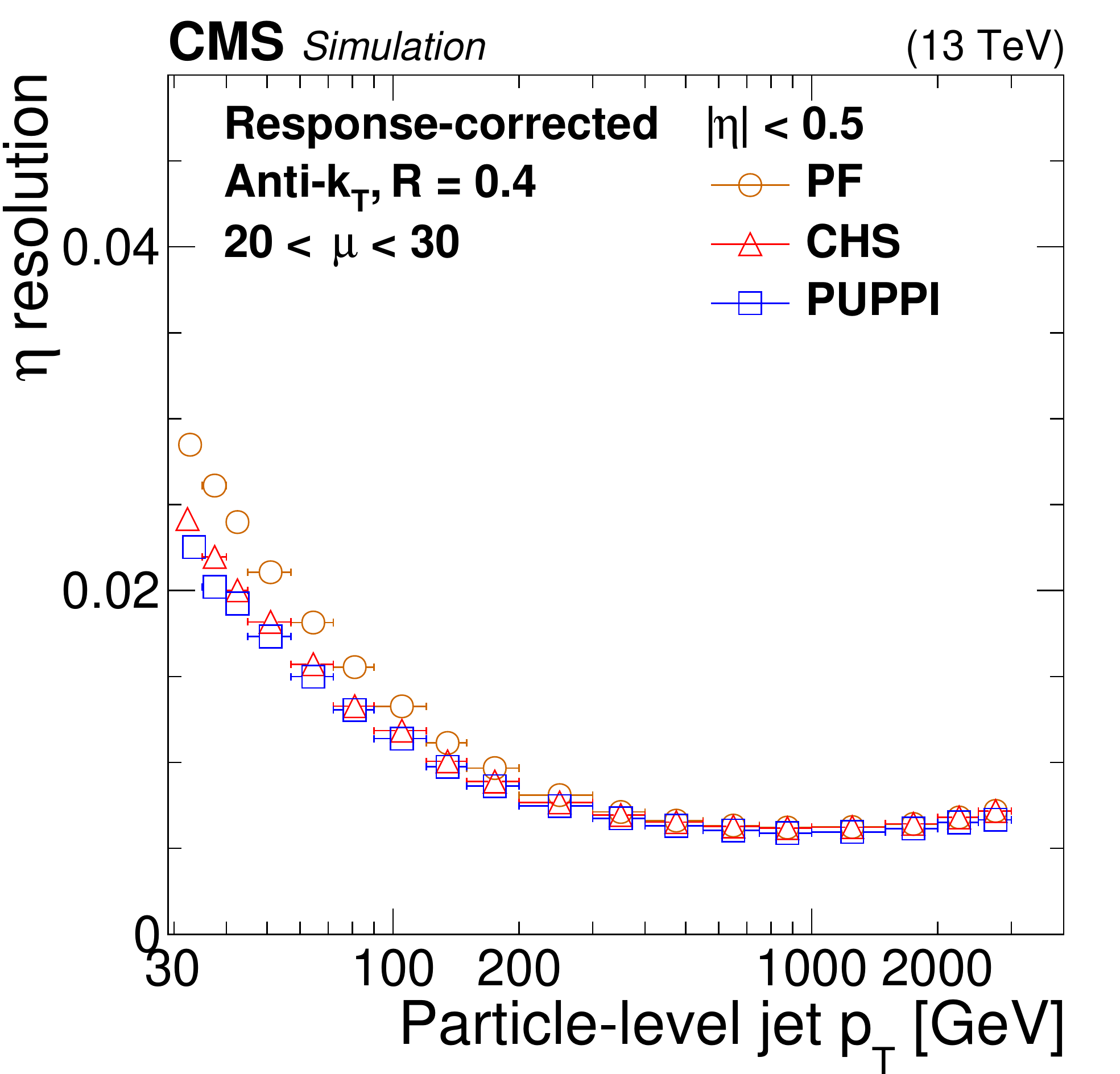}
    \includegraphics[width=0.45\textwidth]{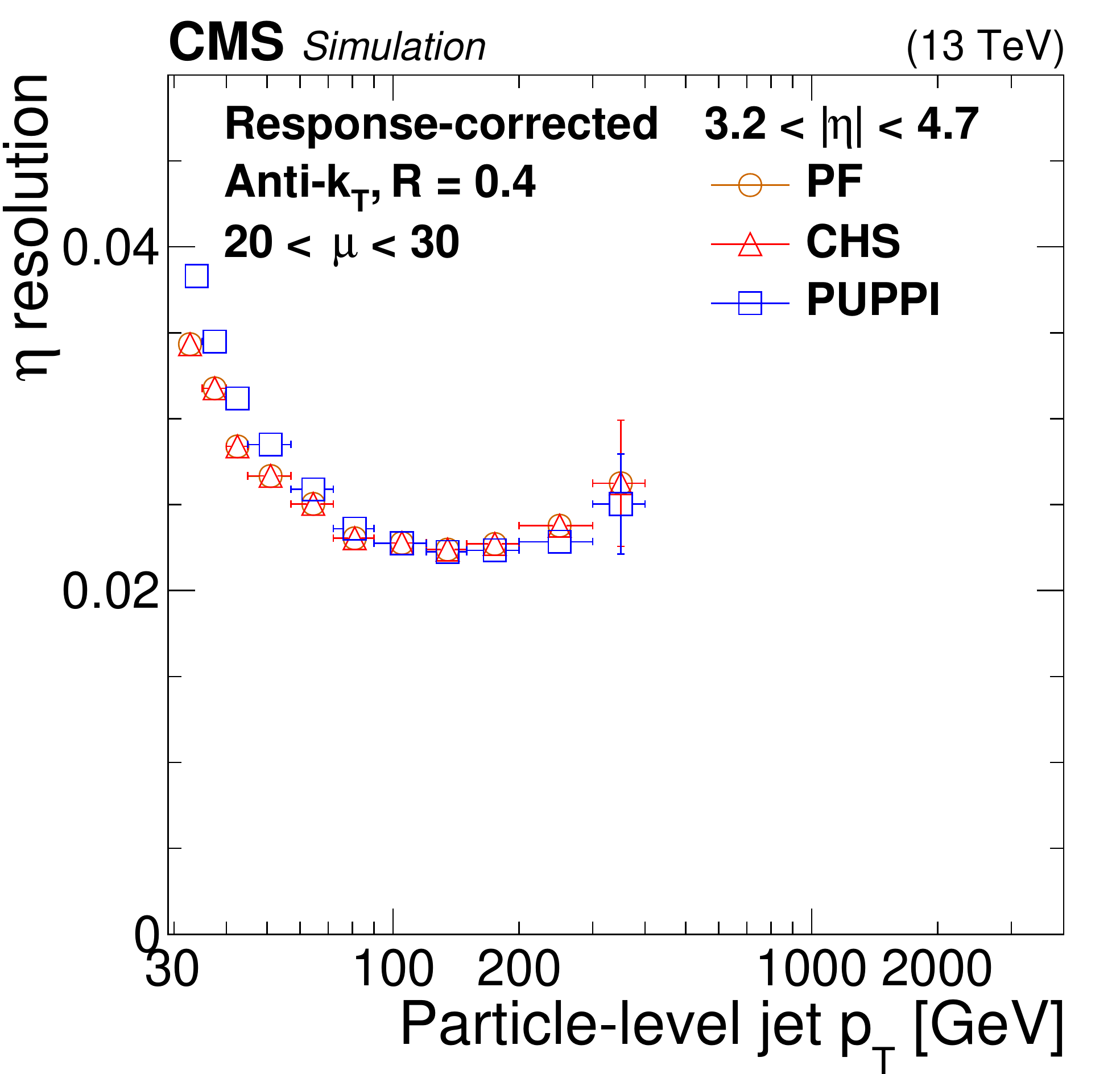}
    \includegraphics[width=0.45\textwidth]{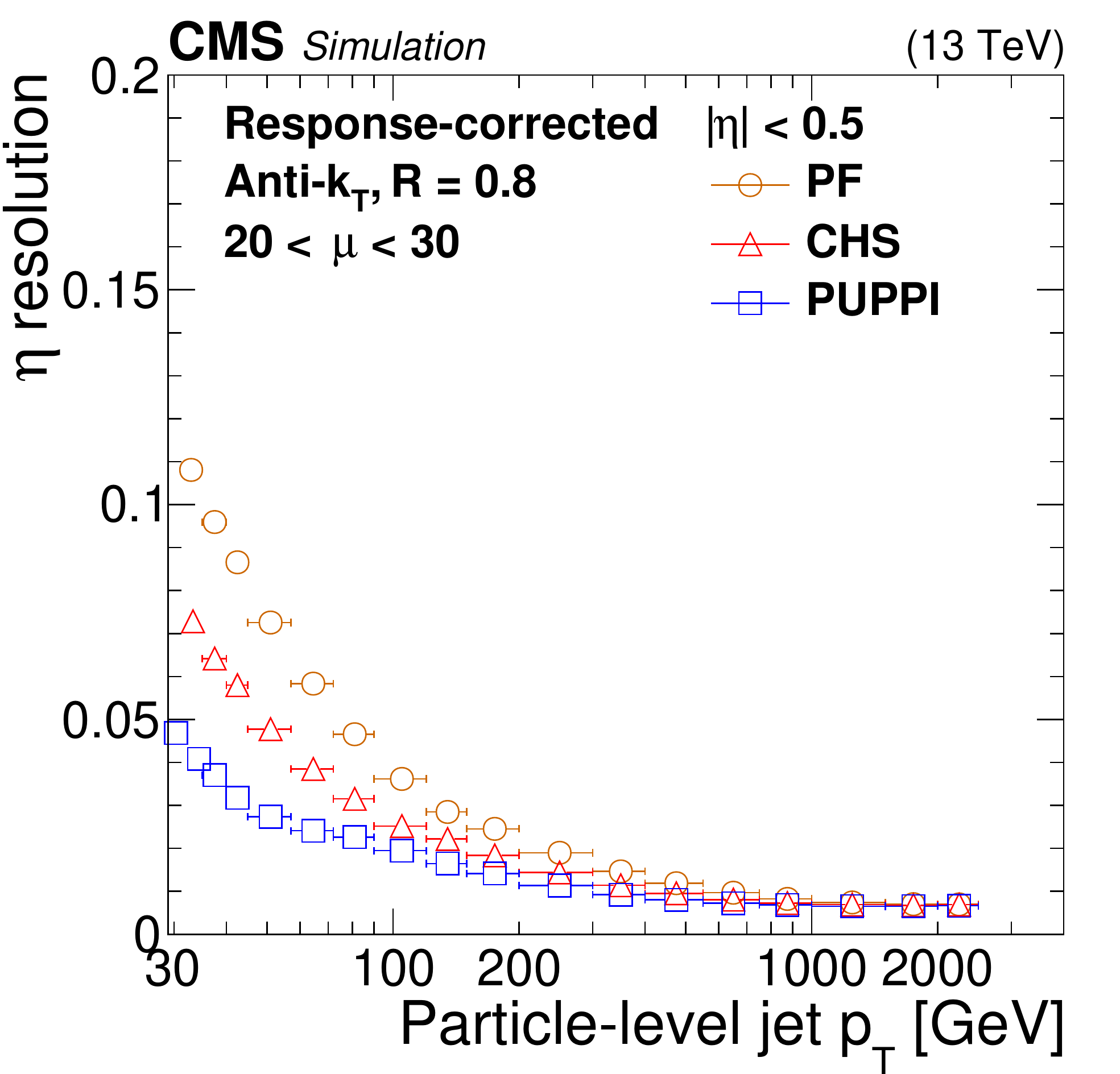}
    \caption{Jet $\eta$ resolution as a function of particle-level jet \pt for PF jets (orange circles), PF jets with CHS applied (red triangles), and PF jets with PUPPI applied (blue squares) in QCD multijet simulation. The number of interactions is required to be between 20 and 30.
    The resolution is shown for AK4 jets with $\abs{\eta}<0.5$ (upper left) and $3.2<\abs{\eta}<4.7$ (upper right) as well as for AK8 jets with $\abs{\eta}<0.5$ (lower). The error bars correspond to the statistical uncertainty in the simulation.}
    \label{fig:jet_resolution3}
\end{figure}

\figrefb{fig:jet_resolution3} shows the jet $\eta$ angular resolution simulated with 20--30 interactions. 
The same qualitative conclusions also hold for the resolution in $\phi$, since $\phi$ and $\eta$ segmentation of the detector are similar. 
The resolution is evaluated as the width of a Gaussian function fit to 
the distribution of the $\eta$-difference between the generator- and reconstruction-level jets.
The same conclusions as for JER also hold for jet angular resolution. 
The CHS and PUPPI algorithms perform similarly for AK4 jets with $\abs{\eta}<0.5$. However, significant improvements from PUPPI are observed for AK8 jets for $\abs{\eta}<0.5$.
Angular resolution of large-size jets is particularly sensitive to PU as the clustered energy from PU particles increases with the jet size. Hence, the improvements are larger when PUPPI jets are considered.

\subsection{Noise jet rejection}
\label{sec_noiseid}

The identification and rejection of jets originating from noise and reconstruction failures are critical
to all CMS analyses where a jet or \ptmiss is used as part of the selection.
To further reject noise after detector signal processing and jet clustering, 
a set of criteria on the PF candidates within a jet are applied~\cite{CMS-PAS-JME-16-003}.  The criteria listed in Table~\ref{table_jetID_criteria} are based on jet constituent energy fractions and multiplicities. They reject residual noise from the HCAL and ECAL, retaining 98--99\% of genuine jets, \ie, jets initiated by genuine particles rather than detector noise.
Although PU mitigation algorithms are not designed to have an effect on detector noise,
they could, in principle, affect the rejection capability of the noise jet ID.

\begin{figure}[hbtp]
  \centering
    \includegraphics[width=0.45\textwidth]{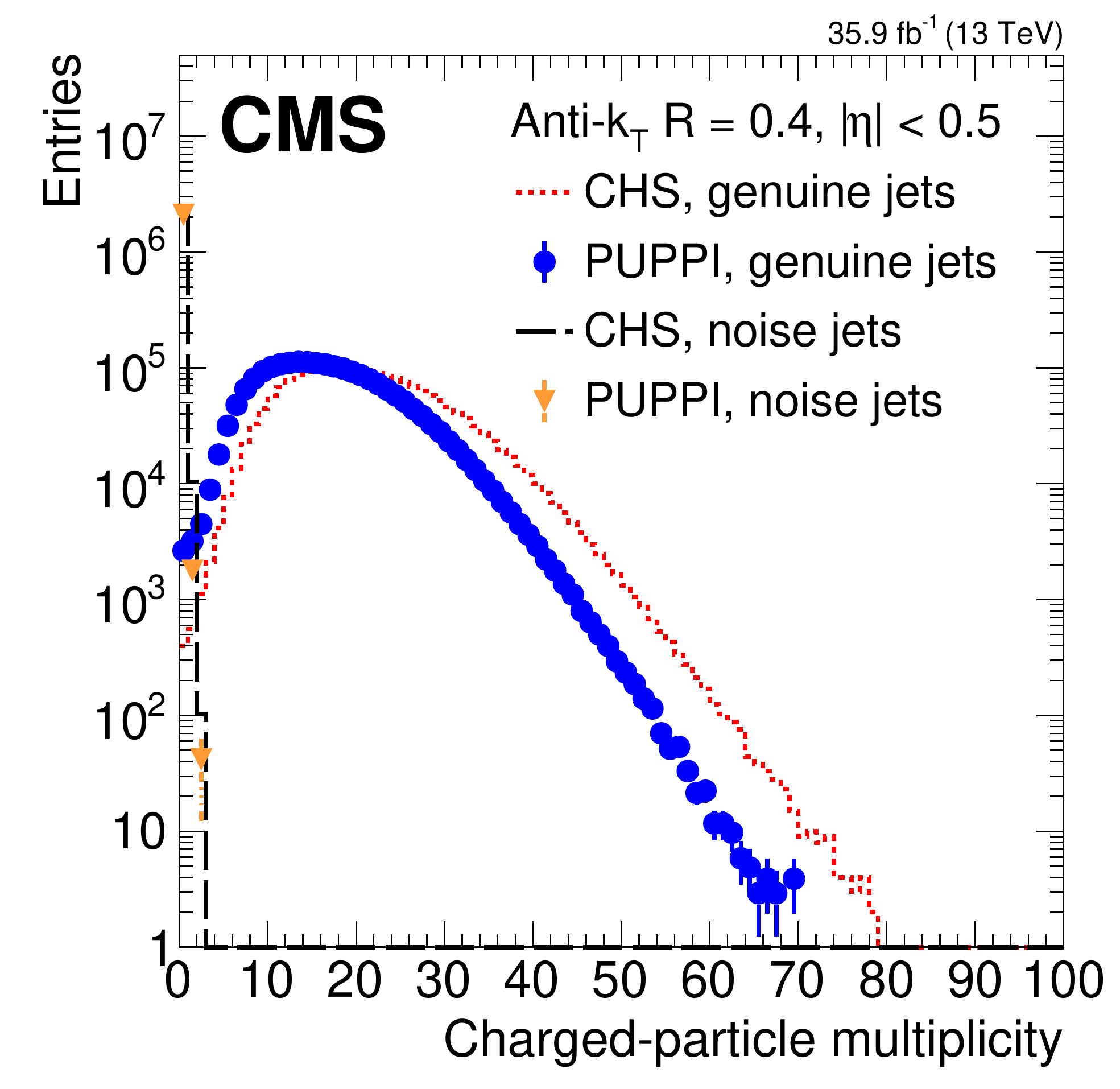}
    \includegraphics[width=0.45\textwidth]{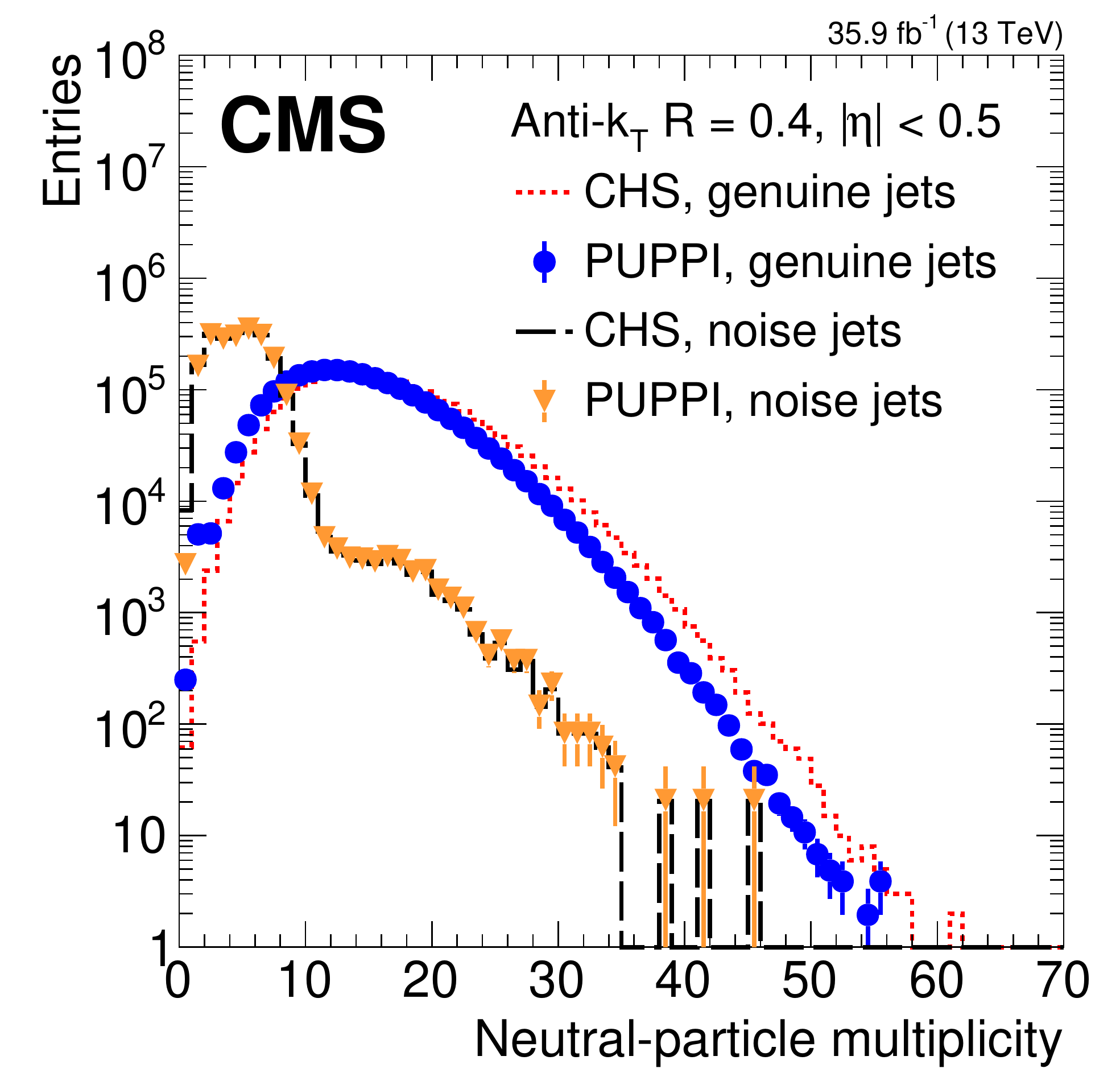}
    \includegraphics[width=0.45\textwidth]{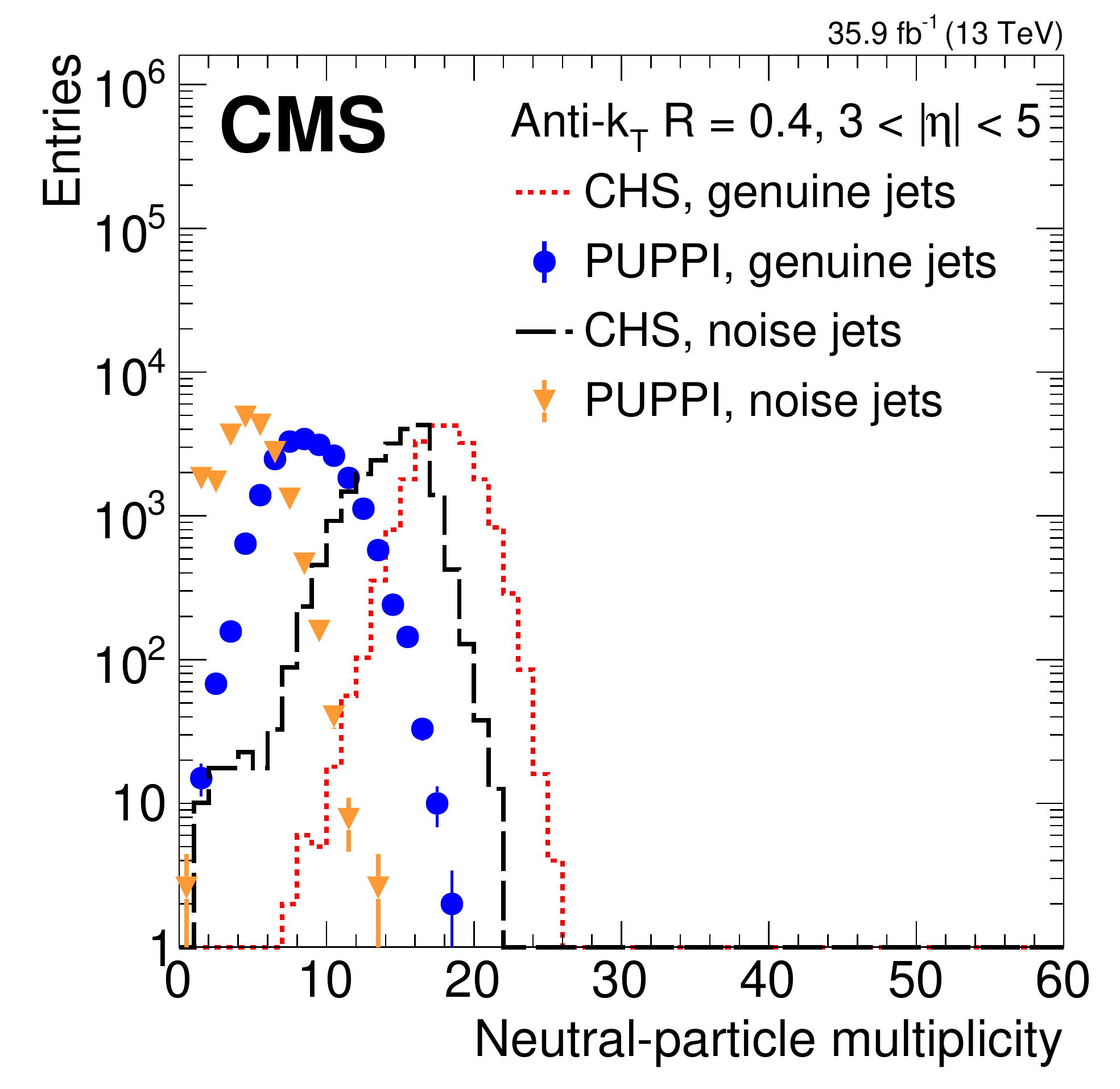}
    \includegraphics[width=0.45\textwidth]{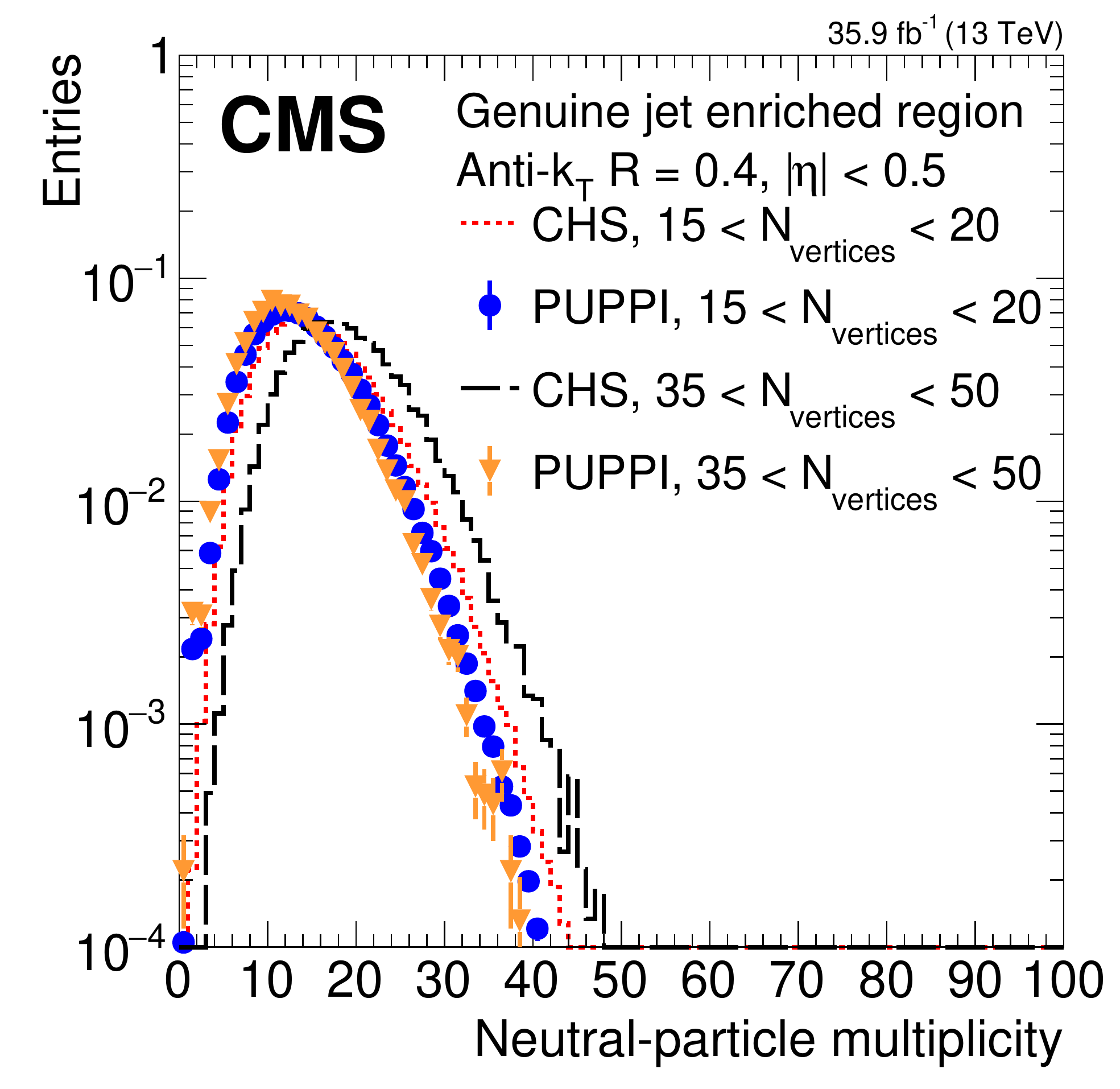}
    \caption{The charged- and neutral-particle multiplicities for CHS and PUPPI in a dijet (genuine jets) and minimum bias (noise jets) selection in data. The multiplicities are shown for AK4 jets using CHS reconstructed real jets (red dashed), CHS reconstructed noise jets (black long dashed), PUPPI reconstructed genuine jets (blue circles), and PUPPI reconstructed noise jets (orange triangles). The upper plots show the charged (left) and neutral particle multiplicities (right) for jets with $\abs{\eta}<0.5$. The lower left plot shows the neutral particle multiplicity for jets with $3<\abs{\eta}<5$. The lower right plot shows the neutral particle multiplicity of AK4 jets with $\abs{\eta}<0.5$ in a dijet selection in data using CHS and PUPPI for 15--20 and 35--50 interactions. The error bars correspond to the statistical uncertainty.}
    \label{fig:jet_noise}
\end{figure}

\figrefb{fig:jet_noise} (upper left/right and lower left) shows the distribution of the charged and neutral constituent multiplicities comparing genuine jet enriched (dijet) and noise jet enriched (minimum bias) data, demonstrating the separation power.
For the dijet selection, data are selected with an HLT requirement of at least one jet having a $\pt >  400 \GeV$, two offline reconstructed jets with \pt greater than 60 and 30\GeV, respectively, and an opening in azimuthal angle greater than 2.7.
For the minimum bias selection, jets with $\pt > 30 \GeV$  passing the minimum bias trigger path are used.
The noise jet ID requires at least one charged constituent for jets with $\abs{\eta}<2.4$ and at least two constituents (neutral or charged) for $\abs{\eta}<2.7$.
The charged constituent multiplicity is smaller for PUPPI than for CHS jets because PUPPI rejects additional charged particles by applying a $d_z$ requirement on tracks not associated with any vertex. 
The PUPPI weighted neutral constituent multiplicity, defined as the sum of PUPPI weights of all neutral particles in the jet, is also smaller than the neutral constituent multiplicity for CHS.
In $3<\abs{\eta}<5$, the PUPPI neutral constituent multiplicity is significantly lower than for CHS. Thus, the ability to separate noise is reduced. 
With CHS, noise jets are rejected by requiring a minimum of 10 neutral particles.  
With PUPPI, a minimum of 3 is required for the PUPPI scaled neutral multiplicity.
\figrefb{fig:jet_noise} (lower right) demonstrates the PU dependence of the neutral constituent multiplicity.
While for CHS, the average multiplicity changes by 30--40\% going from 20--30 to 50--60 reconstructed vertices, the PUPPI scaled multiplicities do not change significantly, making noise jet rejection independent of PU.

The efficiency of the jet ID criteria for genuine jets is measured in data using a tag-and-probe procedure in dijet events~\cite{CMS-PAS-JME-16-003}.
The background rejection is estimated using a noise-enriched minimum bias event selection.
The fraction of rejected noise jets after applying jet ID criteria that yield a 99\% efficiency for genuine jets is summarized in Table~\ref{table_jetID_performance} for different regions in $\eta$. The number of noise jets reconstructed with the CHS and PUPPI algorithms is not the same, because the PUPPI reconstruction criteria reject particles that would otherwise give rise to a fraction of noise jets before jet ID criteria are applied. The absolute number of noise jets remaining after PU mitigation and jet ID together differs by less than 20\% between CHS and PUPPI jets.

\begin{table}
\centering
\topcaption{Jet ID criteria for CHS and PUPPI jets yielding a genuine jet efficiency of 99\% in different regions of $\abs{\eta}$.\\}
\label{table_jetID_criteria}
\begin{tabular}[t]{cccc}
Region of $\abs{\eta}$ & Variable & Requirement (CHS) & Requirement (PUPPI)\\
\hline
& & & \\
\multirow{2}{*}{ $\abs{\eta} < 2.4$} &
Charged hadron energy fraction & $>$0 &$>$0 \\
& Charged multiplicity& $>$0 &$>$0 \\
& & & \\
\multirow{3}{*}{ $\abs{\eta} < 2.7$}
& 
Neutral hadron energy fraction	& $<$0.90 &$<$0.90 \\
&Neutral EM energy fraction	& $<$0.90 &$<$0.90 \\
&Number of constituents	& $>$1 &$>$1 \\
& & & \\
\multirow{3}{*}{$2.7 < \abs{\eta} < 3$}
&
Neutral EM energy fraction & $>$0.02 and $<$0.99 & \NA \\
&Number of neutral particles & $>$2 & \NA \\
&Neutral hadron energy fraction & \NA & $<$0.99\\
& & &\\
\multirow{3}{*}{$\abs{\eta} > 3$}
& 
Neutral EM energy fraction & $<$0.90 & $<$0.9\\
&Neutral hadron energy fraction & $>$0.02 & $>$0.02 \\
&Number of neutral particles & $>$10 & $>$3\\
\end{tabular}
\end{table}

\begin{table}
\centering
\topcaption{Fraction of noise jets rejected when applying jet ID criteria to PUPPI and CHS jets yielding a genuine jet efficiency of 99\% in different regions of $\abs{\eta}$.\\}
\label{table_jetID_performance}
\begin{tabular}[t]{r @{\ $\abs{\eta}<$\ }  l  c  c}
\multicolumn{2}{c}{Region of $\abs{\eta}$} & \multicolumn{2}{c}{Fraction of noise jets rejected}\\
\hline
 & $2.7$ & \multicolumn{2}{c}{99.9\%} \\
$2.7 <$& $3.0$ & \multicolumn{2}{c}{97.6\%}  \\
 $3<$ &$5$ & 15\% (PUPPI) & 35\% (CHS)\\
\end{tabular}
\end{table}

\subsection{Pileup jet rejection}
\label{sec_puid}

Particles resulting from PU collisions will introduce additional jets that do not originate from the LV. These jets are referred to as PU jets. 
PU jets can be classified in two categories: QCD-like PU jets, originating from PU particles from a single PU vertex, and stochastic PU jets, originating from PU particles from multiple different PU vertices. Both PU mitigation techniques, PUPPI and CHS, remove the charged tracks associated with PU vertices, reducing the \pt{} of QCD-like PU jets to roughly 1/3 of their original \pt, such that they can be largely reduced by selections on the jet \pt.
In CMS, a multivariate technique to reject the remaining PU jets (dominated by stochastic PU jets) has been developed
and applied for CHS jets~\cite{CMS-PAS-JME-16-003}, whereas PUPPI intrinsically suppresses PU jets better by rejecting more charged and neutral particles from PU vertices before jet clustering. Both techniques suppress both QCD-like and stochastic PU jets, though the observables used for neutral particle rejection are primarily sensitive to stochastic PU jets.

The performance of the PU jet rejection for both PUPPI and CHS is evaluated in \PZ{}+jets events in data and simulation.
The jet recoiling against the \PZ boson provides a pure sample of LV jets, whereas additional jets are often from PU collisions.
The \PZ{}+jets events are selected by requiring two oppositely charged muons with $\pt>20 \GeV$ and $\abs{\eta}<2.4$ whose combined invariant mass is between 70 and 110\GeV.
Jets that overlap with leptons within $\Delta R (\text{lepton},\,\text{jet})<0.4$ from the \PZ boson decay are removed from the collections of particle- and reconstruction-level jets.

In simulation jets are categorized into four groups based on the separation from particle-level jets and their constituents.
If a reconstruction-level jet has a particle-level jet within $\Delta R < 0.4 $, 
 it is regarded as originating from the LV.
 Jet flavors are defined by associating generated particles to reconstructed jets.
This is done by clustering a new jet with the generated and reconstructed particles together where, in this case,  the four-momenta of generated particles are scaled by a very small number.
Newly reconstructed jets in this way are almost identical to the original jets because
 the added particles, with extremely small energy, do not affect the jet reconstruction.
If a jet originating from the LV contains generated quarks or gluons, it is regarded as a jet of quark or gluon origin, depending on the label of the highest \pt particle-level particle.
If a jet not originating from the LV does not contain any generated particles from the hard scattering, it is regarded as a jet originating from a PU vertex, \ie, a PU jet.
The remaining jets, which do not have nearby particle-level jets but contain particle-level particles (from LV), are labeled as unassigned.

\begin{table}
\centering
\topcaption{List of variables used in the PU jet ID for CHS jets.\\}
\label{table_PUID_inputvariables}
\begin{tabular}[t]{p{2cm} p{10cm}}
Input \newline variable & Definition \\
\hline 
 $\text{LV} \sum \pt$ fraction   & Fraction of \pt of charged particles associated with the LV, defined as $ \sum_{i \in \text{LV} } p_{\mathrm{T},\, i} / \sum_{i} p_{\mathrm{T},\, i} $
 where $i$ iterates over all charged PF particles in the jet\\
 & \\
 $N_{\text{vertices}}$ & Number of vertices in the event \\
 & \\
$ \langle \Delta R^2\rangle $    &
   Square distance from the jet axis scaled by $\pt ^2$ average of jet constituents:
$    \sum _i  \Delta R^2  p_{\mathrm{T},\, i} ^ 2 / \sum _i  p_{\mathrm{T},\, i} ^ 2  $ 
 \\
 & \\
$f_{\text{ringX}}$, $ X =  1, 2, 3, \text{ and }4$ &
Fraction of \pt of the constituents
($\sum  p_{\mathrm{T},\, i} /  \pt^{\text{jet}} $)
 in the region $ R_i < \Delta R < R_{i+1} $ around the jet axis,
 where $R_i = 0, 0.1, 0.2,$ and 0.3 for $X= 1, 2, 3$, and 4 \\
 & \\
$\pt^{\text{lead}}/\pt^{\text{jet}}$       & \pt fraction carried by the leading PF candidate \\
 & \\
$\pt^{\text{l. ch.}}/\pt^{\text{jet}}$    & \pt fraction carried by the leading charged PF candidate \\
 & \\
$\abs{\vec{m}}$   & Pull magnitude, defined as $\abs{ (\sum_{i} \pt ^i \abs{r_i} \vec{r}_i) } / \pt ^{\text{jet}} $ 
 where $\vec{r_i}$ is the direction of the particle $i$ from the direction of the jet
\\
 & \\
$N_{\text{total}}$ & Number of PF candidates \\
 & \\
$N_{\text{charged}}$   & Number of charged PF candidates \\
 & \\
$\sigma_{1}$       &  Major axis of the jet ellipsoid in the $\eta$-$\phi$ space \\
 & \\
$\sigma_{2}$       &  Minor axis of the jet ellipsoid in the $\eta$-$\phi$ space \\
 & \\
 $\pt^{\text{D}}$ &
Jet fragmentation distribution, 
 defined as 
$ \sqrt{  \sum _i p^2 _{\mathrm{T},\, i}}  /  \sum _i p_{\mathrm{T},\, i}  $
 \\ 
\end{tabular}
\end{table}

This identification of PU jets is based on two observations: (i) the majority of tracks associated with PU jets do not come from the LV, and (ii) PU jets contain particles originating from multiple PU collisions and therefore tend to be more broad and diffuse than jets originating from one single quark or gluon.
Table~\ref{table_PUID_inputvariables} summarizes the input variables for a multivariate analysis.
Track-based variables include the $\text{LV} \sum \pt$ fraction and $N_{\text{vertices}}$, where the $\text{LV} \sum \pt$ fraction  is the summed \pt of all charged PF
candidates in the jet originating from the LV, divided by the summed \pt of
all charged candidates in the jet.
The $\text{LV} \sum \pt$ fraction  variable provides the strongest discrimination of any variable included in the discriminator, but is available only within the tracking volume.
The inclusion of the $N_{\text{vertices}}$ variable allows the multivariate analysis to determine the optimal discriminating variables as the PU is increased.
Jet shape variables included in the multivariate discriminant are as follows: 
$\langle\Delta R^2 \rangle$,  
$f_{\text{ring0}}$,  
$f_{\text{ring1}}$,  
$f_{\text{ring2}}$,  
$f_{\text{ring3}}$,
$\pt^{\text{lead}}/\pt^{\text{jet}}$,   
$\abs{\vec{m}}$, 
$N_{\text{total}}$,  
$N_{\text{charged}}$,  
major axis ($\sigma_{1}$),   
minor axis ($\sigma_{2}$), and  
$\pt^\mathrm{D}$, with their definitions given in Table~\ref{table_PUID_inputvariables}.
Pileup jets tend to have $ \langle \Delta R^2\rangle $ of 
large value relative to genuine jets.
For the set of $f_{\text{ringX}}$,
 PU jets tend to have large values for variables with large $R$,
 which represents the characteristic of PU jets
 having a large fraction of energy deposited in the outer annulus.
Most of the other variables are included to distinguish quark jets from gluon jets, and thus enhance the separation from PU jets.
In particular, the variable $\pt^\mathrm{D}$ tends to be larger for quark jets than for gluon jets, and smaller than both quark jets and gluon jets for PU jets.
The $N_\text{total}$, $\pt^\mathrm{D}$ and $\sigma_{2}$ variables have previously been used for a dedicated quark- and gluon-separation technique; more details on their definition and performance are found in Ref.~\cite{CMS-PAS-JME-16-003}.

\begin{figure}[hbtp]
  \centering
    \includegraphics[width=0.48\textwidth]{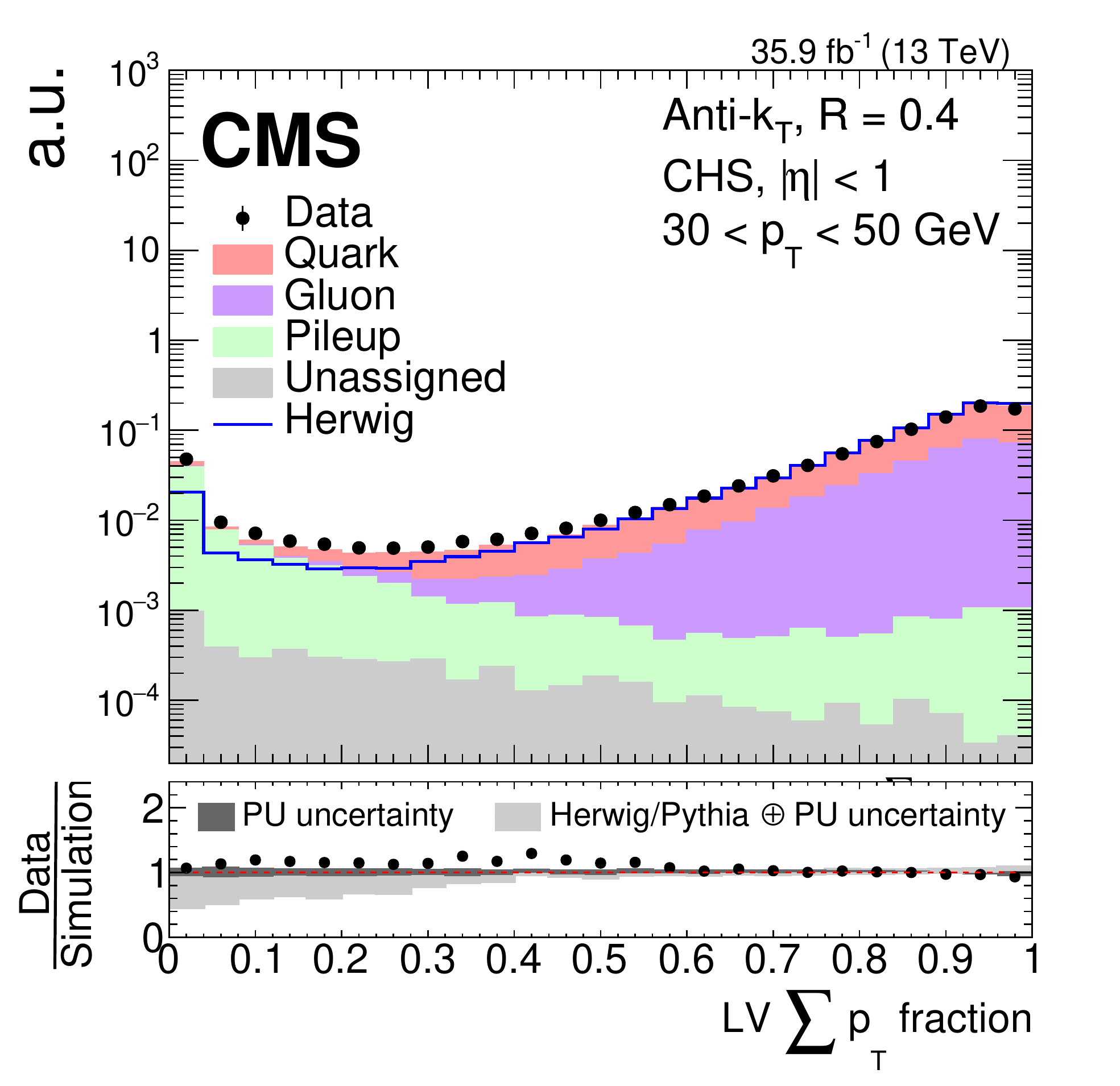}
    \includegraphics[width=0.48\textwidth]{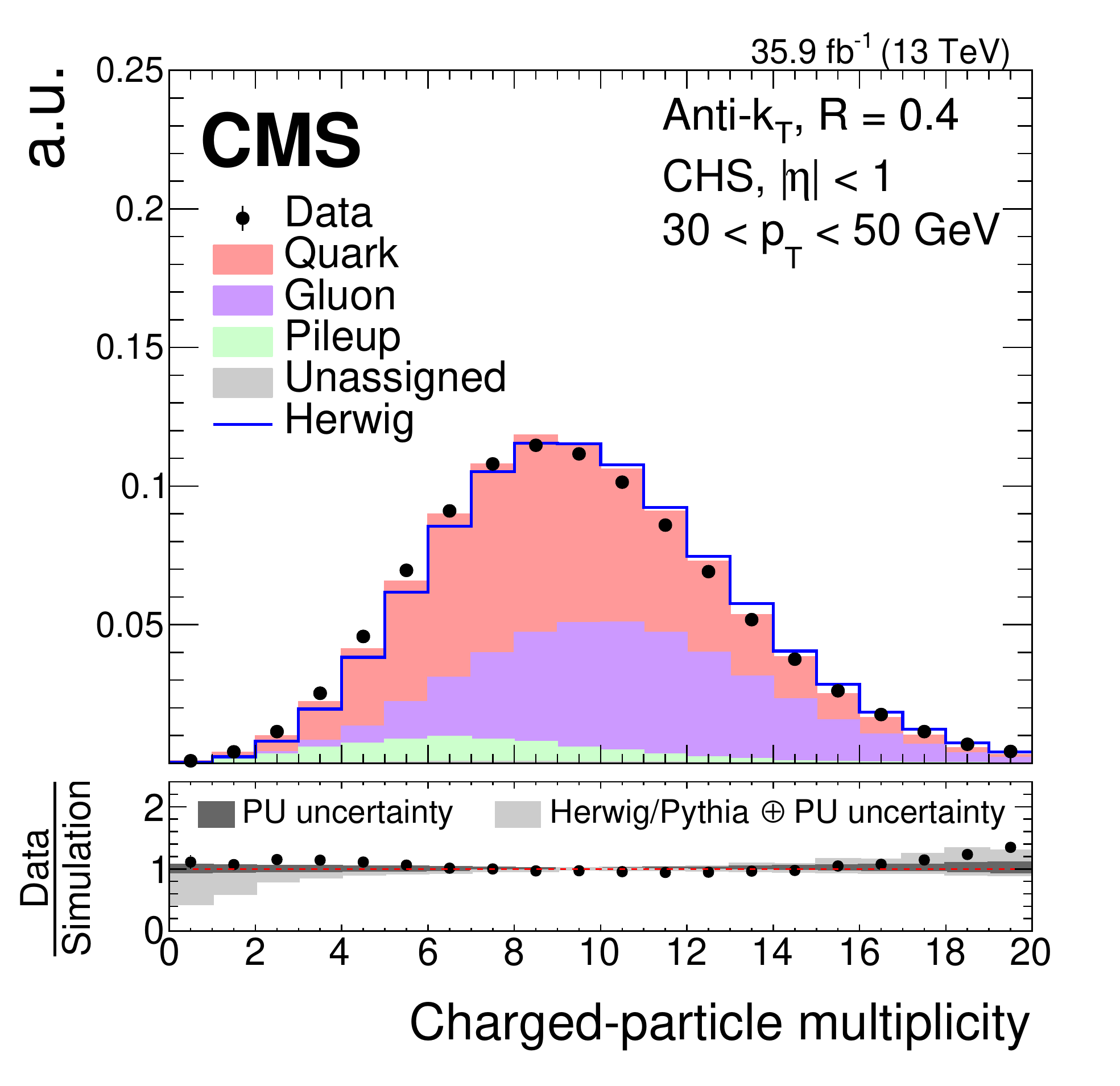}
    \caption{Data-to-simulation comparison for two input variables to the PU jet ID calculation for CHS jets with $30 < \pt < 50 \GeV$: the $\text{LV} \sum \pt$ fraction (left) and charged-particle multiplicity (right). Black markers represent the data while the colored areas are \PZ{}+jets simulation events. The simulation sample is split into jets originating from quarks (red), gluons (purple), PU (green), and jets that could not be assigned (gray). The distributions are normalized to unity. The shape of a sample showered with \HERWIG{}++ is superimposed.
The lower panels show the data-to-simulation ratio along with a gray band corresponding to the one-sided uncertainty, which is the difference between simulated Z+jets events showered with the PYTHIA parton shower and those showered with the HERWIG++ parton shower.    
     Also included in the ratio panel is the PU rate uncertainty (dark gray).} 
    \label{fig:puid_examplevariables}
\end{figure}

\figrefb{fig:puid_examplevariables} shows the distribution of
 the $\text{LV} \sum \pt$ fraction  and the charged-particle multiplicity 
 of jets with $30 < \pt < 50 \GeV$ and $\abs{\eta} < 1 $
in data and simulation.
The distributions of the variables in selected data events agree with simulation within the uncertainties, with a clear separation in the discriminating variables between LV and PU jets.

The set of 15 variables listed in Table~\ref{table_PUID_inputvariables} is used to train a boosted decision tree (BDT) algorithm,
 and to distinguish between jets from the LV and PU jets. For the BDT training, \MGvATNLO \PZ{}+jets simulation events are used. 
To perform the training, reconstruction-level jets that are within a distance of $\Delta R < 0.4$  from any particle-level jet are regarded as jets from the LV, and the remaining jets are identified as PU jets.
A jet is considered to satisfy the PU jet ID if it passes certain thresholds on the output of the BDT discriminator. This output is dependent on the $\eta$ and $\pt$ of the jet. Three working points are considered in the following resulting in different efficiencies and misidentification rates.
These working points are defined by their average efficiency on quark-initiated jets. The definitions are: 
 \begin{itemize}
 \item tight working point: 80\% efficient for quark jets,
 \item medium working point: 90\% efficient for quark jets,
 \item loose working point: 99\% efficient for quark jets in $\abs{\eta}<2.5$, 95\% efficient for quark jets in $\abs{\eta}>2.5$.
 \end{itemize}
Since 92\% of the PU jets tend to occur at $\pt < 50 \GeV$, the contamination from PU jets with $\pt>50 \GeV$ is small. Thus, the PU jet ID is designed to act only on jets with $\pt<50 \GeV$. 

\begin{figure}[hbtp]
	\centering
		\includegraphics[width=0.48\textwidth]{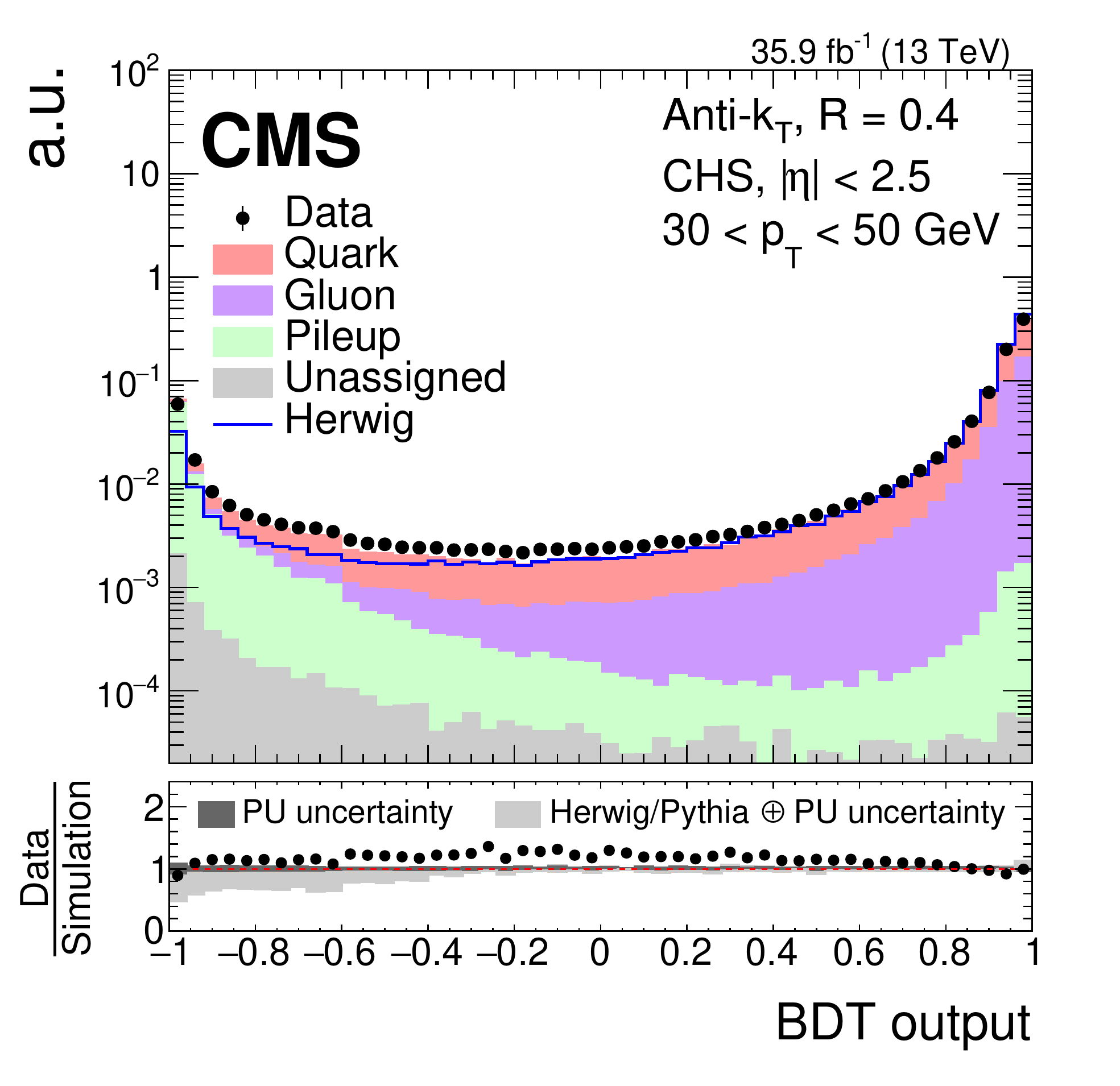}
		\includegraphics[width=0.48\textwidth]{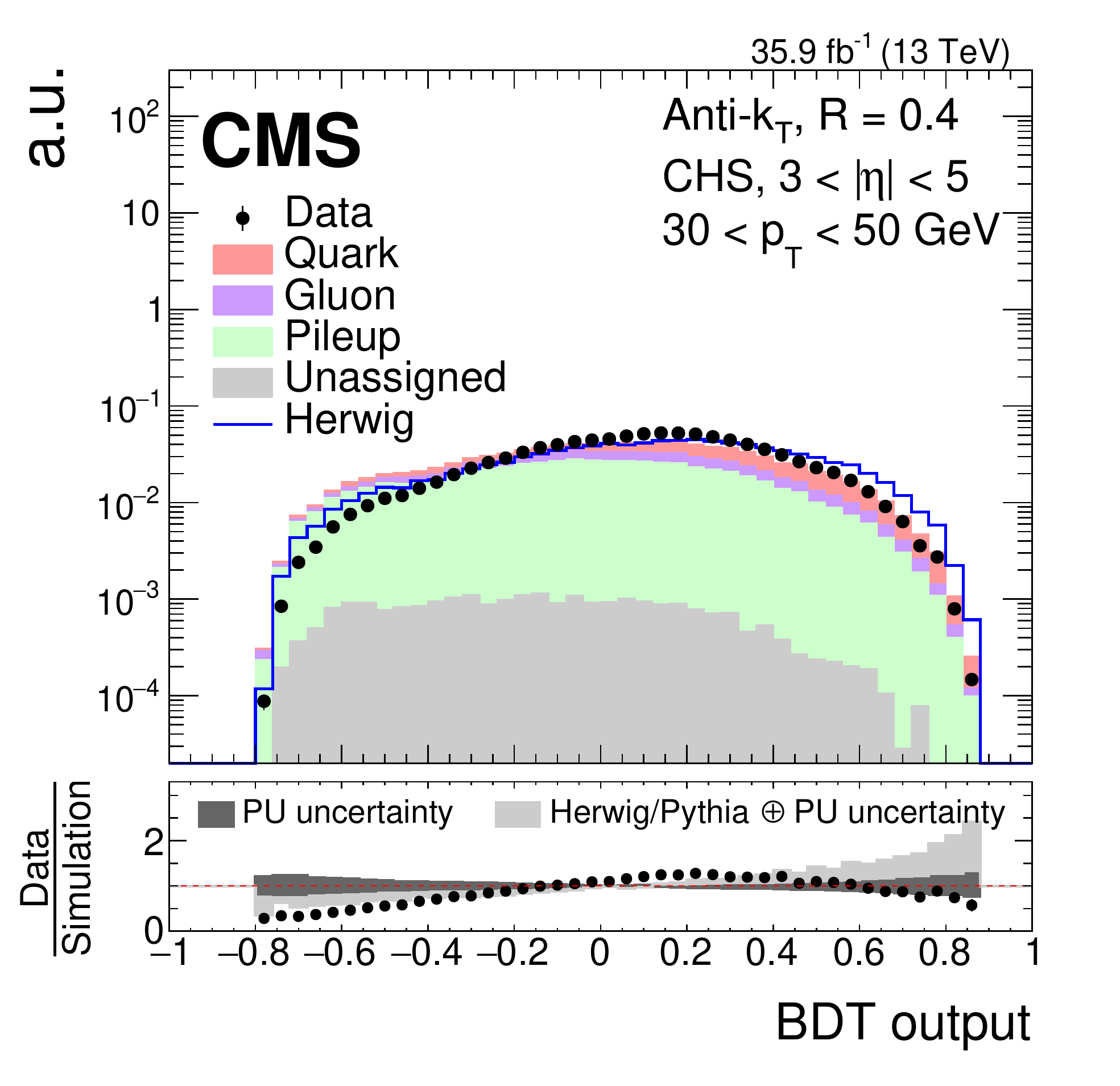}
		\caption{Data-to-simulation comparison of the PU jet ID boosted decision tree (BDT) output for AK4 CHS jets with $30 < \pt <50 \GeV$ for the detector region within the tracker volume (left) and $3 < \abs{\eta}<5$ (right). Black markers represent the data while the colored areas are \PZ{}+jets simulation events. The simulation sample is split into jets originating from quarks (red), gluons (purple), PU (green), and jets that could not be assigned (gray). The distributions are normalized to unity. The shape of a sample showered with \HERWIG{}++ is superimposed
The lower panels show the data-to-simulation ratio along with a gray band corresponding to the one-sided uncertainty that is the difference between simulated Z+jets events showered with the PYTHIA parton shower to those showered with the HERWIG++ parton shower.   		
Also included in the ratio panel is the PU rate uncertainty (dark gray).} 
		\label{fig:puid_bdt}
\end{figure}

The fraction of PU jets in simulation passing this kinematic event selection is 10\% for $\abs{\eta}<2.5$, 48\% for $2.50<\abs{\eta}<2.75$, 59\% for $2.75 <\abs{\eta}<3.00$, and 65\% for $3 < \abs{\eta}<5$.
The distribution of the output BDT discriminator in selected data events and simulation is shown in \figref{fig:puid_bdt}.
Some disagreement is present between the data and simulation. This disagreement is largest for $\abs{\eta}>2.5$ and at low discrimination values, where PU jets dominate. The difference between data and simulation is roughly comparable to the total uncertainty in simulation, considering the uncertainty in the number of interactions and the difference to an alternative \HERWIG{}++-based parton shower prediction.

\begin{figure}[hbtp]
	\centering
		\includegraphics[width=0.45\textwidth]{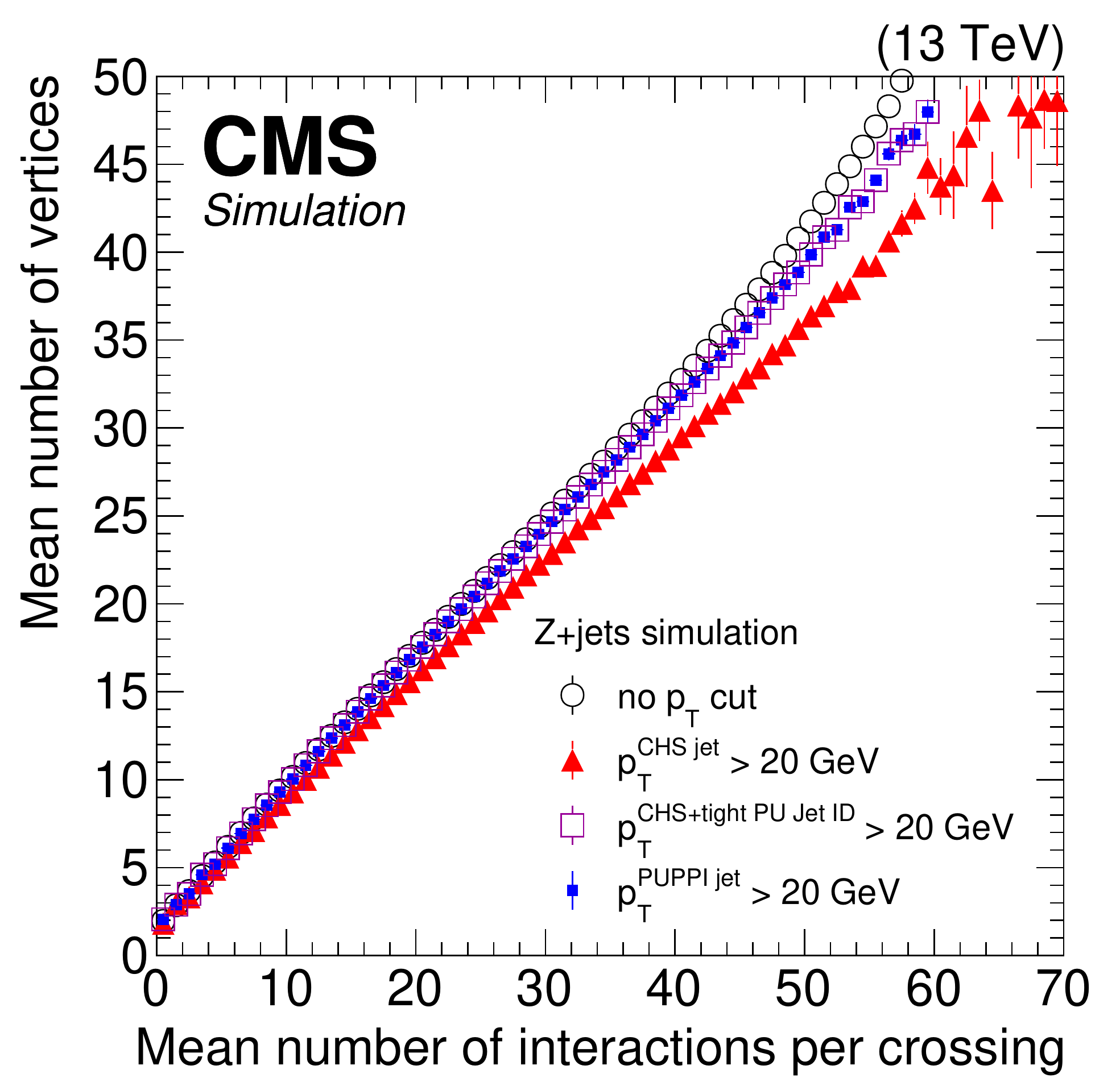}
		\includegraphics[width=0.45\textwidth]{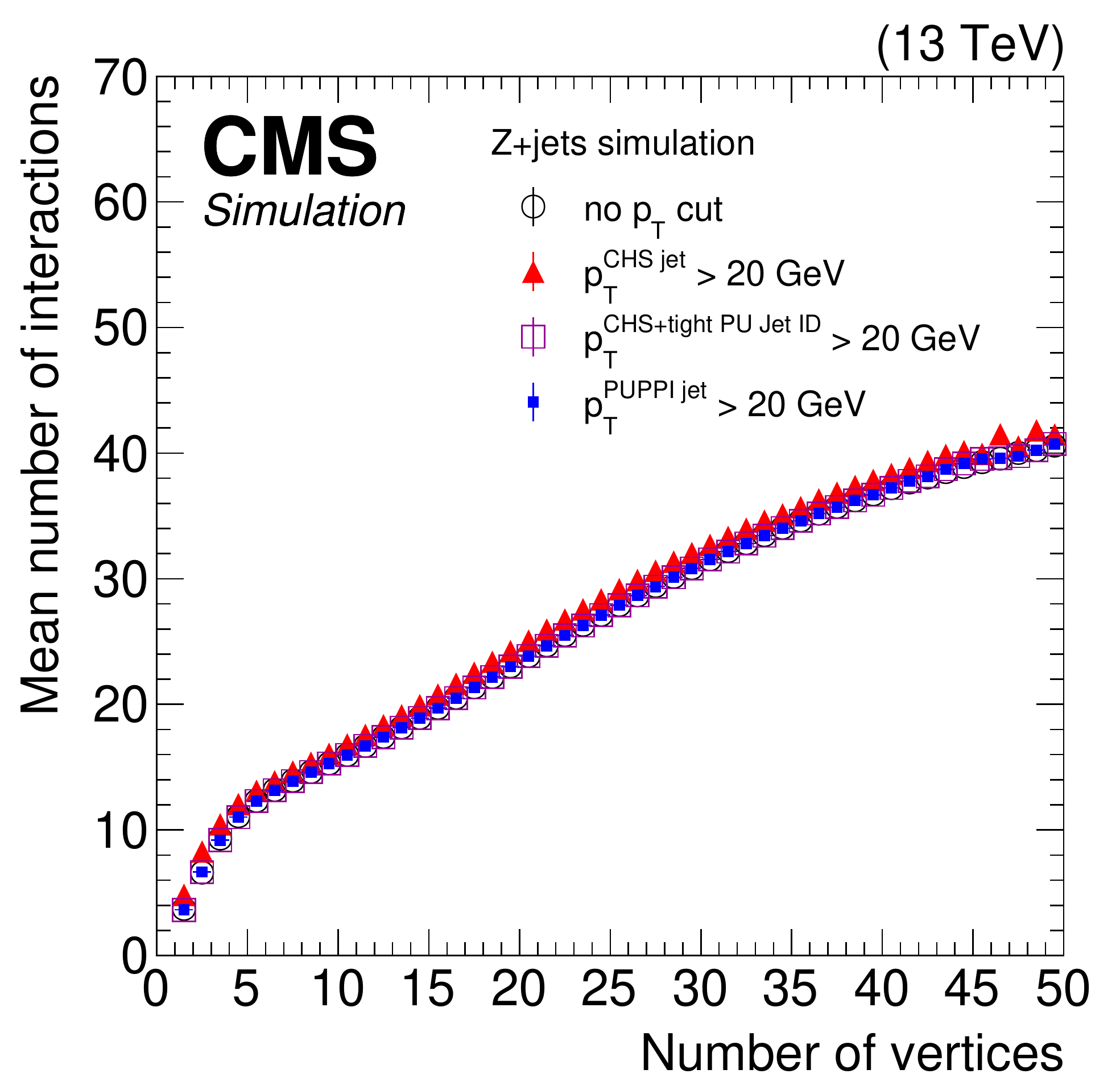}
		\caption{Left: distribution of mean number of reconstructed vertices as a function of the mean number of interactions in \PZ{}+jets simulation. 
			Right: distribution of the mean number of interactions as a function of the number of vertices in \PZ{}+jets simulation. The black open circles show the behavior without applying any event selection, while for the other markers a selection on jets of $\pt > 20\GeV$ is applied using the CHS (full red triangles), CHS+tight PU jet ID (violet open squares), and PUPPI (full blue squares) algorithms. The error bars correspond to the statistical uncertainty in the simulation.}
		\label{fig:pu_conditions2}
\end{figure}

When studying jet performance with PU, it is clear that jet reconstruction and selection, including PU mitigation, affect the relationship between the number of reconstructed vertices and the mean number of interactions per crossing. The mean number of vertices as a function of the number of interactions can be seen in Fig.~\ref{fig:pu_conditions2} (left). 
Without jet selection, the number of vertices is on average 30\% smaller~\cite{CMS-DP-2017-015,Chatrchyan:2014fea} than the number of interactions, because the vertex reconstruction and identification efficiency is about 70\% (although it is nearly 100\% for hard-scattering interactions).
When introducing a selection on the jet \pt{}, the mean number of vertices for a given number of interactions is reduced.
This effect is largest for CHS jets, where no treatment of jets composed of mostly PU particles is present.
If a PU vertex is close to or overlaps with the LV, jets composed of PU particles end up in the event reconstruction and cause the observed bias.
When applying a technique to reduce the number of additional jets composed of mostly PU particles (PUPPI or CHS+tight PU jet ID), the relationship shows a behavior more similar to the one without selection.
The mean number of interactions as a function of the number of vertices is presented in Fig.~\ref{fig:pu_conditions2} (right).
This relationship depends on the assumed distribution of pileup interactions in data and is adjusted to match the 2016 data taking.
The largest difference between events with and without a \pt cut is observed for a high number of vertices, while the different PU mitigation techniques show a similar behavior.

\begin{figure}[hbtp]
  \centering
    \includegraphics[width=0.45\textwidth]{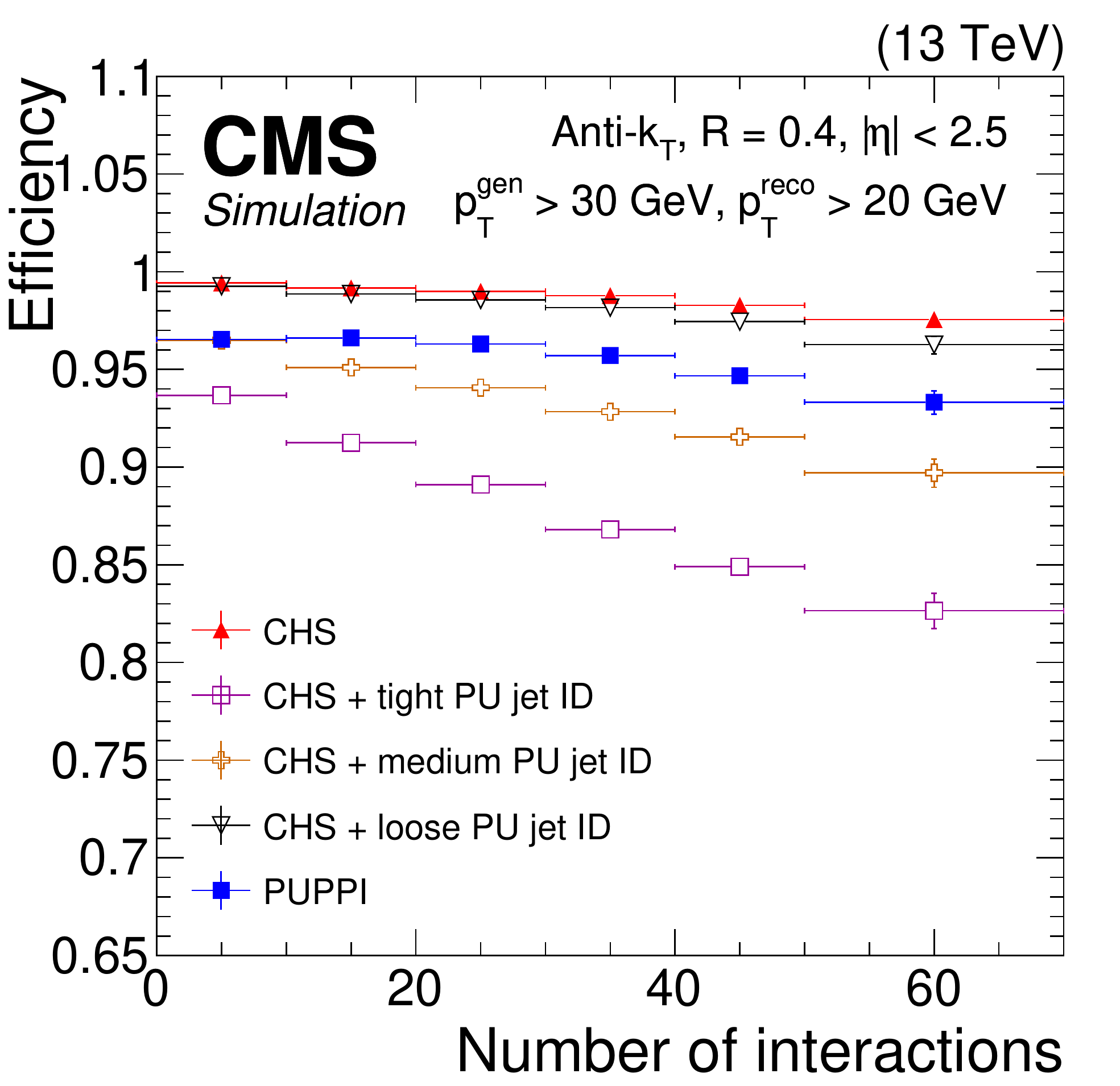}
    \includegraphics[width=0.45\textwidth]{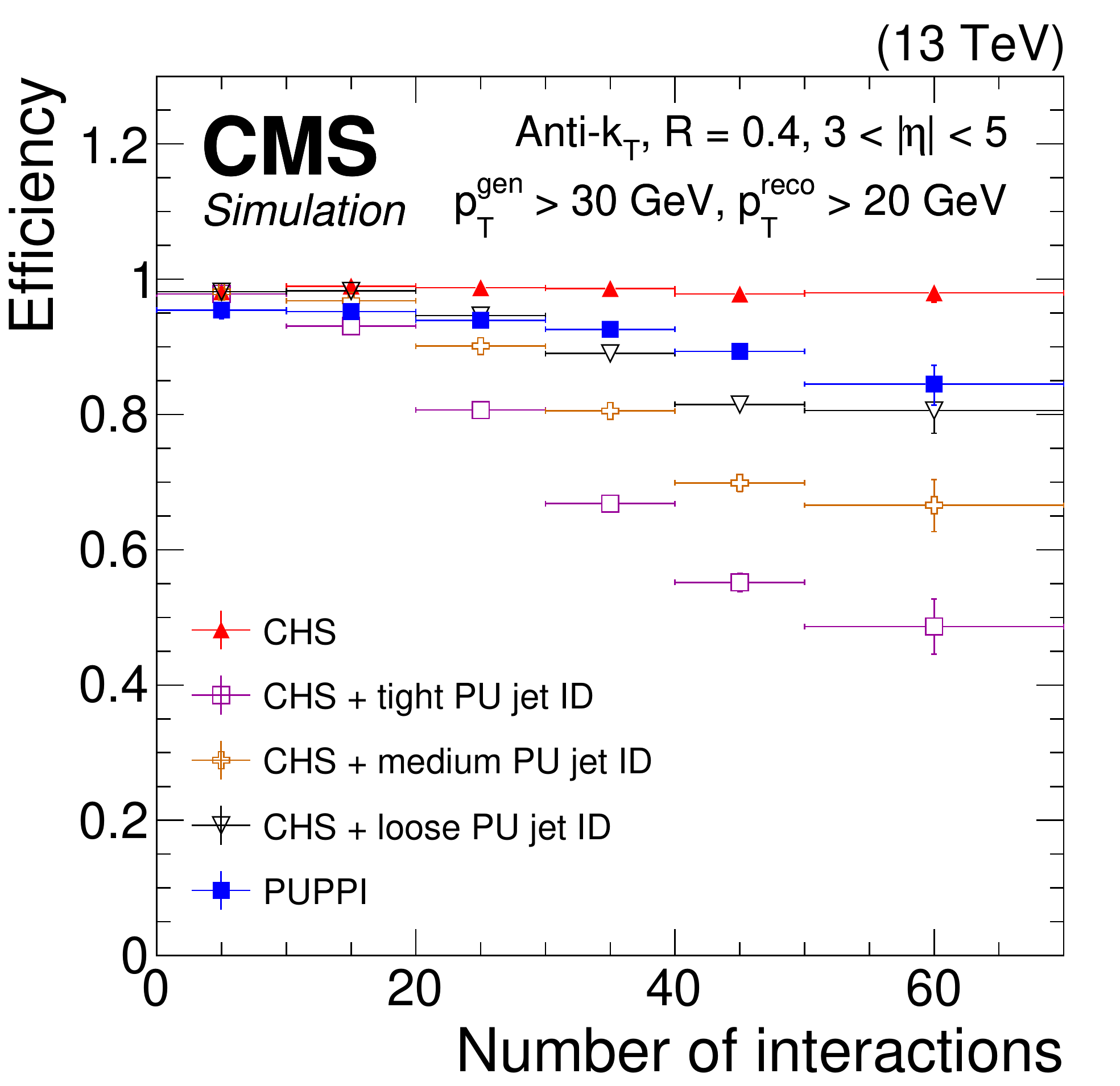}
    \includegraphics[width=0.45\textwidth]{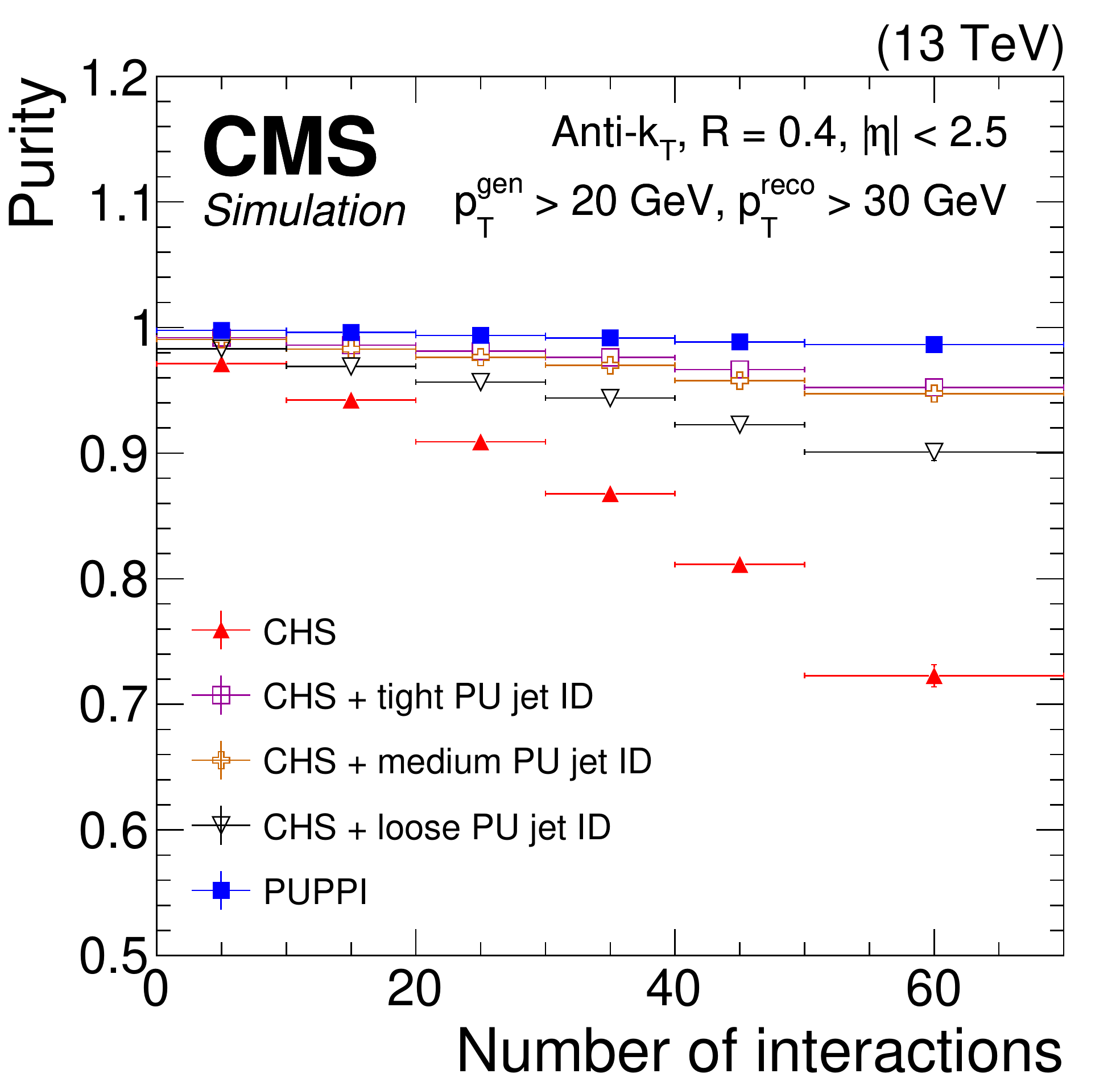}
    \includegraphics[width=0.45\textwidth]{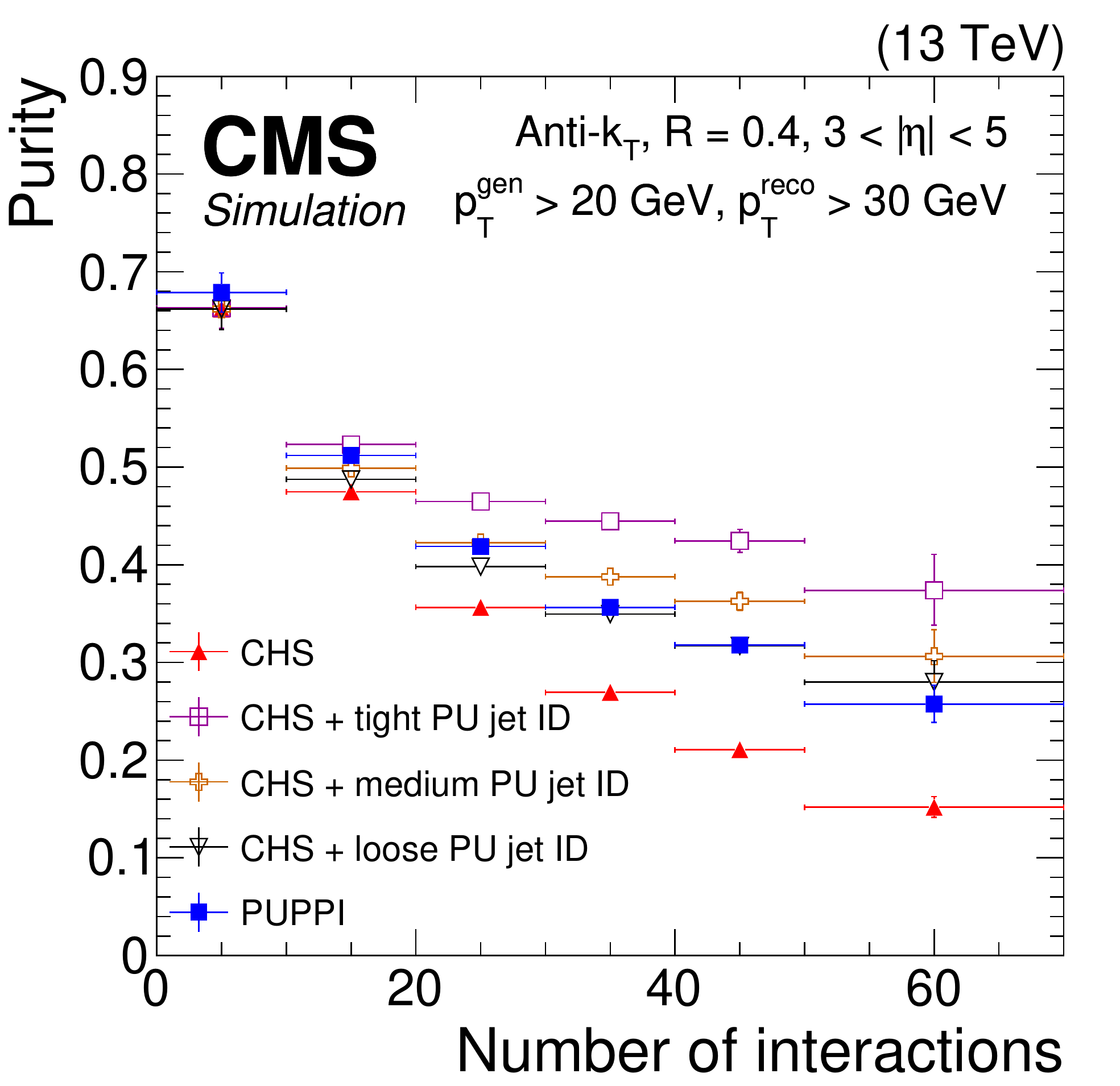}
    \caption{The LV jet efficiency (upper) and purity (lower) in \PZ{}+jets simulation as a function of the number of interactions for PUPPI (blue closed squares), CHS (red closed triangles), CHS+tight PU jet ID (magenta open squares), CHS+medium PU jet ID (orange crosses), and CHS+loose PU jet ID (black triangles).
      Plots are shown for AK4 jets $\pt>20 \GeV$, and (left) $\abs{\eta} < 2.5$ and (right) $\abs{\eta} > 3$.
The LV jet efficiency is defined as the number of matched reconstruction-level jets with $\pt>20 \GeV$ divided by the number of
particle-level jets with $\pt>30 \GeV$ that originate from the main interaction. For the lower plots, the purity is defined as the number of matched particle-level jets with $\pt > 20\GeV$ divided by the number of reconstructed jets that have $\pt>30 \GeV$. The error bars correspond to the statistical uncertainty in the simulation.}
    \label{fig:jet_pu2}
\end{figure}

\figrefb{fig:jet_pu2} shows the LV jet efficiency and purity in \PZ{}+jets simulation as a function of the number of interactions for CHS jets, CHS jets with a PU jet ID applied, and PUPPI jets.
The efficiency is defined as the fraction of particle-level jets with $\pt>30 \GeV$ that match within $\Delta R<0.4$ with a reconstruction-level jet with $\pt>20 \GeV$.
The purity is defined as the fraction of reconstruction-level jets with $\pt>30 \GeV$ that match within $\Delta R<0.4$ with a particle-level jet with $\pt>20 \GeV$ from the main interaction.
The \pt cuts at reconstruction and generator level are chosen to be different to remove any significant JER effects on this measurement.

For CHS jets, the efficiency is larger than 95\% in entire detector region up to $\abs{\eta}<5$ regardless of the number of interactions.
However, the purity drops strongly with the number of interactions down to 70 and 18\% at 50 interactions for the regions of $\abs{\eta}<2.5$ and $\abs{\eta}>2.5$, respectively.
The PU jet ID applied on top of CHS reduces the efficiency with respect to using only CHS, but at the same time improves the purity, especially for low-\pt jets.
In $\abs{\eta}<2.5$, the loose working point has only a slightly reduced efficiency compared to CHS alone. In $\abs{\eta}>2.5$, the efficiency drops to roughly 80\% at high PU for the loose working point.
In $\abs{\eta}<2.5$, the purity remains constant at around 98\% over the whole range of PU scenarios. In $\abs{\eta}>2.5$, the purity is PU-dependent, but improves over CHS alone by a factor of 1.7 at high PU for the loose working point.
The tight PU jet ID achieves the best purity in $\abs{\eta}>2.5$ at 40\% with collisions at 50 interactions and a jet efficiency of 45\%.
PUPPI also reduces the efficiency with respect to CHS by removing neutral particles. At the same time, PUPPI improves the purity by removing PU jets from the event without the need of a PU jet ID.
At low PU (below 10 interactions), the purity of PUPPI jets is equal to that of CHS.  At high PU, the purity of PUPPI jets with respect to CHS jets is significantly higher than that of CHS jets. 
PUPPI has a constant efficiency above 95\% in $\abs{\eta}<2.5$, and a purity compatible with the tight PU jet ID working point at high PU.
In $\abs{\eta}>2.5$, above 30 interactions the efficiency of PUPPI is better than the loose PU jet ID, whereas the purity is compatible to within a few percent to the loose PU jet ID.
In summary, PUPPI shows an intrinsic good balance between efficiency and purity compared to CHS, but if purity in $\abs{\eta}>2.5$ is crucial to an analysis, CHS+tight PU jet ID yields better performance.
Using variables designed to distinguish quark jets from gluon jets results in a $<1\%$ difference for $20<\mathrm{PU}<30$ in efficiency for PUPPI and CHS in $\abs{\eta}<2.5$ and range up to 5\% (12\%) in $\abs{\eta}>3$ for PUPPI (CHS) with tight PU ID.

\begin{figure}[hbtp]
  \centering
    \includegraphics[width=0.45\textwidth]{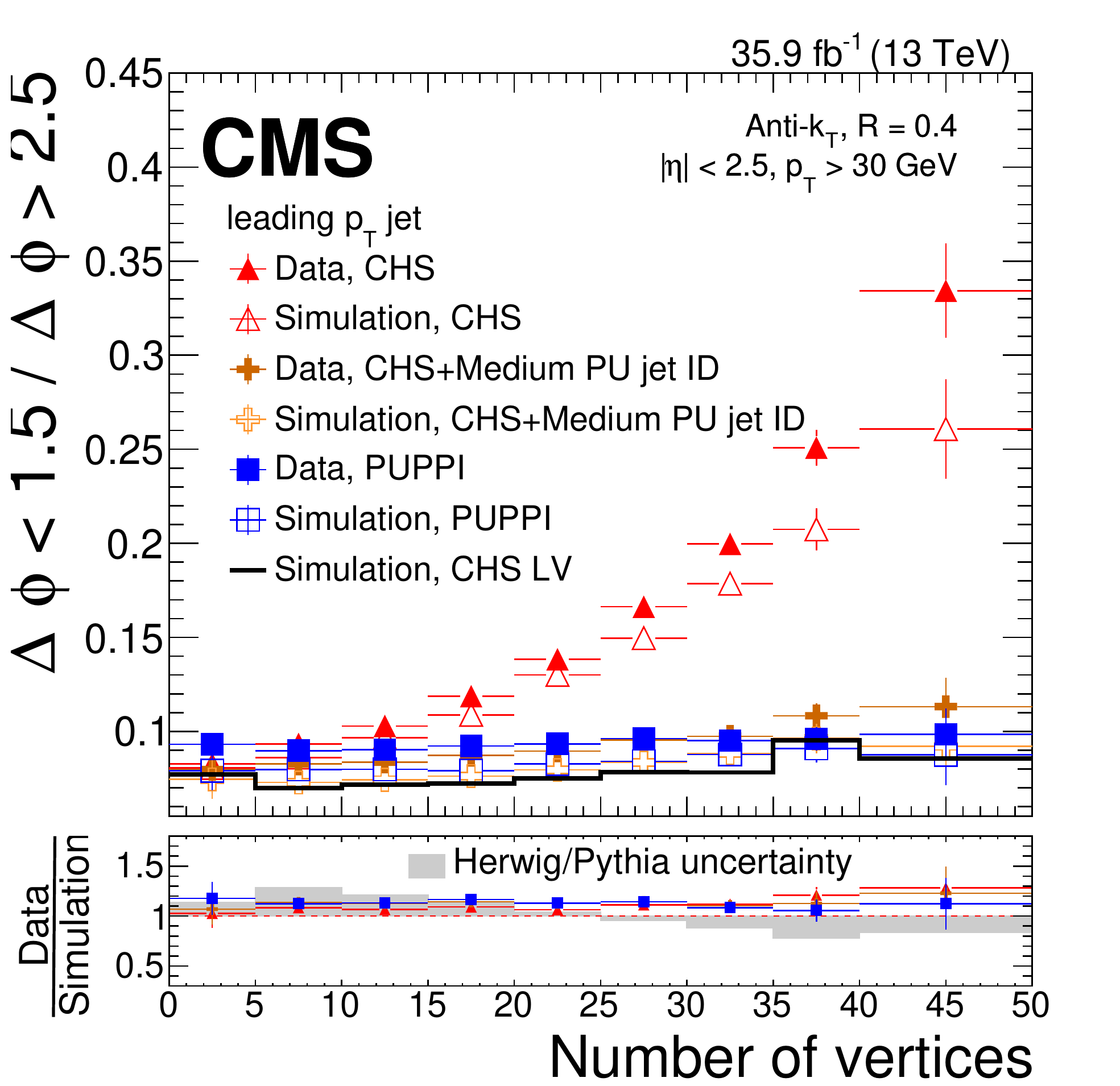}
    \includegraphics[width=0.45\textwidth]{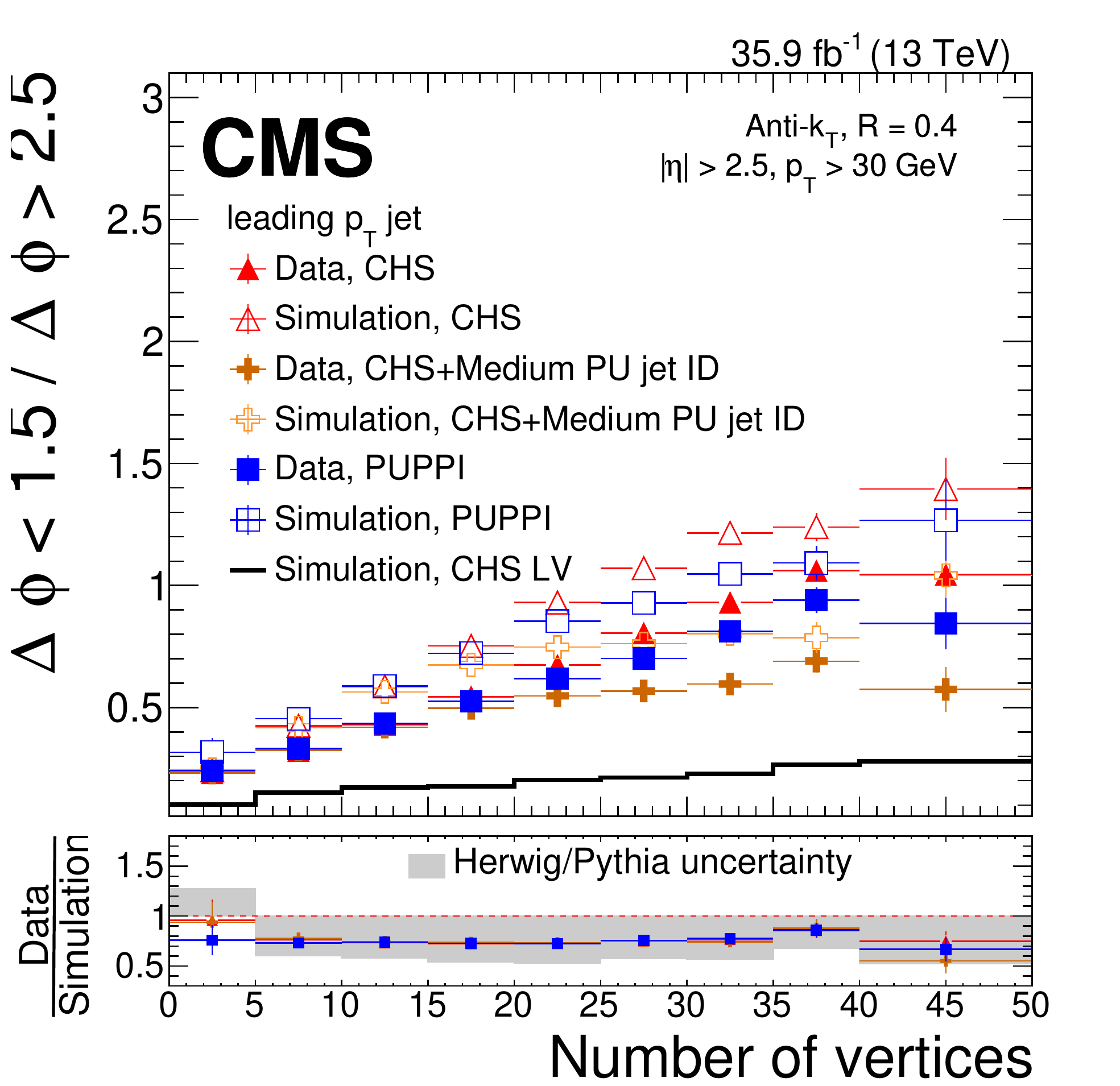}
    \caption{Rate of jets in the PU-enriched region divided by the rate of jets in the LV-enriched region as a function of the number of vertices for CHS jets (red triangles), CHS jets with medium PU jet ID applied (orange crosses) and PUPPI jets (blue squares) in \PZ{}+jets simulation (open markers), and data (full markers). 
For reference, the rate of jets that are matched to a particle-level jet in simulation is also shown for CHS jets (black solid line).    
    The plots show the ratio for events with $\abs{\eta}<2.5$ (left) and $\abs{\eta}>2.5$ (right). The lower panels show the data-to-simulation ratio along with a gray band corresponding to the one-sided uncertainty that is the difference between simulated \PZ{}+jets events showered with the \PYTHIA parton shower to those showered with the \HERWIG{}++ parton shower.}
    \label{fig:jet_pu3}
\end{figure}

To evaluate the performance of PU jet identification in data, the ratio of PU jets to genuine jets for the leading \pt jet in the event is studied. 
Events are split into two categories to compare both PU and LV jets. The categorization is performed utilizing 
the difference between the azimuths  $\phi$ of the leading \pt jet and the \PZ boson.
The PU-enriched events are required to have $\Delta \phi(\PZ\,\text{boson},\text{jet})<1.5$, while events enriched in LV jets are required to have $\Delta \phi(\PZ\, \text{boson},\text{jet})>2.5$.
\figrefb{fig:jet_pu3} shows the rate of events in the PU-enriched region divided by the rate of events in the LV-enriched region, as a function of the number of vertices for CHS jets, CHS jets with medium PU jet ID applied, and PUPPI jets in \PZ{}+jets simulation and data.
The rate of PU-enriched events selecting CHS jets alone exhibits a strong dependence on the number of vertices in detector regions where $\abs{\eta}<2.5$. This dependence increases from 8 to 25\% when going from 5 to 40 vertices. The dependence is strongly reduced when the PU jet ID is applied or PUPPI is utilized.
PUPPI shows a stable behavior across the whole range in $\abs{\eta}<2.5$ for both data and simulation.
For $\abs{\eta}>2.5$, all three algorithms show a PU dependence with CHS jets having the worst performance.
Furthermore, categorization with PUPPI jets has a PU-enriched rate between that of events categorized with CHS and CHS+medium PU jet ID.
For reference, the rate of jets that are matched to a particle-level jet in simulation is also shown for CHS jets (simulation, CHS LV). This line shows the expected ratio of events in the two regions when only the LV jets are used for the categorization. This curve shows a slight PU dependence because of the high matching parameter of generator- with reconstruction-level jets ($\Delta R < 0.4$).

Scale factors for the efficiency of data and simulation for both matched jets from the LV and PU jets for various PU jet ID working points are derived using the event categories enriched in genuine jets and PU jets. Scale factors are within a few percent of unity in the detector region where $\abs{\eta}<2.5$. In $\abs{\eta}>2.5$, they are farther from unity, with differences up to 10\% for jets with $2.5 < \abs{\eta} <3.0$ and the tight working point applied. The scale factor for PU jets is significantly larger and leads to a visible disagreement in \figref{fig:jet_pu3}. This disagreement is found to be as large as 30\% for low \pt jets with $\abs{\eta}>2.5$. 
The difference in modeling when using \HERWIG{}++ instead of \PYTHIA for parton showering shown in the lower panel of \figref{fig:jet_pu3} is considered as an additional uncertainty.
The difference of data with respect to \PYTHIA showered jets is contained within the total variation when considering both \HERWIG{}++ and \PYTHIA based parton showers. 

\section{\texorpdfstring{\PW{}}{W}, \texorpdfstring{\PZ{}}{Z}, Higgs boson, and top quark identification}
\label{sec_fatjet}

\subsection{Jet substructure reconstruction}

In various searches for new physics phenomena and measurements of standard model properties, 
top quarks, \PW{}, \PZ{}, and Higgs bosons are important probes.
They can be produced with a large Lorentz boost, $\gamma$, such that the direction of their decay particles becomes very collinear.
The spatial separation between the decay products in the $\eta$-$\phi$ plane is approximately
$\Delta R \approx 2 / \gamma$.
In such cases, it is difficult to reconstruct the decay products of the hadronically decaying objects of interest properly with traditional jets of size 0.4.
As a result, techniques to reconstruct all decay products within one jet with a larger size of 0.8 have been widely studied and used~\cite{Khachatryan:2014vla,CMS-PAS-JME-18-002}.
The invariant mass and substructure of the reconstructed jets are typically used to identify the different bosons and top quarks.
These larger cone size jets tend to collect more PU, hence PU mitigation techniques are relevant across a larger \pt range, extending to well beyond $\pt > 100\GeV$. In addition, the jet mass and substructure variables are particularly affected by soft and wide-angle radiation.
A grooming technique is applied on top of CHS and PUPPI to remove soft radiation from the jet-clustering algorithm and thereby mitigate the effects from PU, underlying event, and initial-state radiation.
The main grooming algorithm used in CMS is the soft drop or modified mass drop tagger~\cite{Larkoski:2014wba,Dasgupta:2013ihk}.
It reclusters a jet with the Cambridge--Aachen algorithm~\cite{Dokshitzer:1997in}, and 
 splits the jet in two subjets by undoing the last step of the jet clustering.
It regards the jet as the final soft drop jet if the two subjets satisfy the condition 
\begin{linenomath}
\begin{equation}
 \frac{ \min ( \pt^1, \pt^2 ) }{\pt^1 + \pt^2  } > z_{\text{cut}} \Big(  \frac{ \Delta R_{12} }{ R_ 0  } \Big) ^ \beta,
\end{equation}
\end{linenomath}
 where $\pt^{1}$ and $\pt^{2}$ are the transverse momenta of the two subjets, $R_0$ is the size parameter of the jet, 
$\Delta R_{12} = \sqrt{\smash[b]{(\Delta \eta_{12})^2 + (\Delta \phi_{12})^2}}$ is the distance between the two subjets, and $z_{\text{cut}} $ and $\beta$ are tunable parameters of soft drop set to $z_{\text{cut}} = 0.1$ and $\beta = 0$ here.
If the soft drop condition is not met, the declustering procedure is repeated with the subjet that has the larger \pt of the two, and the other subjet is rejected.
The soft drop jet mass is computed from the sum of the four-momenta of the constituents passing the grooming algorithm. The mass is then corrected by a factor derived in simulated \PW boson samples to ensure a $\pt$- and $\eta$-independent jet mass distribution~\cite{CMS-PAS-JME-16-003}.

Additional separation of boosted \PW{}, \PZ{}, and Higgs bosons, and top quarks from quark and gluon jet background can be achieved with a substructure observable.
A commonly used observable in CMS is $N$-subjettiness~\cite{Thaler:2011gf}, defined as 
\begin{linenomath}
\begin{equation}
\tau_{N} = \frac{1}{d_{0}}\sum_{k} p_{\mathrm{T} k}\, \min(\Delta R_{1,k}, \Delta R_{2,k},\ldots, \Delta R_{N,k}),
\end{equation}
\end{linenomath}
 with the normalization factor $d_0$:
\begin{linenomath}
\begin{equation}
d_0 = \sum_{k}p_{\mathrm{T} k}\, R_0,
\end{equation}
\end{linenomath}
where $R_0$ is the size parameter used in the clustering process, 
$p_{\mathrm{T} k}$ is the transverse momentum of the $k$-th constituent
 of the jet, and $\Delta R_{n,k} $ estimates
 the angular separation of the constituents of the jet to the closest subjet axis.
We use a one-step optimization of the exclusive $\kt$ axes as a definition for the subjet axes.
The ratio $\tau_{2}/\tau_{1}$, which is called $\tau_{21}$, has
 excellent capability in separating jets with bipolar structures, 
originating from boosted \PW{}, \PZ{}, and Higgs bosons, from jets coming from quarks and gluons. 
The ratio $\tau_{32}=\tau_{3}/\tau_{2}$ can be used to discriminate top quark jets from \PW{}, \PZ{}, and Higgs boson jets, or quark and gluon jets.

\subsection{Identification performance and pileup}

\begin{figure}[hbtp]
  \centering
    \includegraphics[width=0.45\textwidth]{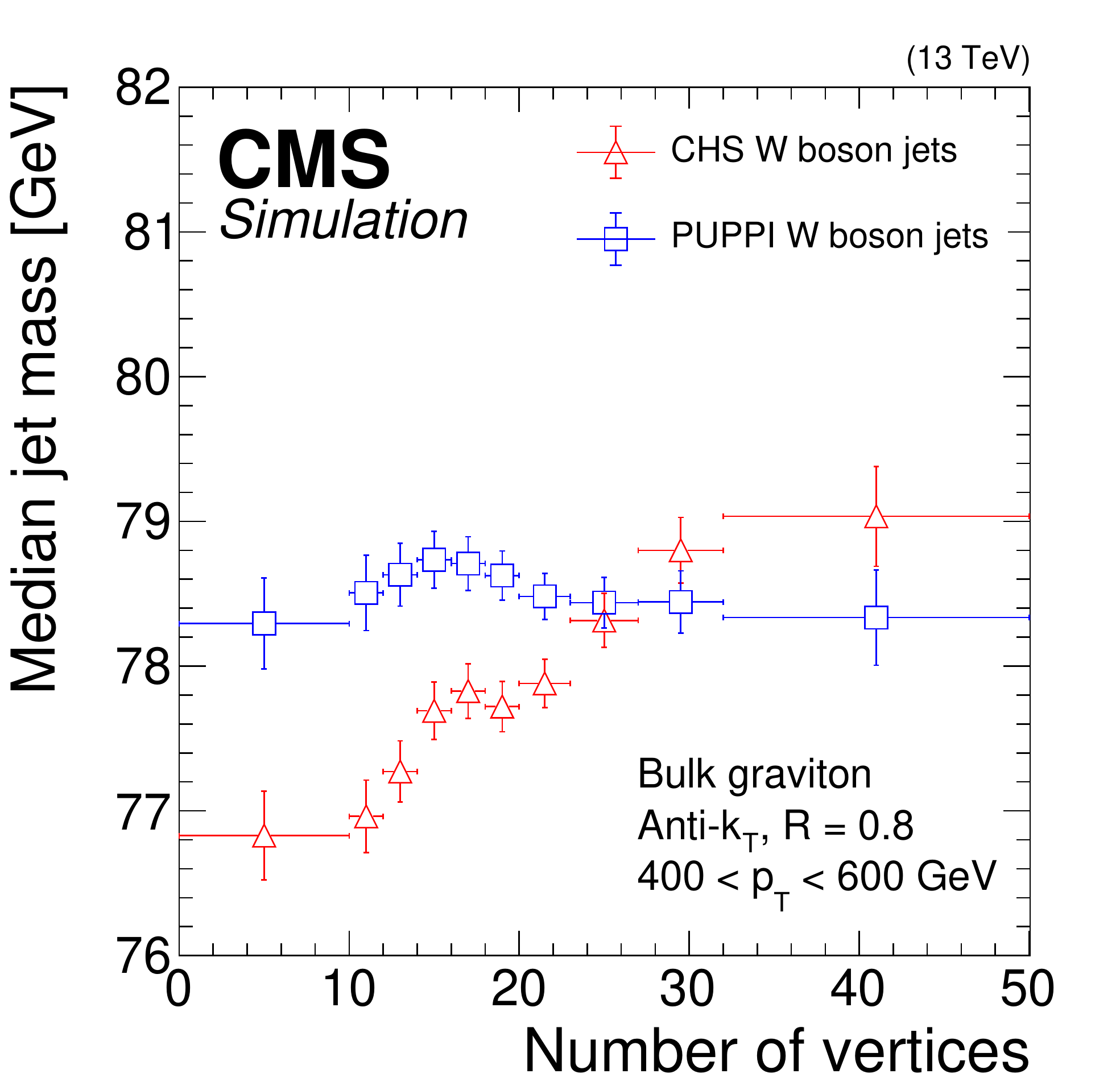}
    \includegraphics[width=0.45\textwidth]{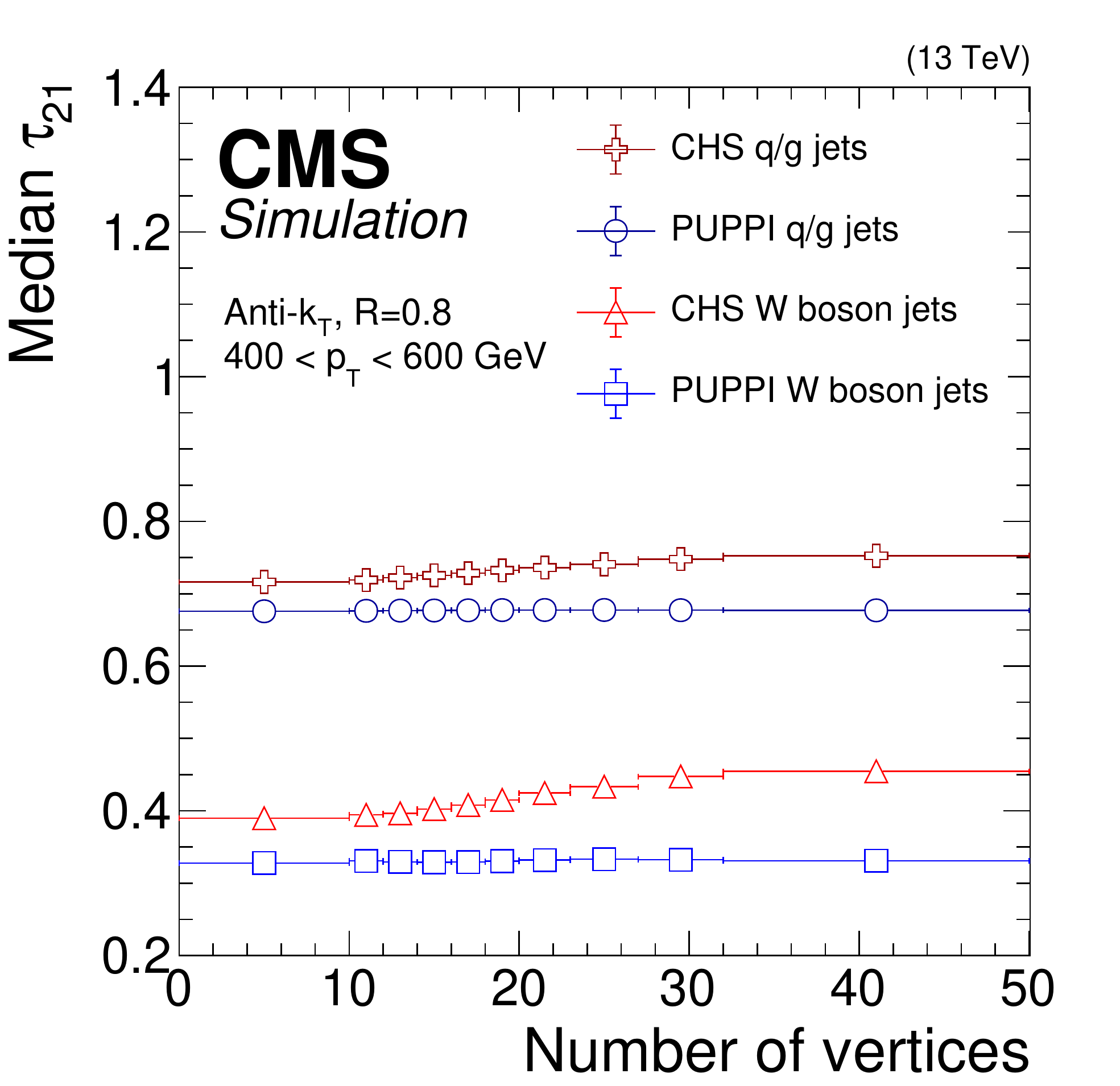}
    \includegraphics[width=0.45\textwidth]{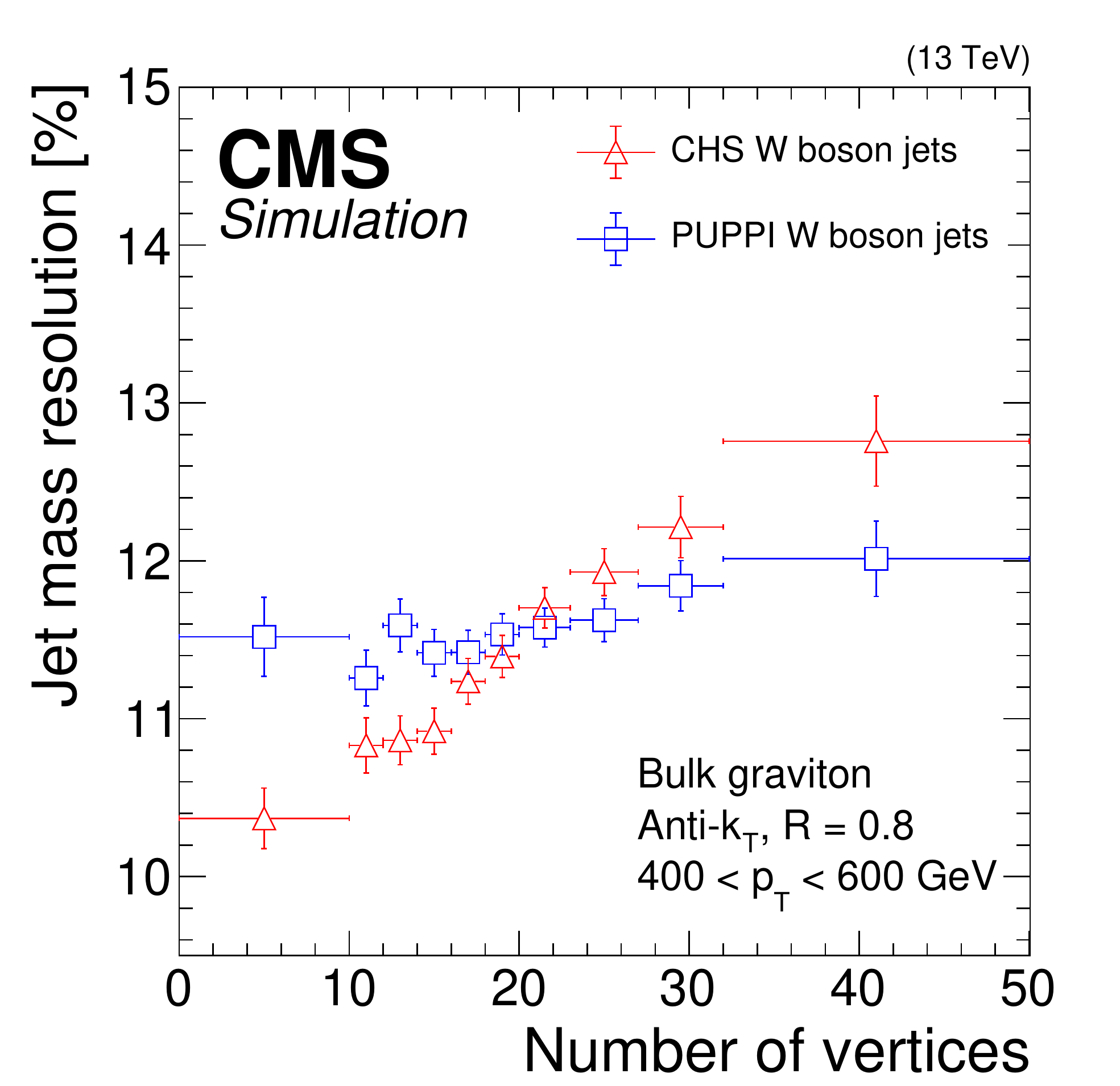}
    \caption{Median soft drop jet mass (upper left), median $\tau_{21}$ (upper right), and soft drop jet mass resolution (lower) for AK8 jets from boosted \PW bosons with $400<\pt<600 \GeV$ for CHS (red triangles) and PUPPI (blue squares) jets in a bulk graviton decaying to \PW{}\PW signal sample, as a function of the number of vertices. The error bars correspond to the statistical uncertainty in the simulation.   
  }
    \label{fig:ztag_scale_pu}
\end{figure}

The variation as a function of pileup of the median soft drop jet mass, median $\tau_{21}$, and the soft drop jet mass resolution is shown in \figref{fig:ztag_scale_pu} for jets from boosted \PW bosons with $400<\pt<600 \GeV$ using simulation of bulk gravitons decaying into $\PW\PW$ pairs.
The soft drop jet mass resolution is defined as the spread of the ratio of reconstruction- and particle-level jet mass (the response) divided by the mean of the response.
The response distribution is, to a very good approximation, Gaussian, and the resolution is determined using the same procedure as for the JER described in Section~\ref{sec_jet}.
The CHS jets exhibit a PU dependence for the soft drop jet mass and $\tau_{21}$ observables. The PUPPI jets, on the other hand, entirely remove the PU dependence of the soft drop jet mass and $\tau_{21}$ medians.
The soft drop jet mass resolution is similar for the CHS and PUPPI algorithms, though a slightly better resolution is observed for the CHS algorithm for fewer than 20 interactions, while the PUPPI algorithm shows less dependence on PU leading to an improved resolution for more than 30 interactions, for which it has been optimized.

\begin{figure}[hbtp]
  \centering
    \includegraphics[width=0.45\textwidth]{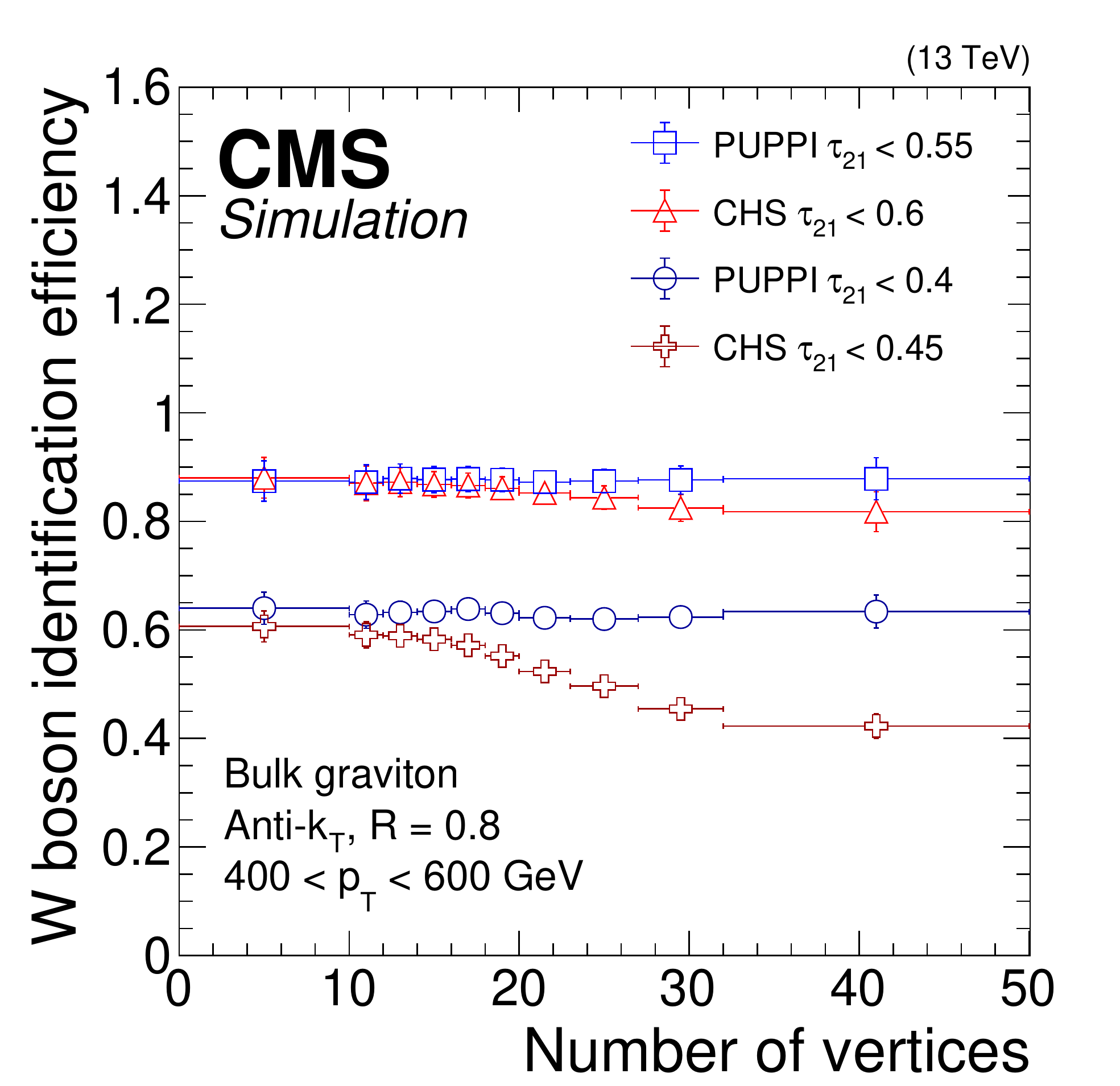}
    \includegraphics[width=0.45\textwidth]{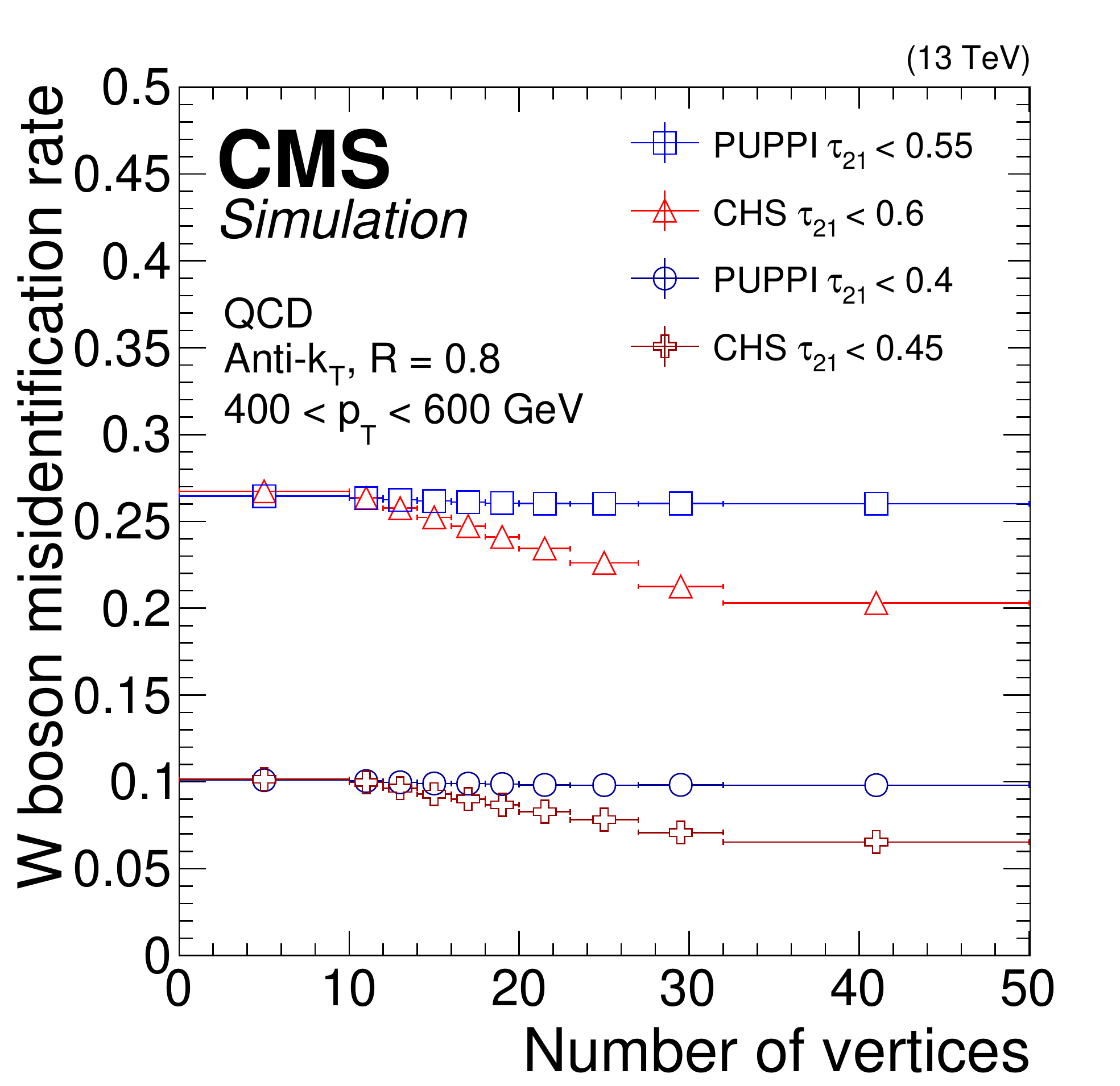}
    \caption{ \PW boson identification performance using a selection on $\tau _{21}$ for CHS (red triangles and dark red crosses) and PUPPI (blue squares and circles) AK8 jets as a function of the number of vertices for loose and tight selections, respectively. Shown on the left is the \PW boson identification efficiency evaluated in simulation for a bulk graviton decaying to a \PW{}\PW boson pair and on the right the misidentification rate evaluated with QCD multijet simulation. The error bars correspond to the statistical uncertainty in the simulation.}
    \label{fig:wtag_pu}
\end{figure}

The performance of a typical \PW or \PZ boson tagger with respect to the PU contamination is studied using
 simulation of bulk gravitons decaying into $\PW\PW$ pairs for tagging efficiency and QCD multijet production for misidentification rate.
Reconstructed jets are required to have \pt larger than 200\GeV and $\abs{\eta} < 2 $,
 and not to overlap with any well-reconstructed leptons.
In addition, jets are required to have reconstructed mass compatible with the \PW boson mass (within 65--105\GeV).
\figrefb{fig:wtag_pu}
 shows the evaluated efficiency and misidentification rate of the tagger
 with CHS and PUPPI jets operated at two cutoff values on $\tau_{21}$
(0.6 and 0.45 for CHS jets, and 0.55 and 0.40 for PUPPI jets, which give a comparable efficiency to that for CHS jets).
The tagger with PUPPI provides stable performance for both efficiency and misidentification rate,
 whereas the one with CHS reduces both efficiency and misidentification rate as the PU increases.
 This behavior of the tagger with CHS results from the linear dependence of median $\tau_{21}$ on the number of vertices for both \Pq/\Pg jets and \PW jets (see Fig.~\ref{fig:ztag_scale_pu}).

\begin{figure}[hbtp]
  \centering
    \includegraphics[width=0.45\textwidth]{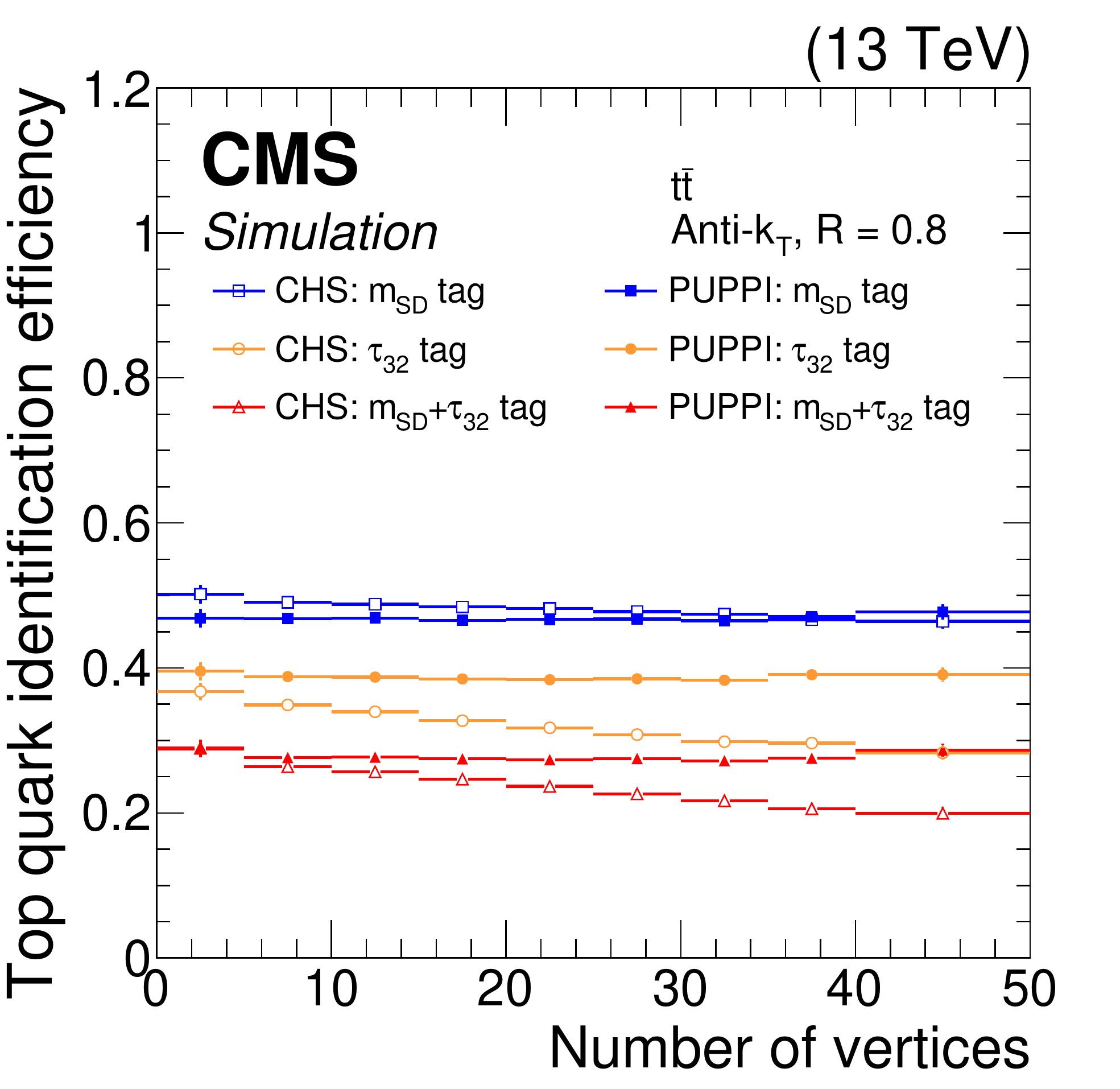}
    \includegraphics[width=0.45\textwidth]{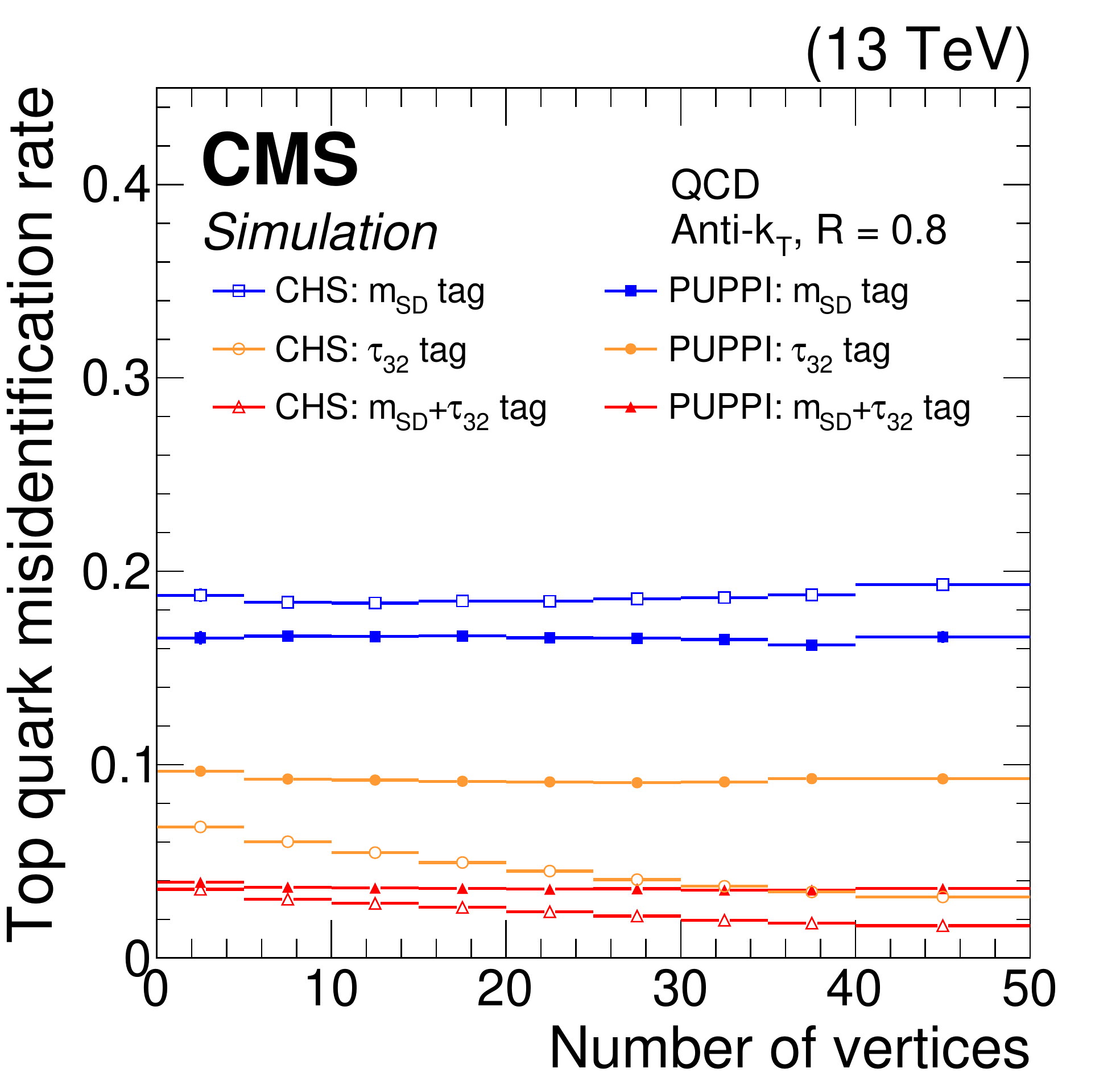}
    \caption{Top quark identification efficiency (left) and misidentification rate (right) as a function of the number of vertices for CHS (open symbols) and PUPPI (closed symbols) jets, using different combinations of substructure variables: soft drop mass cut between 105 and 210\GeV (blue rectangles),  $\tau_{32} < 0.54$ (orange circles), and both requirements together (red triangles). The error bars correspond to the statistical uncertainty in the simulation.}
    \label{fig:toptag_pu}
\end{figure}

The same stability of the PUPPI algorithm is seen in top quark jet identification, which is performed by selecting jets originating from top quarks in simulation that have a soft drop mass within 105--210\GeV and $\tau_{32} < 0.54$.
\figrefb{fig:toptag_pu}  shows the tagging performance using the CHS and PUPPI algorithms with the soft drop mass and $\tau _{32}$ conditions applied separately, and with both of them together. Although the efficiency is slightly different between the application of PUPPI or CHS, the same stability is observed with respect to PU as for \PW tagging.

The performance of the \PW boson tagger with the CHS and PUPPI algorithms is compared in data and simulation following the procedure described in Ref.~\cite{Khachatryan:2014vla}. 
The \PW boson identification efficiency is measured in a region enriched in \ttbar events, 
where one top quark decays to a final state with a lepton, neutrino, and a bottom quark and is used to tag the event. The other top quark is required to decay to a bottom quark and a \PW boson that further decays to a quark-antiquark pair. The AK8 jet with the highest \pt in the event is probed as the \PW boson jet candidate and required to have $\pt >200\GeV$ and $\abs{\eta} < 2.4$. Data collected by single-lepton triggers are compared with simulation samples of top quark pair production and backgrounds from single top, \PW boson, and diboson production.
The soft drop jet mass scale and resolution, as well as the $\tau_{21}$ selection efficiency with the CHS and PUPPI algorithms, are well modeled by the simulation.
The data-to-simulation scale factors for jet mass scale, jet mass resolution, and $\tau_{21}$ selection efficiency are found in Table~\ref{tab:wtagging}.
The leading systematic effects include parton showering and variations of the fit model (treatment of nearby jets) as detailed in Ref.~\cite{Khachatryan:2014vla}.

\begin{table}
\topcaption{Data-to-simulation scale factors for the jet mass scale, jet mass resolution, and the $\tau_{21}$ selection efficiency for the CHS and PUPPI algorithms.}
\begin{tabular}{ccc}
\multirow{2}{*}{Parameter} & \multicolumn{2}{c}{Data/simulation} \\
& CHS & PUPPI \\
\hline
Mass scale & $1.007 \pm 0.009\stat \pm 0.005\syst$ & $0.998 \pm 0.007\stat \pm 0.006\syst$  \\
Mass resolution & $1.15 \pm 0.04\stat \pm 0.04\syst$  &  $1.08 \pm 0.02\stat\pm 0.09\syst$ \\
$\tau_{21}<0.45$ & $1.00 \pm 0.06\stat \pm 0.07\syst$ & \NA \\
$\tau_{21}<0.4$ & \NA &  $1.01 \pm 0.06\stat \pm 0.05\syst$ \\
\end{tabular}
\label{tab:wtagging}
\end{table}

\section{Missing transverse momentum resolution}
\label{sec_met}

The imbalance of momentum for all reconstructed objects in the transverse plane, called missing transverse momentum \vptmiss with magnitude \ptmiss, 
 is a signature of neutrino production. It also plays an important role in searches for unknown stable neutral particles.
In CMS, \vptmiss is calculated as the negative vector \pt sum of all PF candidates (called PF \vptmiss in the following). The \vptmiss thus relies on the accurate measurement of the reconstructed physics objects, namely muons, electrons, photons, hadronically decaying taus, jets, and unclustered energy. The unclustered energy is the contribution from the PF candidates not associated with any of the previous physics objects.
The CHS procedure is not suitable for \vptmiss computation  since it selectively removes only particles within the tracker volume ($\abs{\eta}<2.5$). PU events that spread across the tracker volume boundary are thus partially removed leading to a degradation in the \vptmiss resolution. 
The \vptmiss is corrected with the difference between the vector \pt sum of all reconstructed jets in the event calibrated to the particle level and the vector sum of all uncalibrated jet momenta (called type-1 correction), to account for the detector response of jet objects.
Anomalous high-\ptmiss events can be due to a variety of reconstruction failures, detector malfunctions, or noncollision backgrounds. Such events are rejected by event filters that are designed to identify more than 85--90\% of the spurious high-\ptmiss events with a mistagging rate less than 0.1\%~\cite{Sirunyan:2019kia}.
The performance of the \vptmiss reconstruction in CMS (covering $\PZ \to \Pe\Pe$, $\PZ \to \PGm\PGm$ and $\gamma$+jets data samples) is discussed in detail in Ref.~\cite{Sirunyan:2019kia}.

The PUPPI algorithm can be used for the computation of \vptmiss by scaling the PF candidates by their PUPPI weight (PUPPI \vptmiss), and then applying the type-1 correction using PUPPI jets.
The PUPPI metric as defined in Eq.~\ref{eq:metricalpha} in \secref{sec_puppibasic} treats charged leptons and charged hadrons in the same way, i.e., charged leptons get a weight of 0 or 1 depending on their vertex association and enter into the computation of the weight of their surrounding particles.
This causes prompt leptons, e.g., leptons from the decay of the \PZ boson, to create a PU dependence by giving PU particles around the prompt lepton a higher weight.
Therefore, a second PUPPI metric is defined in which charged leptons are excluded from the calculation.
In this definition, it is assumed that all leptons in the event are prompt.
This results in PU particles surrounding a prompt lepton having a lower weight consistent with the PU hypothesis.
In the following discussion, the metric defined with the default PUPPI weight, including the leptons, is referred to as ``PUPPI-with-lepton'' and the metric, which excludes the leptons, as ``PUPPI-no-lepton.''
For the purpose of the PUPPI \vptmiss computation, PUPPI-no-lepton collection is combined with the collection of leptons given a weight of 1.
In addition, a PUPPI weight of 1 is automatically assigned to photons reconstructed in the tracker region ($\abs{\eta} < 2.5$) with $\pt > 20\GeV$. These photons are required to pass certain identification and isolation criteria ensuring an efficiency of above 80\% and a purity of above 95\%.

The resolution of \ptmiss is quantified by measuring
 the resolution of the hadronic recoil in \PZ boson events.
The recoil is defined as the vector sum of the momenta of all the objects (with the same PU mitigation applied as for \ptmiss) in the event but the \PZ boson.
The transverse momenta of the recoil and of the \PZ boson are balanced against each other,
 such that their difference
allows the determination of the momentum resolution.
The momentum of the \PZ boson decaying into charged leptons 
can be reconstructed with high resolution such that it can serve as a reference for the measurement of the energy resolution of the hadronic recoil.
The momentum of the recoil is projected to axes 
 parallel and perpendicular to the momentum of the reconstructed \PZ boson.
The resolution of the former is sensitive to the energy resolution and the latter to the PU contribution.

The $\Pp\Pp$ collision data collected with a dielectron trigger are used to evaluate the performance.
Events with two isolated electrons, within $\abs{\eta} < 2.5$, with the leading (subleading) electron $\pt > 25\, (20)\GeV$, and the invariant mass of the two electrons within a 20\GeV window centered around the \PZ boson mass are selected.
The four-momentum of the \PZ boson are reconstructed from the four-momentum of the two electrons.
The recoil is calculated as the vector sum of the momenta of all particles, but the two electrons.

\begin{figure}[hbtp]
  \centering
    \includegraphics[width=0.45\textwidth]{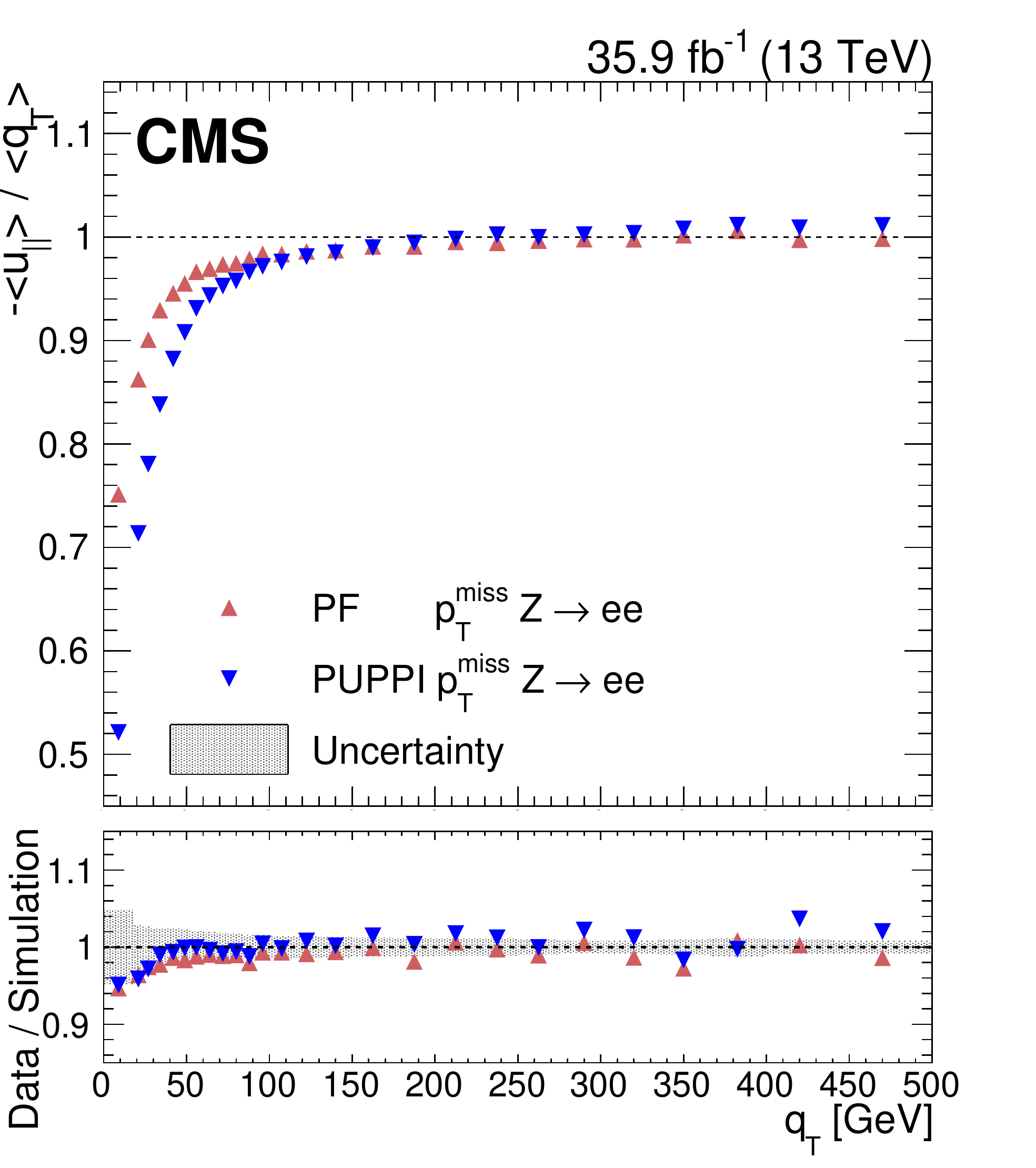}
    \caption{ 
    The hadronic recoil response ($-\langle u_{\parallel} \rangle/\langle q_\mathrm{T} \rangle$) of the \PZ boson computed for PUPPI and PF \ptmiss, as a function of $q_\mathrm{T}$ in \Zee events in collision data. The lower panel shows the  data-to-simulation ratio. A gray shaded band is added in the lower panel showing the systematic uncertainties resulting from jet energy scale and jet energy resolution variations, and variations in the unclustered energy added in quadrature.}
    \label{fig:met_responce}
\end{figure}

\figrefb{fig:met_responce}
 shows the ratio of the recoil to the \PZ boson transverse momentum ($u_{\parallel}$)
 as a function of the \PZ boson transverse momentum ($q_\mathrm{T}$) for PUPPI \vptmiss and PF \vptmiss.
The PUPPI \ptmiss tends to have a smaller response in events with low momentum recoil.
This is because of the removal of PF candidates that are wrongly assigned to the PU vertex by the PUPPI algorithm.
Deviations from unity indicate imperfect calibration of the hadronic energy scale.

\begin{figure}[hbtp]
  \centering
    \includegraphics[width=0.45\textwidth]{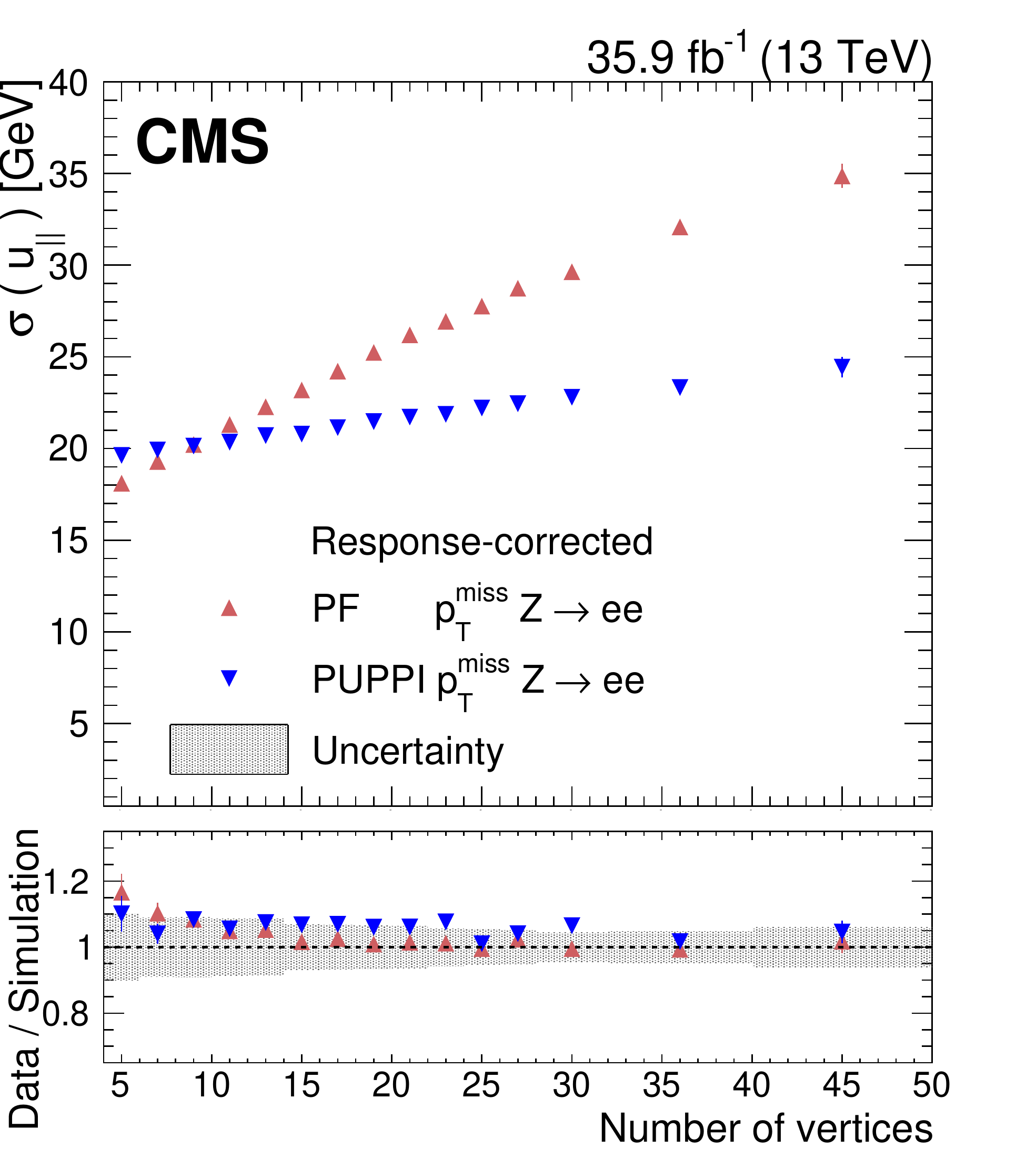}
    \includegraphics[width=0.45\textwidth]{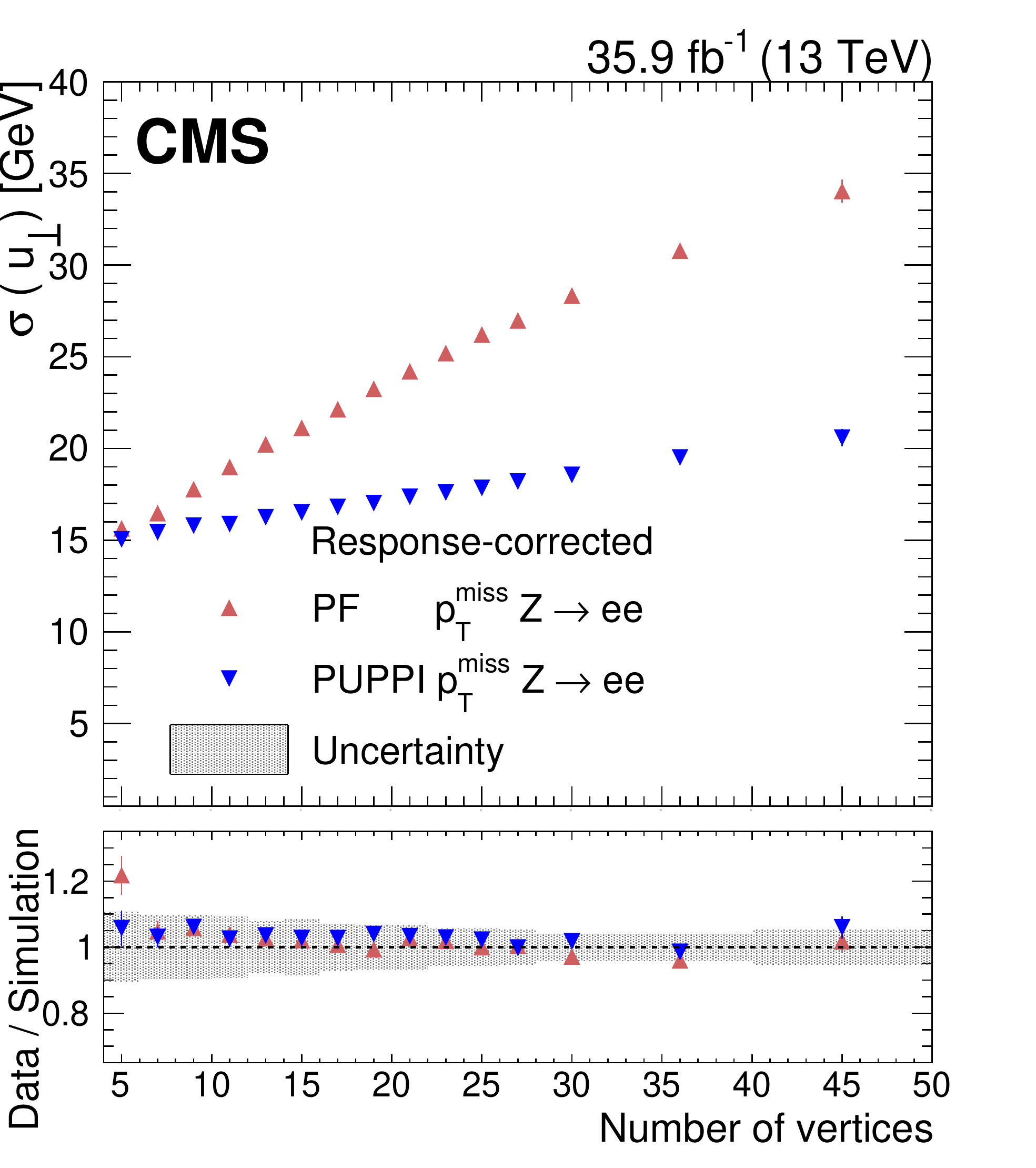}
   \caption{The hadronic recoil components  \upar{}(left) and \uper{}(right) for PUPPI and PF \ptmiss resolution as a function of the number of vertices in \Zee events in collision data. The lower panel shows the data-to-simulation ratio. A gray shaded band is added in the lower panel showing the systematic uncertainties resulting from jet energy scale and jet energy resolution variations, and variations in the unclustered energy added in quadrature.
}
    \label{fig:met_recoil}
\end{figure}

\figrefb{fig:met_recoil}
 shows the resolution of the recoil, parallel ($\sigma_{\parallel}$), and perpendicular ($\sigma _{\perp}$) to the \PZ boson momentum, as a function of the number of vertices
 for PUPPI \vptmiss and PF \vptmiss.
The scale of the recoil is corrected as a function of the \PZ boson momentum for comparison.
The PUPPI \vptmiss resolution for both components is consistently better than the PF \vptmiss resolution above a number of vertices of 10.
In addition, PUPPI \vptmiss provides a more stable performance with respect to PU than \vptmiss, up to at least 50 vertices.

\section{Muon isolation}
\label{muoniso}

Muons are reconstructed through a fit to hits in both the inner tracking system and the muon spectrometer~\cite{Sirunyan:2018fpa,Chatrchyan:2012xi}.
Muons must satisfy identification and reconstruction requirements on the impact parameters of the track, the number of hits reconstructed in both the silicon tracker and the muon detectors, and the uncertainty in the \pt measurement.
These quality criteria ensure a precise measurement of the four-momentum, and rejection of badly reconstructed muons.

To distinguish prompt charged leptons from those originating from semileptonic decays of hadrons, the  lepton isolation provides a powerful handle. Lepton isolation is defined as the \pt sum of all surrounding particles in a cone around the lepton. 
In this study, PUPPI is investigated in the context of muon isolation and compared with other techniques commonly used in CMS.
While not shown here, these techniques are also applicable to electron isolation.

Various techniques exist to limit the impact of PU on isolation. A widely used variable within CMS is the \deltabetacorrected isolation~\cite{Sirunyan:2018fpa}.
This variable is used to estimate the contribution of neutral particles based on the nearby contributions of charged particles, defined by:
\begin{linenomath}
\begin{equation}
\label{PUPPIMUONISOLATION}
\text{\deltabeta-Iso}^{\mu ^ i } = \sum_{ \Delta R(i,j) < 0.4 }^\text{Ch-LV} \pt^j +\max \left( 0 ,  \sum_{ \Delta R(i,j) < 0.4 }^\mathrm{Nh} \pt^j + \sum_{ \Delta R(i,j) < 0.4 }^\mathrm{Ph} \pt^j - \frac{1}{2} \sum_{ \Delta R(i,j) < 0.4 }^\text{Ch-PU} \pt^j \right) ,
\end{equation}
\end{linenomath}
where each sum runs over the particles, indexed with $j$, with $\Delta R < 0.4$ of the muon,
$\pt^j$ is the transverse momentum of each surrounding particle, Ch-LV and Ch-PU are charged particles associated with the 
LV and PU vertices, respectively,
and Nh and Ph are neutral hadrons and photons reconstructed with the PF algorithm, respectively.
The subtraction by one half of the amount of Ch-PU corresponds to the subtraction of the PU contamination. It is motivated by isospin symmetry, yielding the ratio of charged to neutral pion production of two, which is responsible for the fact that jets are composed of roughly one-third neutral pions and two-thirds charged pions~\cite{Khachatryan:2016kdb}.
An alternative isolation can be constructed using PUPPI. The simplest definition of PUPPI muon isolation 
 is:
\begin{linenomath}
\begin{equation}
\label{PUPPIMUONISOLATION2}
\text{Iso}^{\mu ^ i } = \sum_{ \Delta R(i,j) < 0.4 } \pt^j \omega ^j,
\end{equation}
\end{linenomath}
where $ \pt^{j} $ and $\omega ^ j $ are the transverse momentum and the PUPPI weight of particle $j$, respectively. The PUPPI weight is either determined from PUPPI-with-lepton or PUPPI-no-lepton as described in \secref{sec_met}.
 In addition, a combined isolation defined as the mean of the two isolation quantities is referred as ``PUPPI-combined'':
\begin{equation}
\text{Iso}_{\text{combined}}  = \frac{\text{Iso}_{\text{no-lepton }} + \text{Iso}_{\text{with-lepton}}}{2}.
\end{equation}
The performance of muon isolation is tested using simulated \PZ boson (prompt muons) and QCD multijet (nonprompt muons) events with a PU distribution having a mean of 20 interactions comparable to the 2016 PU conditions.
For comparison, the relative isolation algorithm, defined as the isolation divided by the muon \pt, is used. Muons are selected if the relative isolation is below a certain threshold.
The threshold value for the relative isolation (0.156 for PUPPI-combined and 0.15 for \deltabetacorrected) is defined such that each isolation quantity 
 gives an inclusive misidentification rate of 12\% for the muons selected in QCD multijet simulation.
The fraction of muons passing the criteria is referred to as isolation efficiency for prompt muons and as misidentification rate for nonprompt muons.
The efficiency is calculated with respect to reconstructed prompt muons with $\pt >20\GeV$ and $\abs{\eta} < 2.4$.
 
As explained before, PUPPI-with-lepton has the shortcoming that PU particles around a prompt lepton get too high a weight because of the \pt of the lepton in the $\alpha_i$ calculation. Therefore, the application of the weights from PUPPI-with-lepton for the muon isolation leads to a PU-dependent efficiency for prompt muons and a PU-independent misidentification rate.
The misidentification rate is PU-independent, because LV particles, which drive the isolation of nonprompt leptons, get a reasonable weight.
Conversely, PUPPI-no-lepton has the shortcoming that LV particles near a nonprompt lepton get a reduced weight because the \pt of the nonprompt lepton is excluded when calculating $\alpha_i$ for these particles. The weight of LV particles contributing to the isolation is thus less stable against their surroundings. PU particles around leptons, however, get reasonable weights, resulting in a good estimate of the isolation for prompt leptons. Therefore, using PUPPI-no-lepton for the isolation calculation yields a stable efficiency and a less PU-resilient misidentification rate.
 
\begin{figure}[hbtp]
  \centering
     \includegraphics[width=0.45\textwidth]{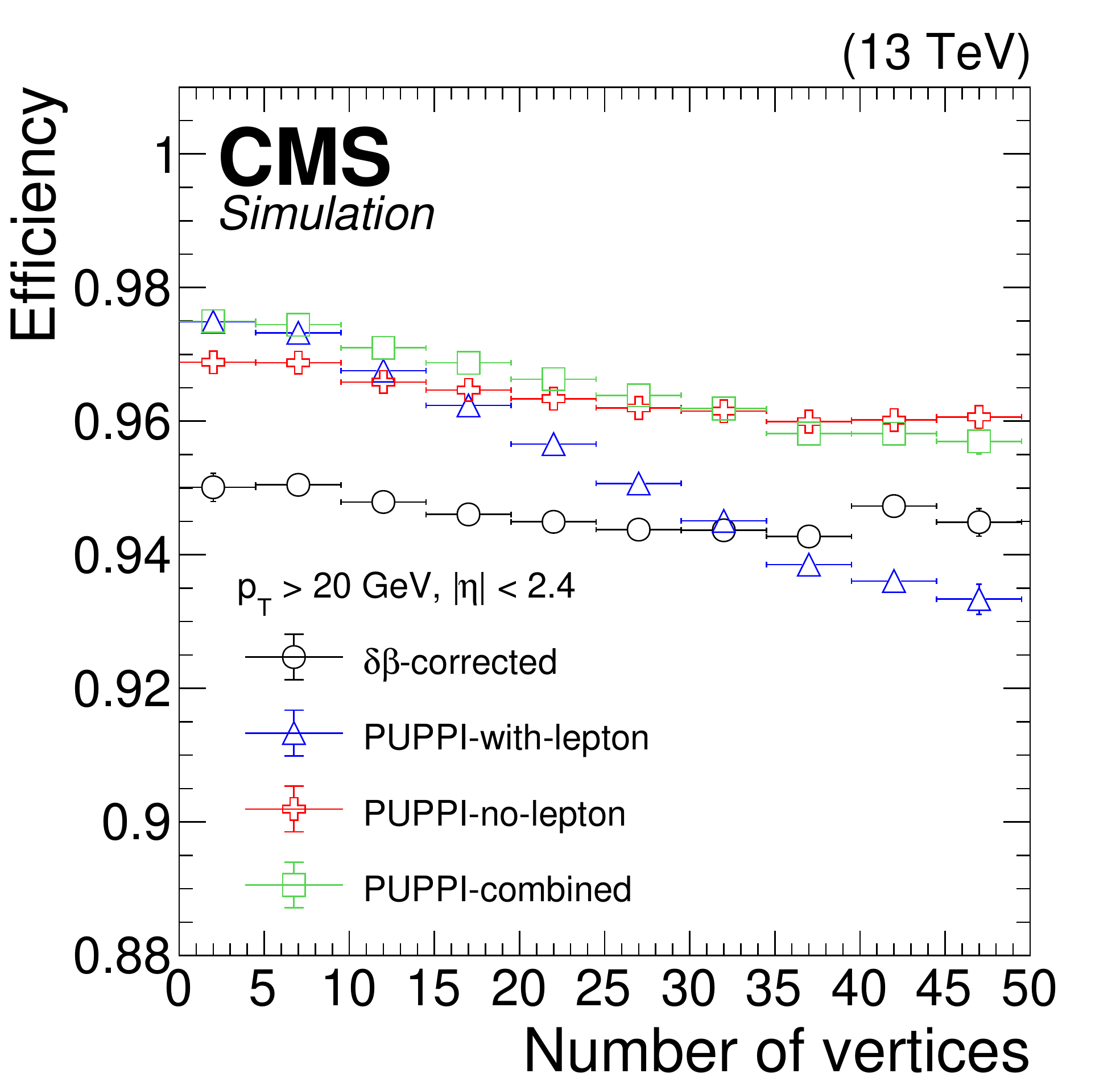}
     \includegraphics[width=0.45\textwidth]{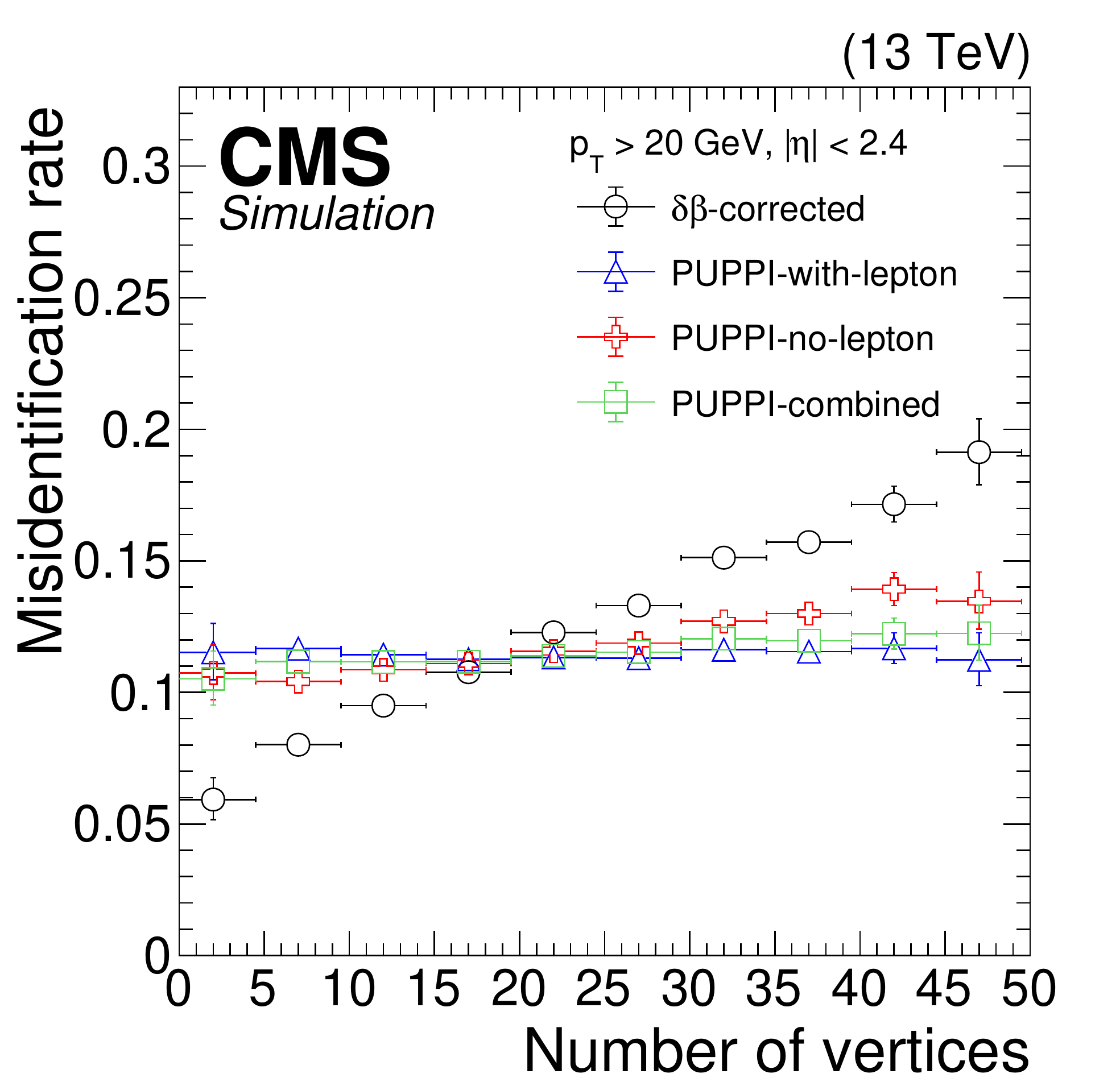}
    \caption{The identification efficiency for prompt muons in simulated \PZ{}+jets events (left) and the misidentification rate for nonprompt muons in QCD multijet simulated events (right) for the different definitions of the isolation: \deltabetacorrected isolation (black circles), PUPPI-with-lepton (blue triangles), PUPPI-no-lepton (red crosses), PUPPI-combined (green squares), as a function of the number of vertices.
The threshold of each isolation is set to yield a 12\% misidentification rate for reconstructed muons in QCD multijet simulation. The error bars correspond to the statistical uncertainty in the simulation.}
    \label{fig:muoniso_performanceNPV}
\end{figure}

\figrefb{fig:muoniso_performanceNPV}
 shows the isolation efficiency and the misidentification rate as a function of the number of vertices.
All three PUPPI isolation quantities are observed to be more stable across PU when compared with the \deltabetacorrected isolation in terms of misidentification rate. In terms of efficiency, the PUPPI-no-lepton shows a more stable behavior compared with \deltabetacorrected isolation whereas PUPPI-with-lepton shows a stronger dependence on the number of vertices. The stability of the PUPPI-combined isolation efficiency is between the two PUPPI isolation variants and similar to the \deltabetacorrected isolation.

\begin{figure}[hbtp]
  \centering
   \includegraphics[width=0.45\textwidth]{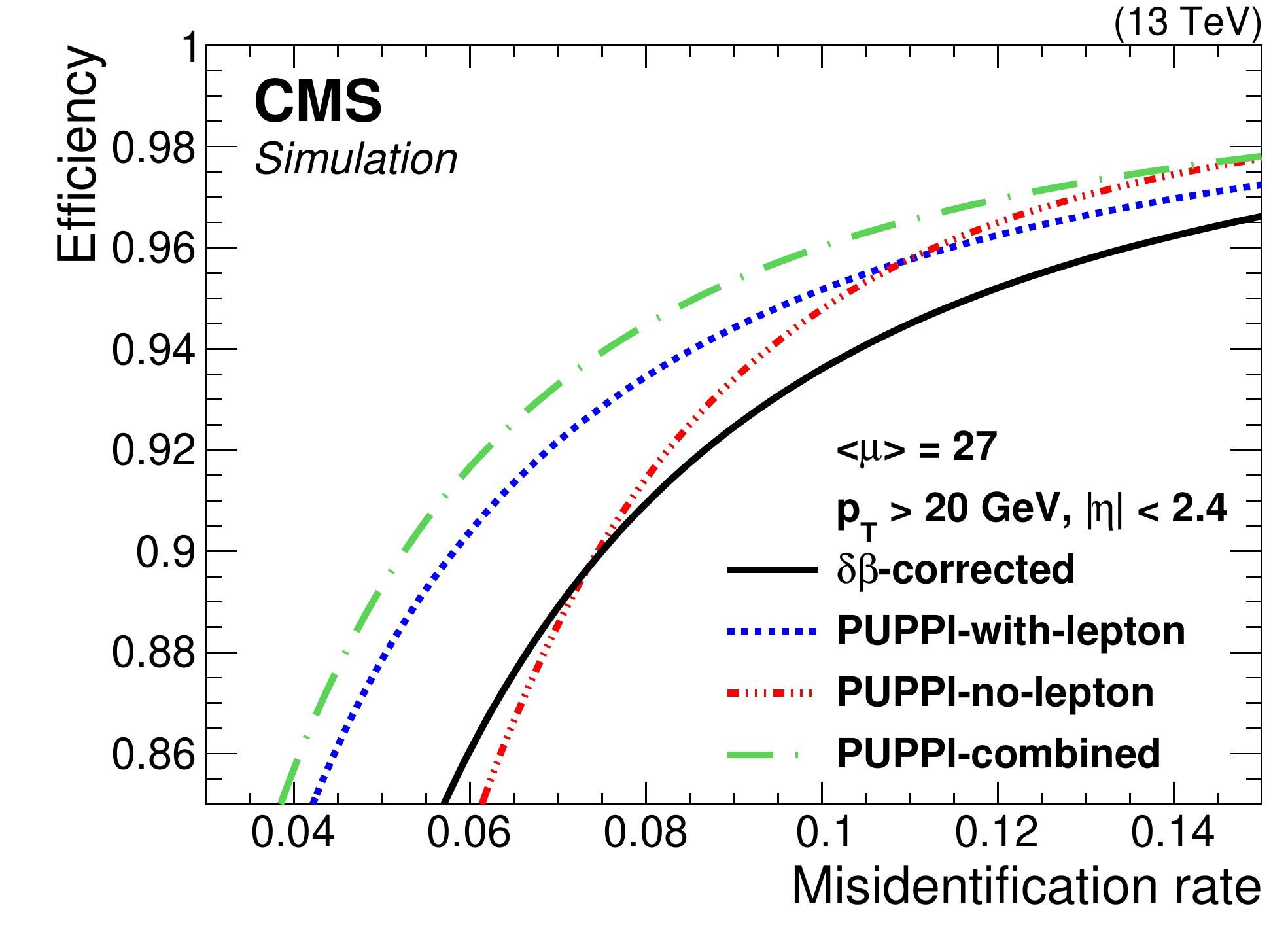}
    \caption{The identification efficiency for prompt muons in simulated \PZ{}+jets events as a function of the misidentification rate for nonprompt muons in QCD multijet simulated events for the different definitions of the isolation: \deltabetacorrected isolation (black solid line), PUPPI-with-lepton (blue dashed line), PUPPI-no-lepton (red mixed dashed), PUPPI-combined (green long mixed dashed). The average number of interactions is 27.}
    \label{fig:muoniso_ROC}
\end{figure}

\figrefb{fig:muoniso_ROC}
 shows a receiver operating characteristic (ROC) curve, \ie, the efficiency as a function of the misidentification rate, when using different definitions of the isolation. 
The combined PUPPI isolation provides the best performance over the typical analysis working points. 

The PUPPI isolation is further investigated in collision data collected with a single-muon trigger path requiring an isolated muon with $\pt > 24 \GeV$.
 Two levels of muons are defined:
  loose muons are required to have $\pt > 15 \GeV$ and $\abs{\eta} < 2.4$ with no isolation requirement
 and tight muons $\pt > 26 \GeV$ and $\abs{\eta} < 2.1$ with a \deltabetacorrected isolation corresponding to an efficiency of 95\%  (threshold of 0.15).
One tight and one loose muon, with the invariant mass of the two muons within a 10\GeV window centered around the \PZ boson mass are selected.
The performance is measured using a tag-and-probe method, with the tight muon as the tag muon and the loose muon as the probe muon.
 The behavior of the isolation variables in data are compared with \PZ{}+jets simulation.  Other backgrounds are neglected.

\begin{figure}[hbtp]
  \centering
    \includegraphics[width=0.45\textwidth]{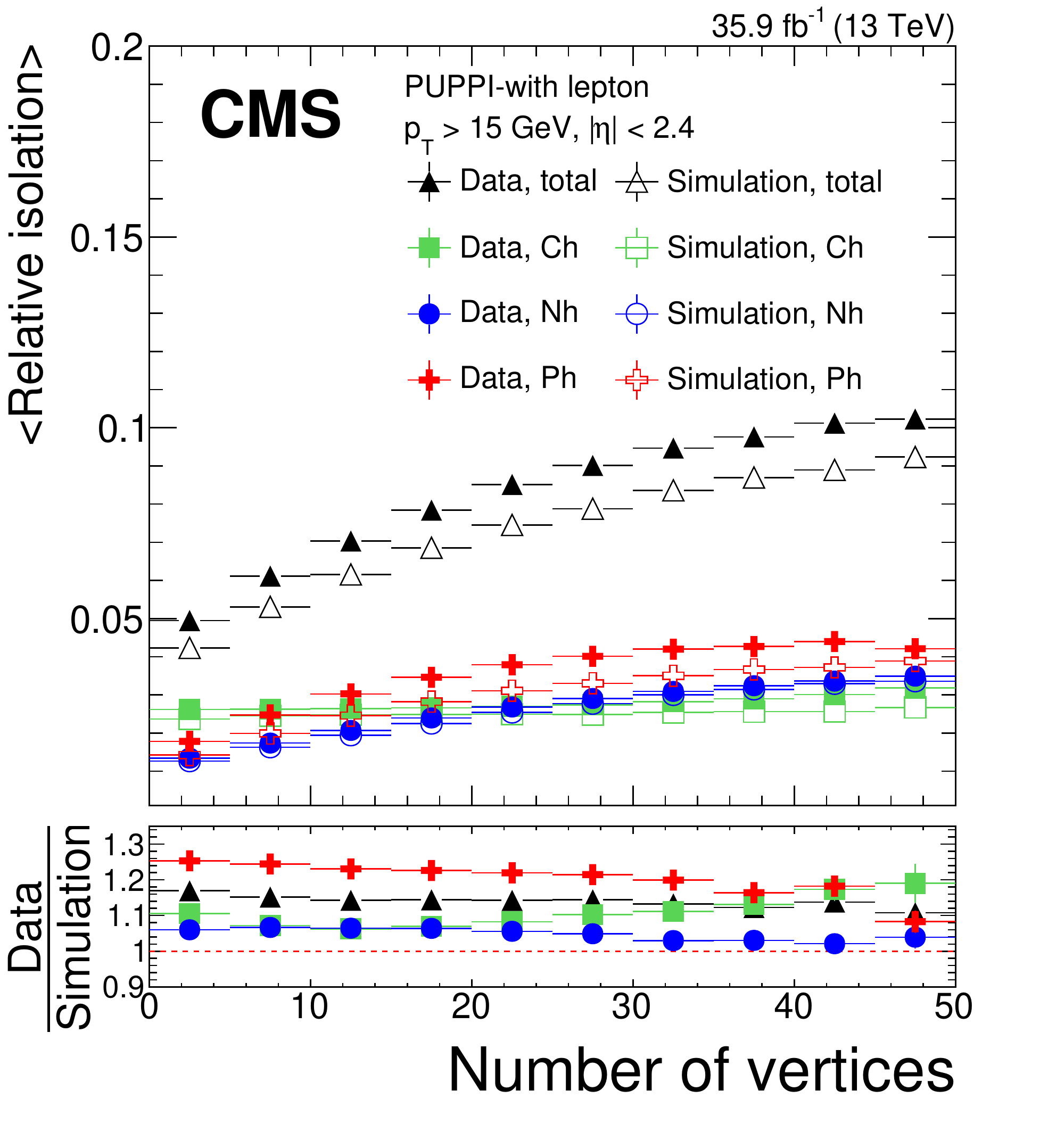}
    \includegraphics[width=0.45\textwidth]{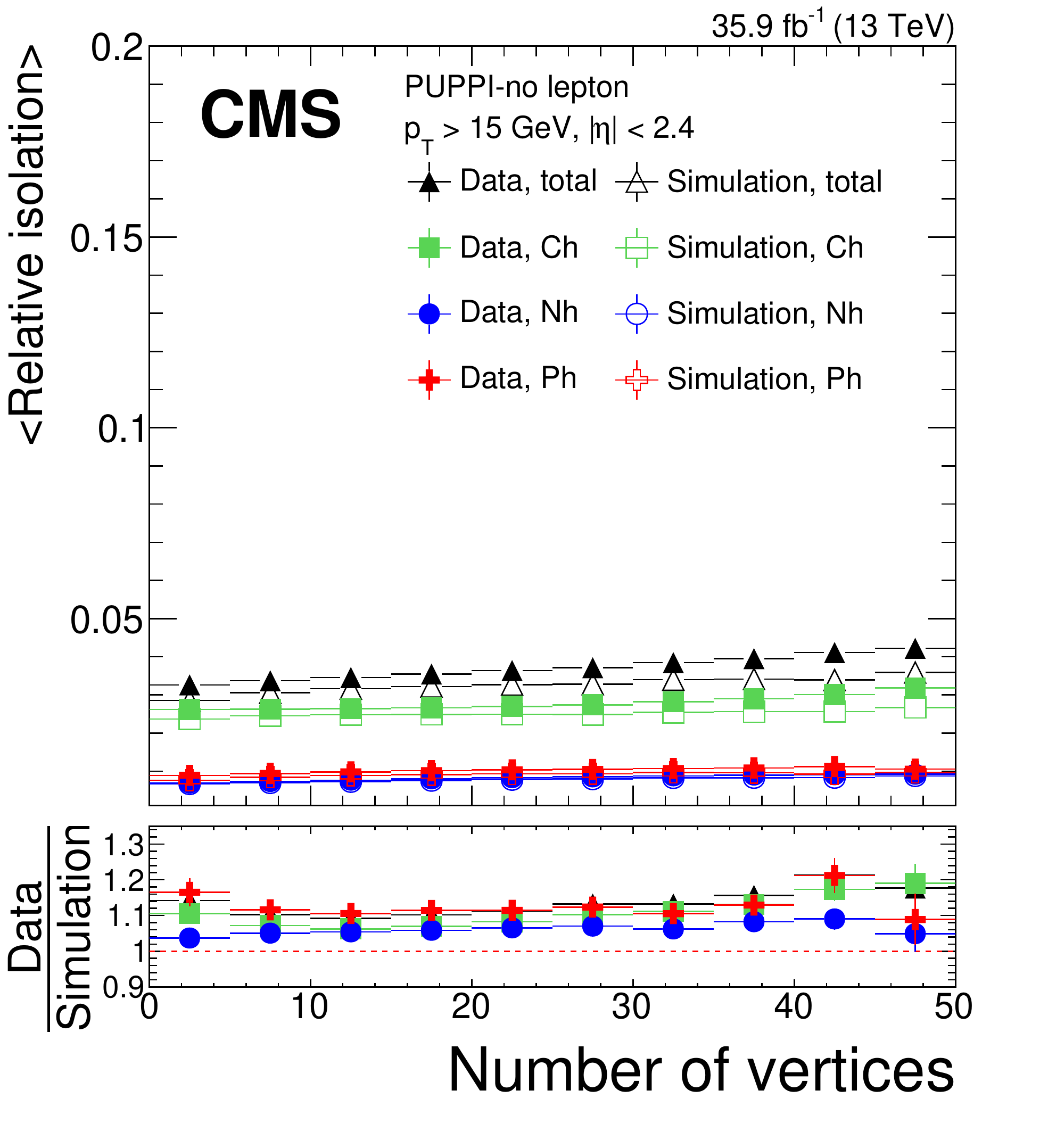}
    \caption{Mean relative isolation for PUPPI-with-lepton (left) and PUPPI-no-lepton (right) in data compared to \PZ{}+jets simulation. The relative isolation is split into separate charged hadron (Ch, green squares), neutral hadron (Nh, blue circles), photon (Ph, red crosses) components, and combined (black triangles). 
Data and simulation are shown using full and open markers, respectively.
    The lower panels show the data-to-simulation ratio of each component.
    The error bars correspond to the statistical uncertainty.
}
    \label{fig:muonisoProfile}
\end{figure}

\figrefb{fig:muonisoProfile} shows the mean fractions of the contributions 
 of charged hadrons, neutral hadrons, and photons to the relative isolation variable,
as a function of the number of vertices for the two types of PUPPI isolation
in data and \PZ{}+jets simulation.
The neutral hadrons and photons make up a large contribution to the total isolation and show a clear PU dependence for the PUPPI-with-lepton isolation, whereas this is not the case for the PUPPI-no-lepton isolation.
The trend in data is well described by simulation.

The isolation efficiency of the PUPPI-combined isolation is evaluated using the same tag-and-probe method,
 and is compared to the \deltabetacorrected isolation.
The threshold for PUPPI-combined isolation (0.15) is chosen such that the isolation efficiencies are roughly equal
 for muons with $15 < \pt < 20 \GeV$, where \deltabetacorrected isolation is applied.

\begin{figure}[hbtp]
  \centering
    \includegraphics[width=10cm,keepaspectratio]{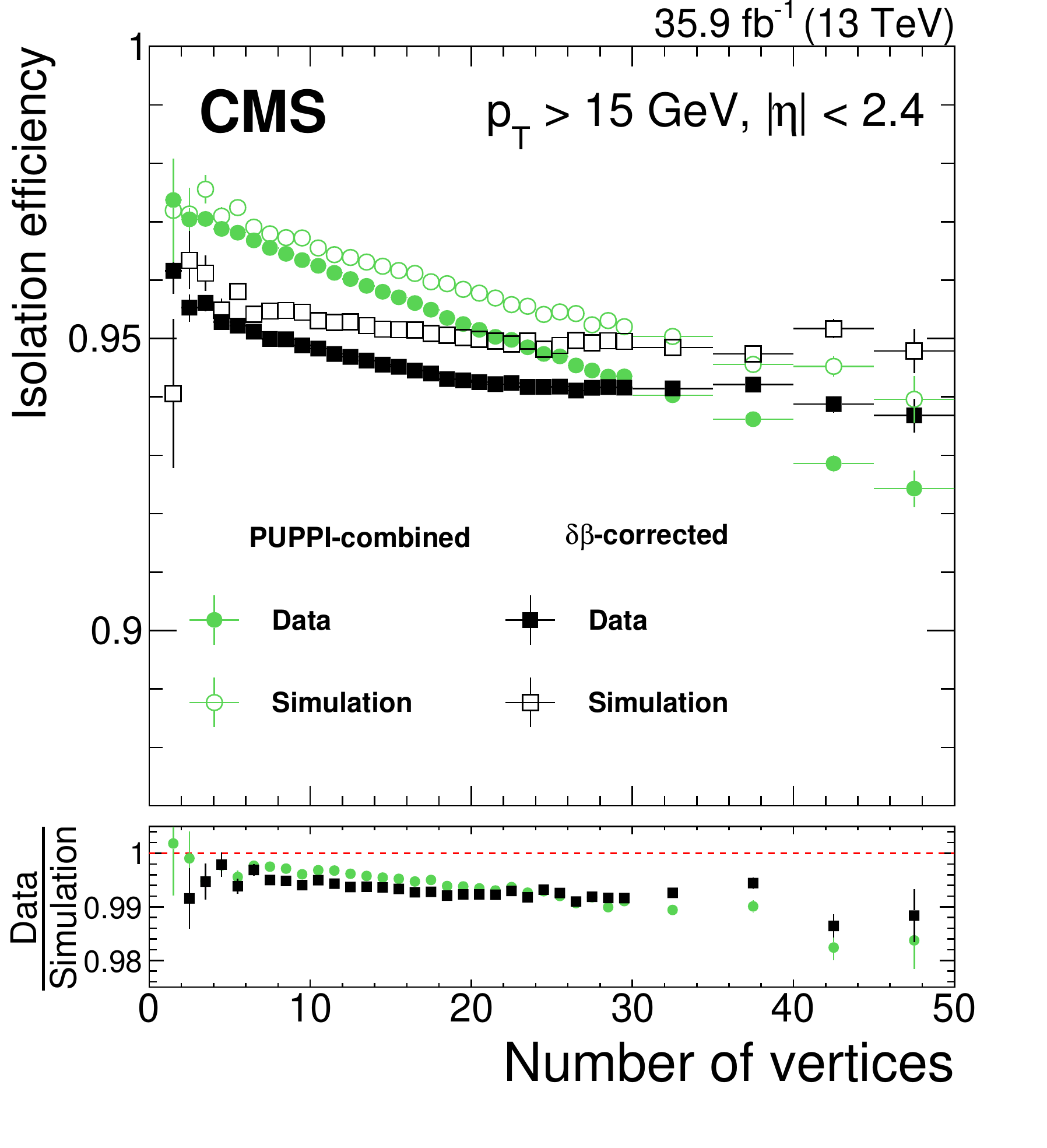}
    \caption{ 
The identification efficiency for prompt muon isolation selection in \Zmumu events in data compared to \PZ{}+jets simulation, as a function of the number of vertices for PUPPI-combined (green circles) and \deltabetacorrected isolation (black squares).
Data and simulation are shown using full and open markers, respectively.
The lower panel shows the data-to-simulation ratio. 
The error bars correspond to the statistical uncertainty.
The threshold for PUPPI-combined isolation (0.15) is chosen such that the isolation efficiencies are roughly equal for muons with $15 < \pt < 20\GeV$, where \deltabetacorrected isolation is applied, 
leading to an approximately 1\% higher efficiency for $\pt > 15 \GeV$ with variations as a function of the number of vertices.
    }
    \label{fig:muon_iso_in_data}
\end{figure}

\begin{figure}[hbtp]
  \centering
    \includegraphics[width=7cm,keepaspectratio]{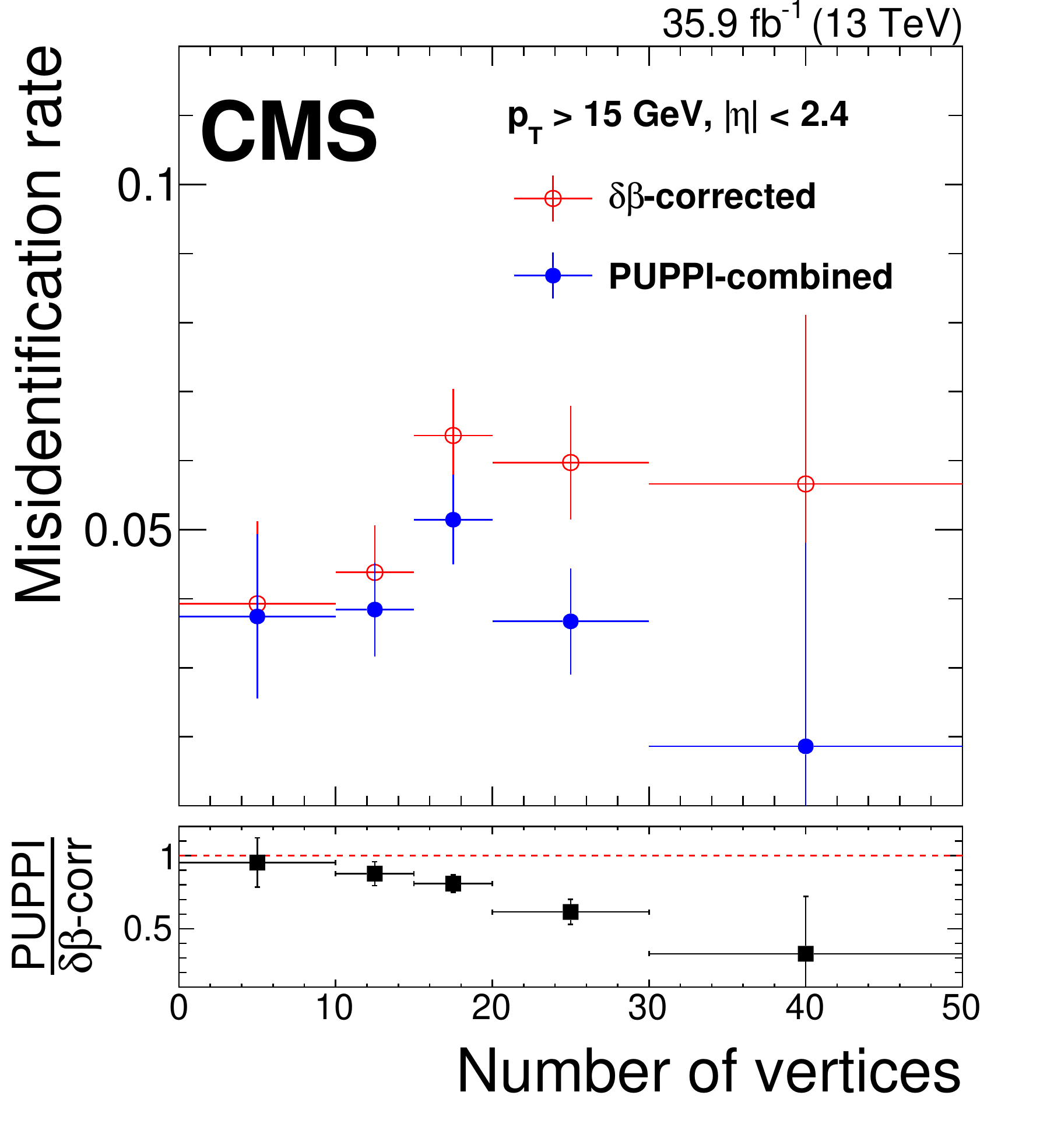}
    \caption{The misidentification rate defined as
 the number of events with a third isolated muon divided
 by the total number of events with a third muon in \Zmumu data for PUPPI-combined (blue closed circles) and \deltabetacorrected isolation (red open circles). 
 The lower panel shows the ratio of PUPPI-combined and \deltabetacorrected isolation, taking the correlation of their uncertainties into account.
 The threshold for PUPPI-combined isolation (0.15) is chosen such that the isolation efficiencies are roughly equal for muons with $15 < \pt < 20\GeV$, where \deltabetacorrected isolation is applied.
    }
    \label{fig:muon_fake_in_data}
\end{figure}

\figrefb{fig:muon_iso_in_data} shows the efficiency of the chosen PUPPI and \deltabetacorrected isolation variables as a function of the number of vertices.
 The ratio of efficiency in data to that in simulation is 0.99.
Although the PU dependence of the efficiency of the PUPPI-combined isolation is 
 stronger than that of the \deltabetacorrected isolation,
 this does not mean PUPPI-combined isolation is more susceptible to PU, because the misidentification rate is stable against PU (see \figref{fig:muon_fake_in_data}).
The PUPPI-combined isolation outperforms \deltabetacorrected isolation across the PU conditions studied.

The misidentification rate of the PUPPI isolation is evaluated in data by selecting \Zmumu events passing a dimuon trigger path ($\pt > 17$ and $ 8 \GeV$ for the leading and subleading muons, respectively). 
To obtain the \PZ boson candidates, two oppositely charged muons are selected within a 10\GeV window centered around the \PZ boson mass and passing loose isolation criteria. 
In addition to the two muons from the \PZ boson decay, a third muon is required and labeled as the misidentified muon.
This additional muon is either a third prompt muon initiated by leptonic decays of \PW{}\PZ and \PZ{}\PZ 
processes or, as is usually the case, a nonprompt muon from a semileptonic hadron decay. 
To further reduce the prompt-muon contribution from \PW{}\PZ production, the transverse mass (as defined in Ref.~\cite{Sirunyan:2019kia}) obtained from the muon with third-highest \pt and \ptmiss needs to be less than 40\GeV.
Both \PW{}\PZ and \PZ{}\PZ production are well measured and generally well modeled. The difference in agreement between data and simulation is thus ascribed to nonprompt-lepton events.

 The misidentification rate shown in \figref{fig:muon_fake_in_data}  is defined as
 the number of events with a third isolated muon divided
 by the total number of events after subtracting the background.
 The misidentification rate of the \deltabetacorrected isolation is 
$(5.4 \pm 0.4)\%$ while that of PUPPI-combined isolation is $(4.2 \pm 0.4)\%$.
The uncertainty is statistical only.
The ratio of the misidentification rate of PUPPI isolation to the \deltabetacorrected isolation is $(77 \pm 4)\%$, where the correlation is included in the uncertainty computation. 
The performance improvements from PUPPI-combined isolation expected from simulation studies are thus confirmed by data measurements.

\section{Summary}

The impact of pileup (PU) mitigation techniques on object reconstruction performance in the CMS experiment has been presented.
The main techniques under study are charged-hadron subtraction (CHS) and pileup per particle identification (PUPPI), which both exploit particle-level information. The performance of these techniques is evaluated in the context of the reconstruction of jets and missing transverse momentum (\ptmiss),  lepton isolation, and the calculation of jet substructure observables for boosted object tagging. The CHS and PUPPI algorithms are further compared with other algorithmic approaches that act on jet, \ptmiss, and lepton objects. While CHS rejects charged particles associated with PU vertices, PUPPI applies a more stringent selection to charged particles and rescales the four-momentum of neutral particles according to their probability to originate from the leading vertex.
Both techniques reduce the dependence on PU interactions across all objects.
A stronger reduction is achieved with PUPPI, especially for events with more than 30 interactions.
The PUPPI algorithm provides the best performance for jet mass and substructure observables, \ptmiss resolution, and rejection of misidentified muons.
With respect to jet-momentum resolution and PU jet rejection, the preferred algorithm depends on the physics process under study:
the PUPPI algorithm provides a better jet momentum resolution for jets with $\pt<100\GeV$, whereas CHS does so for $\pt>100\GeV$.
The highest rejection rate for jets originating purely from PU is obtained when using a dedicated PU jet identification in addition to CHS. However, when a looser working point for the PU jet identification is chosen such that its efficiency for selecting jets coming from the leading vertex is similar to that of PUPPI, both provide a similar rejection power.
The PU suppression techniques studied in this paper are proven to maintain reasonable object performance up to 70 interactions. Their use will be crucial for future running of the LHC, where even more challenging PU conditions up to 200 interactions per bunch crossing are expected.

\begin{acknowledgments}
We congratulate our colleagues in the CERN accelerator departments for the excellent performance of the LHC and thank the technical and administrative staffs at CERN and at other CMS institutes for their contributions to the success of the CMS effort. In addition, we gratefully acknowledge the computing centers and personnel of the Worldwide LHC Computing Grid for delivering so effectively the computing infrastructure essential to our analyses. Finally, we acknowledge the enduring support for the construction and operation of the LHC and the CMS detector provided by the following funding agencies: BMBWF and FWF (Austria); FNRS and FWO (Belgium); CNPq, CAPES, FAPERJ, FAPERGS, and FAPESP (Brazil); MES (Bulgaria); CERN; CAS, MoST, and NSFC (China); COLCIENCIAS (Colombia); MSES and CSF (Croatia); RPF (Cyprus); SENESCYT (Ecuador); MoER, ERC IUT, PUT and ERDF (Estonia); Academy of Finland, MEC, and HIP (Finland); CEA and CNRS/IN2P3 (France); BMBF, DFG, and HGF (Germany); GSRT (Greece); NKFIA (Hungary); DAE and DST (India); IPM (Iran); SFI (Ireland); INFN (Italy); MSIP and NRF (Republic of Korea); MES (Latvia); LAS (Lithuania); MOE and UM (Malaysia); BUAP, CINVESTAV, CONACYT, LNS, SEP, and UASLP-FAI (Mexico); MOS (Montenegro); MBIE (New Zealand); PAEC (Pakistan); MSHE and NSC (Poland); FCT (Portugal); JINR (Dubna); MON, RosAtom, RAS, RFBR, and NRC KI (Russia); MESTD (Serbia); SEIDI, CPAN, PCTI, and FEDER (Spain); MOSTR (Sri Lanka); Swiss Funding Agencies (Switzerland); MST (Taipei); ThEPCenter, IPST, STAR, and NSTDA (Thailand); TUBITAK and TAEK (Turkey); NASU (Ukraine); STFC (United Kingdom); DOE and NSF (USA). 
  
\hyphenation{Rachada-pisek} Individuals have received support from the Marie-Curie program and the European Research Council and Horizon 2020 Grant, contract Nos.\ 675440, 752730, and 765710 (European Union); the Leventis Foundation; the A.P.\ Sloan Foundation; the Alexander von Humboldt Foundation; the Belgian Federal Science Policy Office; the Fonds pour la Formation \`a la Recherche dans l'Industrie et dans l'Agriculture (FRIA-Belgium); the Agentschap voor Innovatie door Wetenschap en Technologie (IWT-Belgium); the F.R.S.-FNRS and FWO (Belgium) under the ``Excellence of Science -- EOS" -- be.h project n.\ 30820817; the Beijing Municipal Science \& Technology Commission, No. Z191100007219010; the Ministry of Education, Youth and Sports (MEYS) of the Czech Republic; the Deutsche Forschungsgemeinschaft (DFG) under Germany’s Excellence Strategy -- EXC 2121 ``Quantum Universe" -- 390833306; the Lend\"ulet (``Momentum") Program and the J\'anos Bolyai Research Scholarship of the Hungarian Academy of Sciences, the New National Excellence Program \'UNKP, the NKFIA research grants 123842, 123959, 124845, 124850, 125105, 128713, 128786, and 129058 (Hungary); the Council of Science and Industrial Research, India; the HOMING PLUS program of the Foundation for Polish Science, cofinanced from European Union, Regional Development Fund, the Mobility Plus program of the Ministry of Science and Higher Education, the National Science Center (Poland), contracts Harmonia 2014/14/M/ST2/00428, Opus 2014/13/B/ST2/02543, 2014/15/B/ST2/03998, and 2015/19/B/ST2/02861, Sonata-bis 2012/07/E/ST2/01406; the National Priorities Research Program by Qatar National Research Fund; the Ministry of Science and Education, grant no. 14.W03.31.0026 (Russia); the Tomsk Polytechnic University Competitiveness Enhancement Program and ``Nauka" Project FSWW-2020-0008 (Russia); the Programa Estatal de Fomento de la Investigaci{\'o}n Cient{\'i}fica y T{\'e}cnica de Excelencia Mar\'{\i}a de Maeztu, grant MDM-2015-0509 and the Programa Severo Ochoa del Principado de Asturias; the Thalis and Aristeia programs cofinanced by EU-ESF and the Greek NSRF; the Rachadapisek Sompot Fund for Postdoctoral Fellowship, Chulalongkorn University and the Chulalongkorn Academic into Its 2nd Century Project Advancement Project (Thailand); the Kavli Foundation; the Nvidia Corporation; the SuperMicro Corporation; the Welch Foundation, contract C-1845; and the Weston Havens Foundation (USA).
\end{acknowledgments}

\bibliography{auto_generated}   
\appendix

\cleardoublepage \appendix\section{The CMS Collaboration \label{app:collab}}\begin{sloppypar}\hyphenpenalty=5000\widowpenalty=500\clubpenalty=5000\vskip\cmsinstskip
\textbf{Yerevan Physics Institute, Yerevan, Armenia}\\*[0pt]
A.M.~Sirunyan$^{\textrm{\dag}}$, A.~Tumasyan
\vskip\cmsinstskip
\textbf{Institut f\"{u}r Hochenergiephysik, Wien, Austria}\\*[0pt]
W.~Adam, F.~Ambrogi, T.~Bergauer, M.~Dragicevic, J.~Er\"{o}, A.~Escalante~Del~Valle, M.~Flechl, R.~Fr\"{u}hwirth\cmsAuthorMark{1}, M.~Jeitler\cmsAuthorMark{1}, N.~Krammer, I.~Kr\"{a}tschmer, D.~Liko, T.~Madlener, I.~Mikulec, N.~Rad, J.~Schieck\cmsAuthorMark{1}, R.~Sch\"{o}fbeck, M.~Spanring, W.~Waltenberger, C.-E.~Wulz\cmsAuthorMark{1}, M.~Zarucki
% ...
\vskip\cmsinstskip
\dag: Deceased\\
1:  Also at Vienna University of Technology, Vienna, Austria\\
% ...
\end{sloppypar}
\end{document}